\documentclass[a4paper,10pt]{article}
 \usepackage{graphicx}
 \usepackage{authblk}
\usepackage{todonotes}
\usepackage{multirow}
 \usepackage{cleveref}
\newcommand{\etal}{{\em et al.}}

\usepackage{epstopdf}

\newcommand{\be}{\begin{equation}}
\newcommand{\ee}{\end{equation}}

\usepackage{vmargin}

\setpapersize{A4}
\setmargins{2cm}       % margen izquierdo
{1.5cm}                        % margen superior
{17.5cm}                      % anchura del texto
{23.42cm}                    % altura del texto
{10pt}                           % altura de los encabezados
{1cm}                           % espacio entre el texto y los encabezados
{0pt}                             % altura del pie de página
{2cm}                           % espacio entre el texto y el pie de página

\begin{document}

\title{Dynamic of plumes and scaling during the melting of a Phase Change Material heated from below}

\author[1]{Santiago Madruga}

 %\email[]{santiago.madruga@upm.es}
%\affiliation{E.T.S.I. Aeron\'autica y del Espacio, Universidad Polit\'ecnica de Madrid, Plaza Cardenal Cisneros 3, 28040 Madrid, Spain}

\author[2,3]{Jezabel Curbelo}
%\email[]{jezabel.curbelo@uam.es}
%\affiliation{Departamento de Matem\'aticas, Facultad de Ciencias, Universidad Auton\'oma de Madrid, 28049 Madrid, Spain}
%\affiliation{Instituto de Ciencias Matem\'aticas, CSIC-UAM-UC3M-UCM. C/ Nicol\'as Cabrera 15, Campus de Cantoblanco UAM, 28049 Madrid, Spain.}

\affil[1]{\footnotesize E.T.S.I. Aeron\'autica y del Espacio, Universidad Polit\'ecnica de Madrid, Plaza 
Cardenal Cisneros 3, 28040 Madrid, Spain}
\affil[2]{\footnotesize Departamento de Matem\'aticas, Facultad de Ciencias, Universidad Auton\'oma de Madrid, 28049 Madrid, Spain}
\affil[3]{\footnotesize Instituto de Ciencias Matem\'aticas, CSIC-UAM-UC3M-UCM. C/ Nicol\'as Cabrera 15, Campus de Cantoblanco UAM, 28049 Madrid, Spain.}

%\date{\today}
\maketitle

\begin{abstract}

We identify and describe the main dynamic regimes occurring during the melting of the PCM n-octadecane in horizontal layers of several sizes heated from below. 
This configuration allows to cover a wide range of effective Rayleigh numbers on the liquid PCM phase, up to $\sim 10^9$, without changing any external parameter 
control.
We identify four different regimes as time evolves: (i) the conductive regime, (ii) linear regime, (iii) coarsening regime and (iv) turbulent regime. 
The first two regimes appear at all domain sizes. However the third and fourth regime require a minimum advance of the solid/liquid interface to develop, and we observe them only for large enough domains.
 The transition to turbulence takes places after a secondary instability that forces the coarsening of the thermal plumes. 
Each one of the melting regimes creates a distinct solid/liquid front that characterizes the internal state of the melting process.  
We observe that  most of the magnitudes  of the melting process are ruled by power laws, although not all of them. Thus the number of plumes, some regimes of the Rayleigh number as a function of time, the number of plumes after the primary and secondary instability, the thermal and kinetic boundary layers follow simple power laws. 
In particular, we find that the Nusselt number scales with the Rayleigh number as $Nu \sim Ra^{0.29}$ in the turbulent regime, consistent with 
theories and experiments on Rayleigh-B\'enard convection
that show an exponent $2/7$.

\end{abstract}

%\pacs{}
%\keywords{PCM, n-octadecane, thermal plumes, convection,  melting, turbulence}


\section{Introduction}

The high latent heat involved in the solid/liquid phase change  allows  Phase Change Materials (PCM) 
to store  or release a significant amount of energy during  melting or solidification  barely changing 
the temperature.   Many technological  applications take advantage of this stability on external temperature variations  and  thermal storage capacity to use these materials  in electronic cooling, air conditioning in buildings, waste  heat recovery,  to compensate the time offset between energy production and consumption in solar power plants, or combining with construction materials to increase the thermal energy storage capacity of with lighter structures \cite{Ahmad2006,Hyun2014,Memon2014,Zhang2016}.
 
The usage of PCM for latent heat storage is attracting much interest and promotion of regulatory authorities  during the last years  due to environmental issues and requirements of efficiency of renewable energy sources  \cite{Sharma2009,teller2013joint}. The melting temperature allows classifying  the PCM in low temperature ($<200^\circ C$) and high temperature ($>200^\circ C$)  categories \cite{Ge2013}. Organic PCM such as paraffins are common among the low temperature category due to their stability in melting and freezing cycles, non-corrosive and suffering no under-cooling.  An enormous amount of experimental, numerical and modeling work has been carried out to understand the heat transfer performance of these materials taking into account only conductive heat transport within the liquid phase of the PCM and  including as well convective transport \cite{Celentano2001,Jones2006,Tan2009}. 

The operation cycle of PCM consist of two phases: (i) a charging phase when the PCM melts, releasing the latent heat of the solid to liquid  transition and developing a solid/liquid front that moves from hotter to colder regions, (ii) a discharging phase when the PCM solidifies creating a front moving in the opposite direction. We will focus on this work in the melting phase of the cycle.
When only conductive transport is involved, the symmetry of the system dictates the form of these fronts. However, when convective transport appears the coupling between hydrodynamics,  temperature field, phase change and  propagation of the interface results in corrugated fronts. The shape of the fronts is more readily available on experiments than velocity and temperature fields in the liquid phase of the PCM. Hence this shape can provide information on the melting state of a PCM if a  relation with the front shape can be established. We will show in this work how melting regimes have distinct front shapes.   

After the selection of the PCM material, the geometry of the PCM container is the most influential factor in the heat transport performance of PCM systems. 
Many of the  studies have been concerned with the heat transfer in cylinders, cylinder shells, rectangular cavities heated from the side \cite{GaVi86,Ghasemi1999,Ben-David2013,Caldwell2000,Khodadadi2001,Tan2009,Hosseinizadeh2013} or even irregular non-symmetric geometries \cite{Sharma2014,Dhaidan2015,MaMe17}.
However, the configuration of a PCM within a rectangular container heated from below has received comparably less attention.  From a practical point of view, this
configuration is important to dissipate heat from electronic devices. From a theoretical point of view, it offers the opportunity to understand the effect
of the phase change on convection when comparing with  well established results of the classic Rayleigh-B\'enard  problem
on convective features, heat transport, stability and scaling \cite{Kadanoff2001,Bodenschatz2000}. This problem of a PCM contained within rectangular cavities and heated from below has been studied in simulations with  paraffins,  ice slurries or cyclohexane, etc.  \cite{Gong1998,Kousksou2010a,Wintruff2001a,Batina2014}, and experiments with cyclohexane, n-octadecane or magma, among others \cite{Hale1980,Dietsche1985,Clark1987}. 
% (Kousksou2010->double diffusion in ice slurry  in a rectangular cavity heated from below) (Gong1998 n-octadecane heated from below for a square, showing streamlines and flux at different times)
%Experimental \cite{Dietsche1985}(cyclohexane Pr=18, several narrow layers, heated from below and cooled from above. 

%Often observed experimetannly in RB convection that thermal turbulence organices in large-scale circulation patterns. 

We focus on this work at studying the  dynamic of melting of the  paraffin n-octadecane, wich belongs to the low temperature group of PCM. 
The n-octadecane is  contained in squares with periodic boundary conditions in the horizontal direction of different sizes and heated from below. This configuration makes possible to explore a broad range
of effective Rayleigh numbers, taking as characteristic length the gap of the melted PCM, without changing any external parameter. 
For the largest domains, we reach Rayleigh numbers $\sim 10^9$. At this high numbers,  
the  convective features depend very meaningfully on the existence of persistent thermal plumes \cite{Lappa2009}.
We  study in this work how they emerge and evolve during the charging phase. 
%Most of the experiments on solidification on opaque metals and alloys, few
%on visible liquid part  \cite{StefanKharic20014}
%Simulation on solidification including conduction and capillarity effects\cite{Udaykumar1995SimuktionEffects}
%Figure 13 of Chen1991, the same 'depressions' observed in our simulations. They call them 'chimneys'. They do experiments on directional solidification of a aqueous solution of ammonium chloride
%In particular,  how to improve  the heat transfer rate is a major challenge due to their low conductivity. 

The Rayleigh-B\'enard  problem of a liquid layer heated from below is characterized by three physical magnitudes: the difference of temperature between the plates, the heat flux, and the type of fluid. These
provide three dimensionless numbers:  Rayleigh $Ra$, Nusselt $Nu$ and Prandtl $Pr$; respectively, which characterize the solutions of this problem.   At the regime of high numbers, the relation between Rayleigh and Nussselt numbers $Nu\equiv Nu(Ra,Pr)$ has  been thoroughly studied    and found that mostly matches a power law $Nu\sim Ra^\alpha$ \cite{Ahlers2009}. 
 The exponent $\alpha$ generally is within a range between $1/4$ and $1/3$ and the numerical results show a dependence of the exponent with $Ra$  \cite{Stevens2011,Grossmann2000,Grossmann2001,Stevens2013} as well
as experiments \cite{Castaing1989,Chavanne1997,Niemela2000a,Chavanne2001a}.

%Vertical natural convection \cite{Elder1965,Churchill1975,George1979,Tsuji1988,Versteegh1999,Kis2012,Ng2013,Ng2015}.
A theory by Malkus \cite{Malkus1954} based on arguments of marginal stability of the thermal boundary layer predicts $Nu\sim Ra^{1/3}$  and a large number of  experiments agree with this scaling \cite{Chilla2004}.
However, when heat transfer and turbulence decouple from the thermal and viscosity transport coefficients there is as well a theoretical prediction for 
an asymptotic regime following $Nu\sim Pr^{1/2}\,Ra^{1/2}$  \cite{Kraichnan1962,Spiegel1971}.  
Interestingly, the limit of infinite Prandtl number predicts as well $Nu\sim Ra^{1/3}$ for the regime of asymptotically high Rayleigh numbers \cite{Grossmann2000,Doering2006}. 
Nevertheless, it must be noticed that not all the experimental studies have agreed  with the theoretical scalings.
Thus,  models  based on a mixing layer \cite{Castaing1989}, turbulent boundary layer \cite{Shraiman1990} and extensions to low Prandtl fluids \cite{Cioni1997} %  extensions \cite{Ching1997}
predict an exponent $2/7$. 

%The range of $Ra$ explored in this work $Nu\sim Ra^{2/7}$ \cite{Chilla1993}, adn for larger $Ra$ the exponent get closer to the classic $1/3$ \cite{Niemela2006}.

%However, out-liners and more complex relations  are a subject of interest and controversy \cite{citas_GL_theory}. 
%Enhancement of heat transport charcterized by the Nusselt number

We aim in this work at studying the convective motion in the  presence of a solid/liquid phase change. The presence of  phase change introduces an additional 
number related to the source of latent heat, the Stefan number,  that  complicates the problem. In spite of that,  we find that the averaged value of the 
numeric exponent $0.29$ is consistent with a group of experiment results \cite{Castaing1989,Chilla1993,Cioni1997,Glazier1999,Ditze2017} 
% and numerical simulations \cite{DeLuca1990} 
 on moderately high Rayleigh numbers $Ra<10^{10}$, and theoretical models predicting $2/7$.

%This value by scaling theories \cite{8,15,16 de Nimela_nature_2000}.
%Theoretical modesl 

%Ching 1997 https://journals.aps.org/pre/abstract/10.1103/PhysRevE.55.1189

%--Cioni1997 DOI: https://doi.org/10.1017/S0022112096004491
%We review two recent theoretical models, namely the mixing zone model of Castaing et al. (1989), and a model of the turbulent boundary layer by Shraiman & Siggia (1990). We discuss how these models fail at l%ow Prandtl number, and propose modifications for this case. Specific scaling laws for fluids at low Prandtl number are then obtained, providing an interpretation of our experimental results in mercury, as well as extrapolations for other liquid metals.
%-----

%-> liquid fraction is efectively an evolving aspect ratio when liquid phase advances upwards. 

The article is organized as follows. Section \ref{S:2} describes the governing equations of the mathematical model we use to simulate  the behavior of a PCM in the 
presence of conductive and convective transport. The numeric procedure and code to solve the previous equations are explained in Section \ref{sec:numerics}. We present in  Section \ref{S:3} the results  of our simulations, where we explain the observed melting regimes, transitions between them and establish relations 
between dynamic quantities characteristics of the melting process. Finally, we discuss in Section \ref{S:4} our most important findings  and provide a brief  outline of our work.

\section{Model of Phase Change Materials}\label{S:2}

We consider a PCM whose solid phase is modeled with a heat equation to account for conductive transport of the heat, and the liquid phase  is modeled with the 
Navier-Stokes equations for momentum   coupled to the energy equation to include in this phase the heat transport by convection. The Navier-Stokes equations are 
modified with a Darcy term to model a diffuse interface between the solid and liquid phases, and the energy equation includes a source term to account for the 
release of latent heat during melting.   A comprehensive derivation and explanation of the model used in this work  can be found in
references \cite{Voller1987} and \cite{Voller1989}. We now describe  the main features of the model  equations and their coupling.

\subsection{Momentum equations}

We consider the flow as laminar, incompressible, and  neglect the viscous dissipation due to the small domains involved.
The difference in density between liquid and solid phases of n-octadecane is about $5\%$ and is neglected in the model for simplicity.

The governing equation expressing the balance of momentum has the vectorial form
\begin{eqnarray}
\rho\left[\frac{\partial \vec{u}}{\partial t}+ \left(\vec{u}\cdot \nabla   \right)\vec{u}\right]=-\nabla p
+ \mu \nabla^2 \vec{u}   -\rho\,g\left[1-\alpha 
(T-T_{ref})\right]\vec{e_y}  - \frac{C(1-f_l)^2}{\delta+f_l^3}\vec{u} 
\label{eq:momentum},
\end{eqnarray}
where $\nabla=(\partial_x, \partial_y)$  and $\partial_t$ are the spatial and  temporal operators 
respectively. 
 $T$ is a temperature averaged in a control volume, which  can contain pure solid 
(volume liquid fraction $f_l=0$), liquid  (volume liquid fraction $f_l=1$) or a mixture of both phases ($0<f_l<1$) in local thermal 
equilibrium; $\vec{u}=(u_x,u_y)$ 
is the fluid velocity, being $u_x$ and $u_y$ the
 horizontal and vertical components; $\rho$ is the  PCM density whose variation between  liquid and solid
 phases  is neglected;
$\mu$ is the dynamic viscosity, $p$ is the pressure, $g$ the magnitude of gravity acceleration,  $\vec{e_y}$ a unit 
vector pointing in the vertical direction upwards, $T_{ref}$ is a reference temperature where physical properties are 
given and $\alpha$  the thermal expansion coefficient. The bulk physical properties are supposed to be constant within 
the range of temperatures studied, expect   the 
density in the buoyancy term that follows  a linear dependence with  the temperature according to the Boussinesq approximation.

%----------------
The last term in the momentum equation  provides an empirical proportionality relationship, due to Darcy, between the pressure gradient in a porous medium and the
fluid velocity within it. The functional form corresponds to the Carman-Kozeny equation and is a  convenient way to damp the velocity flow within the mushy region and suppress it at the solid phase.
We set in this term the Darcy coefficient $C=1.6\cdot 10^6 kg\,m^{-3}\,s$, in compliance with previous works (\cite{Abdollahzadeh2015,MaMe17}),  and $\delta<<1$   a tiny constant to avoid division by zero without physical meaning. 
When the cell is completely liquid ($f_l=1$) the Darcy term is null, like in a single phase fluid, when it is completely solid ($f_l= 0$)  the Darcy term becomes dominant, and the velocity of the  liquid becomes null. 
For intermediate values of $f_l$ ($0<f_l<1$), the PCM is in the mushy zone.
Through this formulation, the Darcy  term allows to use the momentum equation in the whole domain and model the phase change  when fluid motion is present without the complication of
tracking the solid/liquid interface (\cite{Voller1987,Voller1989}).

%----------------

\subsection{Energy equation}

The thermal energy of the system comes from the contribution  of the usual sensible heat, due to changes in the temperature 
of the solid and liquid phases of the PCM,  and from the latent heat content. Assuming the same density in each phase,  
it can be expressed as a function of the temperature as follows

\begin{eqnarray}
 \left[\frac{\partial }{\partial t}+\vec{u}\cdot\nabla \right] \rho 
\left((1-f_l)C_{s}+f_l C_{l}\right)T =\nabla  \left((1-f_l)\lambda_{s}+f_l\lambda_{l}\right)  \nabla 
T-\rho \Delta h \frac{\partial f_l}{\partial t}
  \label{eq:energy_saldi}
\end{eqnarray}

 where $C_{s}$ ($C_{l}$) and  $\lambda_{s}$ ($\lambda_{l}$) are the specific heats and 
conductivities of the  PCM in the solid (liquid) phase  averaged with the 
liquid fraction, and $\Delta h$ is the latent heat of the solid/liquid phase change of the PCM.

The latent heat released by a control volume during the solid to liquid phase change depends on the melted PCM given by 
liquid fraction as $\Delta H=f_l\rho \Delta h$ being $H$ the enthalpy. 
As a consequence, the coupling between the energy and momentum equations is given through the liquid fraction field $f_l$,
which in turn depends on the temperature, the master variable of the phase change process. We model the liquid fraction 
in the mushy zone using a linear relationship between the solidus and liquidus temperatures 

\begin{eqnarray}
f_l=\Delta H/(\rho \Delta h)=\left\{
\begin{array}{l}
0, \quad T \le  T_s  \\
1, \quad T \ge  T_l  \\
(T-T_s)/(T_l-T_s), \quad T_s<T<T_l 
\end{array}
\right. \label{eq:f_l}
\end{eqnarray}
$\Delta H$  ranges from $0$ (PCM completely 
solid) to $\rho \Delta h$ (PCM completely melted). 
Liquid and solid phases  coexist for intermediate values. 
Notice that the small value $(T_l-T_s)/(T_h-T_m)\sim 0.026$ used in this work creates
a mushy region with a width of only some hundredths of micrometers, leading to neglectable latent heat advection. 
A thorough discussion of this can be  in  \cite{Voller1989}  and shown numerically in \cite{Voller1990b}.

\begin{table}
\begin{center}
%\begin{ruledtabular}
\begin{tabular}{c|ccccccc}
& $\rho (\frac{kg}{m^{3}})$ & $C_{pcm} (\frac{m^2}{s^{2}K})$ & $\lambda_{pcm} (\frac{kgm}{s^{3}K})$ &$\mu 
(\frac{kg}{ms})$ & $\alpha (\frac{1}{K})$ & $T (K)$ & $\Delta h (\frac{m^2}{s^2})$\\ %en el código $\alpha$ se llama
\hline
liquid phase & \multirow{2}{*}{776} & 2196 & 0.13 &\multirow{2}{*}{0.0036} & \multirow{2}{*}{$9.1\cdot 10^{-4}$}& 299.65& 
 \multirow{2}{*}{$243.5\cdot 10^3$}\\
solid  phase & & 1934& 0.358 & & & 298.25\\
\end{tabular}
\end{center}
\caption{Physical properties of n-octadecane \cite{MaMe17}. Liquid and solid values are given when distinguised at the model equations. \label{T:PP_octadecane}}
%\end{ruledtabular} 
\label{table:octadecane}
\end{table}

\subsection{Domain and boundary conditions}

We solve the equations in a set of horizontal layers in the form of squares 
with  periodic boundary conditions along the horizontal direction 
of sides from $L=0.005\,m $ to $L= 0.2\,m$ heated from below.  
The bottom wall is conductive  and held at
constant temperature $T_{h}=353.15\,K$, greater than the  melting  temperature  of n-octadecane 
($T_{l}= 298.65\,K$) which is held initially at solid state at a homogeneous temperature $T_i=298.15\,K$. The upper wall is supposed to be adiabatic, excluding 
heat exchange with the surroundings.

N-octadecane is a representative viscous paraffin with Prandtl number  $Pr=60.3$ and latent heat typical of this group  of materials $L=243.5\cdot 10^{3} \,J\,Kg^{-1}$. The thermo-physical properties within the liquid and solid phases used in the simulations are listed in Table \ref{table:octadecane}.

\subsection{Dimensionless numbers \label{sec:dimensional_numbers}}

We calculate an effective Rayleigh number  in this work  as $Ra=\frac{g\,\alpha \Delta T\,h^3}{\kappa\,\nu} $,  where 
$\Delta T=T_h-T_m$, and $h$ is not the size of the domain but the  averaged height of the melted PCM \cite{Chung1997}. Notice that the Rayleigh number is time 
dependent due to the continued advance of the melting front. 
Since the interface is diffuse, we choose the cells with liquid fraction $f_l=0.5$ as the criterium for the height of the  interface. This liquid fraction  corresponds to the mean temperature of the phase transition $T_m=(T_s+T_l)/2$. 

The local Nusselt number is calculated as $Nu_x=\frac{h(x)\,\left. \partial T/ \partial y \right|_{y=0}}{T_h-T_m}$. 
Notice that this expression is not the same as commonly used in Rayleigh-B\'enard convection. Instead of using the height
of the domain, we use the height of the solid/liquid interface $h(x)$ and the differences of temperature between this interface and the hot wall at the bottom. As a consequence, the Nusselt number is not $1$ when conduction prevails on the liquid phase due to the finite thickness of the interface. 
 This definition of the local Nusselt  provides a more accurate computation of the relative strength between convective and conductive flux.  From now on, we 
will discuss our results using the averaged Nusselt number along the horizontal axis  $Nu= \frac{1}{L}\int_{0}^L Nu_x\,dx.$

Finally, the Stefan number of the melting process is defined as   $Ste=c_l\, (T_h-T_l)/\Delta h$, which throws a fixed value $0.49$ in this work.

\section{Numerical methods \label{sec:numerics}}

We use the open source software under GPL license OpenFOAM \cite{Weller1998}, based on finite volumes to simulate the 
evolution along the time of the model discussed in the previous section.  The conservation of fluxes ensured by the finite volume 
method makes this discretization especially advantageous
when non-linear advective terms appear as in convection dominated PCM dynamics.
The convective terms are discretized using a second order upwind scheme. The time integration is
carried out using a second order Crank-Nicolson scheme. The solver automatically limits the time step using a maximum Courant-Friedrichs-Lewy  condition. 
Also, we have imposed a maximum temporal step of $0.5\,s$ to ensure  robustness for all the tested conditions. The source term in the energy equation requires
particular attention since it couples  the temperature and liquid fraction fields strongly. Following \cite{Voller1991}, it is 
linearized as a function depending on temperature, split in an explicit
(zero order term) and implicit part (first order term), and the liquid fraction updated at every iteration.
A Rhie-Chow interpolation method is used to solve the enthalpy-porosity equations  to avoid
checkerboard solutions. The momentum and continuity equation are solved using the PIMPLE algorithm, which ensures a right pressure-velocity 
coupling by combining SIMPLE and PISO algorithms
\cite{Aguerre2013}. The temperature equation is solved for each PIMPLE outer-iteration, securing the
convergence of the velocity, pressure, and temperature fields. 
%%To improve convergence under-relaxation factors are used in the velocity, pressure and temperature with values $0.4$, $0.3$ and $0.3$ respectively.
%After the discretization and linearization of the equations, the original PDE problem is transformed into
%a system of linear equations solved at each iteration using a multi-grid solver with tolerance $10^{−8}$ for the
%pressure, and $10^{−6}$ for the velocity and temperature fields.

Regarding the validation of the model, the proper implementation of the bulk equations  has 
been extensively verified with numeric and experimental works to ensure the 
stability and accuracy  of the code, 
with special care of the treatment of the source term of the energy equation in  \cite{MaMi17}.  The implementation is very 
successful in modeling the strong coupling between momentum and energy equations typical of phase change problems  in 
rectangular and  geometries in the form of circular sections \cite{MaMe17,MaMe17b,MaMi17}.

We have used a structured and uniform mesh of $400\times400$ cells for domain size from $0.005\,m$ to $0.1\,m$.
This provides a maximum grid size of $2.5\cdot 10^{-4}\,m$ for $L=0.1\,m$.  For $L>0.1\,m$ we use 
$800\times800$, providing  a grid size of  $2.5\cdot 10^{-4}\,m$ for $L=0.2\,m$.   Notice how these grid resolutions resolve
 the  Kolmogorov scale $\eta(y)=\left(\mu^3/\epsilon_u\right)^{1/4}$, where  $\epsilon_u(y)= \frac{\mu}{L}\int_0^L {\left(\partial u'_i/\partial x_j \right)^2}\,dx$ is the turbulent
dissipation and $\vec{u'}$ the fluctuation about a spatial average of the velocity on the horizontal axis \cite{Ng2015}.  Thus, for instance, for $L=0.04\,m$ 
we obtain  from
our simulations $min(\eta) =  8.6\cdot 10^{-4}\,m $, and for $L=0.1\,m$  we obtain $min(\eta) =  6\cdot 10^{-4}\,m $, where the function {\em min} provides 
the minimum value of the Kolmogorov scale from the beginning  up to complete melting of the PCM.

\section{Results \label{S:3}}

\subsection{Regimes}

We identify up to four dynamic regimes during the melting of the n-octadecane. Next, we describe them and show how they depend on the domain size. 

\subsubsection{Conductive regime}

 \begin{figure}[!h]
  \begin{tabular}{ccccc}
  & a) $t=50s$ &  b) $t=100s$ &  c) $t=300s$ &  d) $t=440s$ \\
   \rotatebox{90}{Stream Function}&
  \includegraphics[scale=0.25]{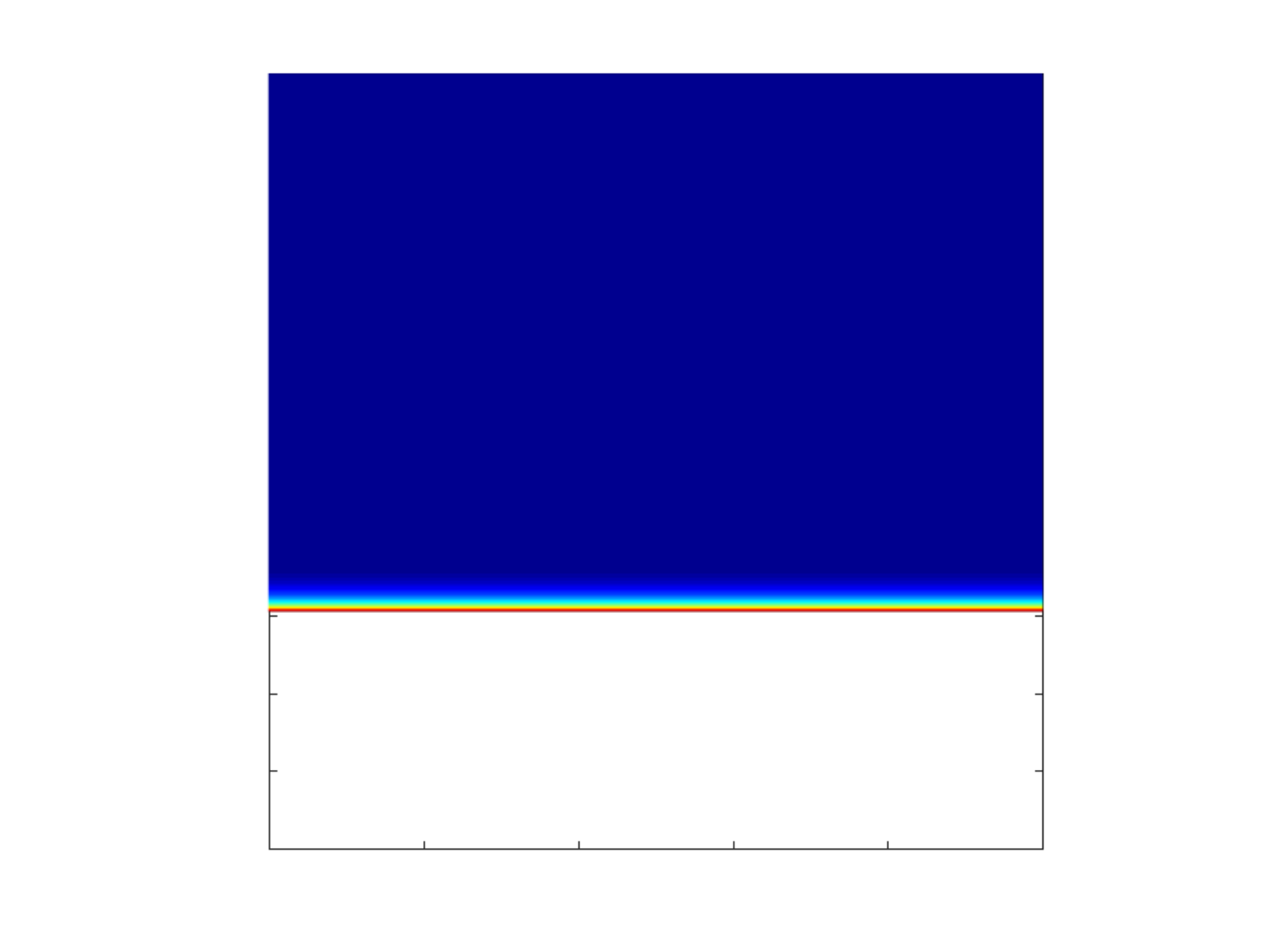}&
  \includegraphics[scale=0.25]{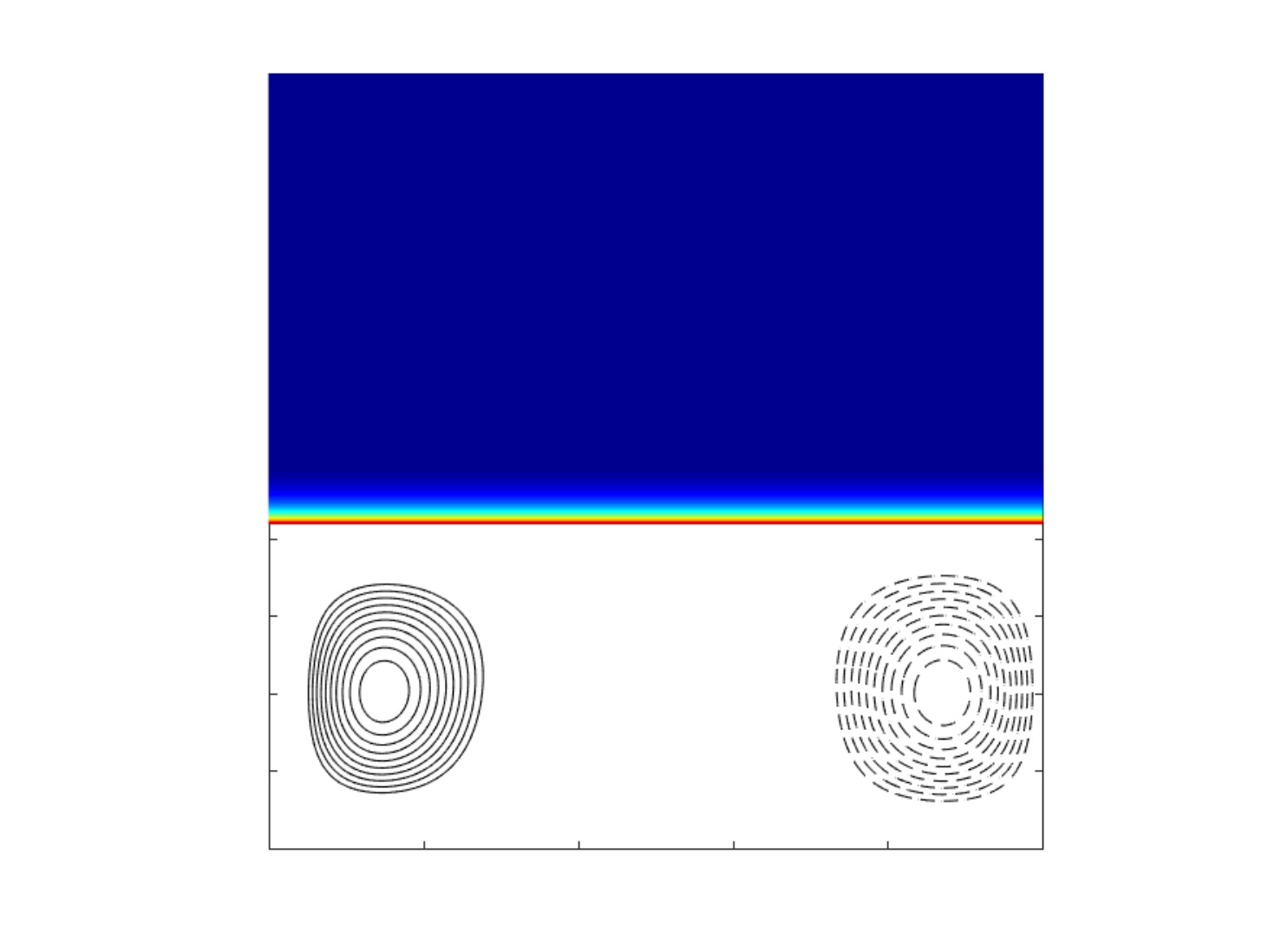}&
  \includegraphics[scale=0.25]{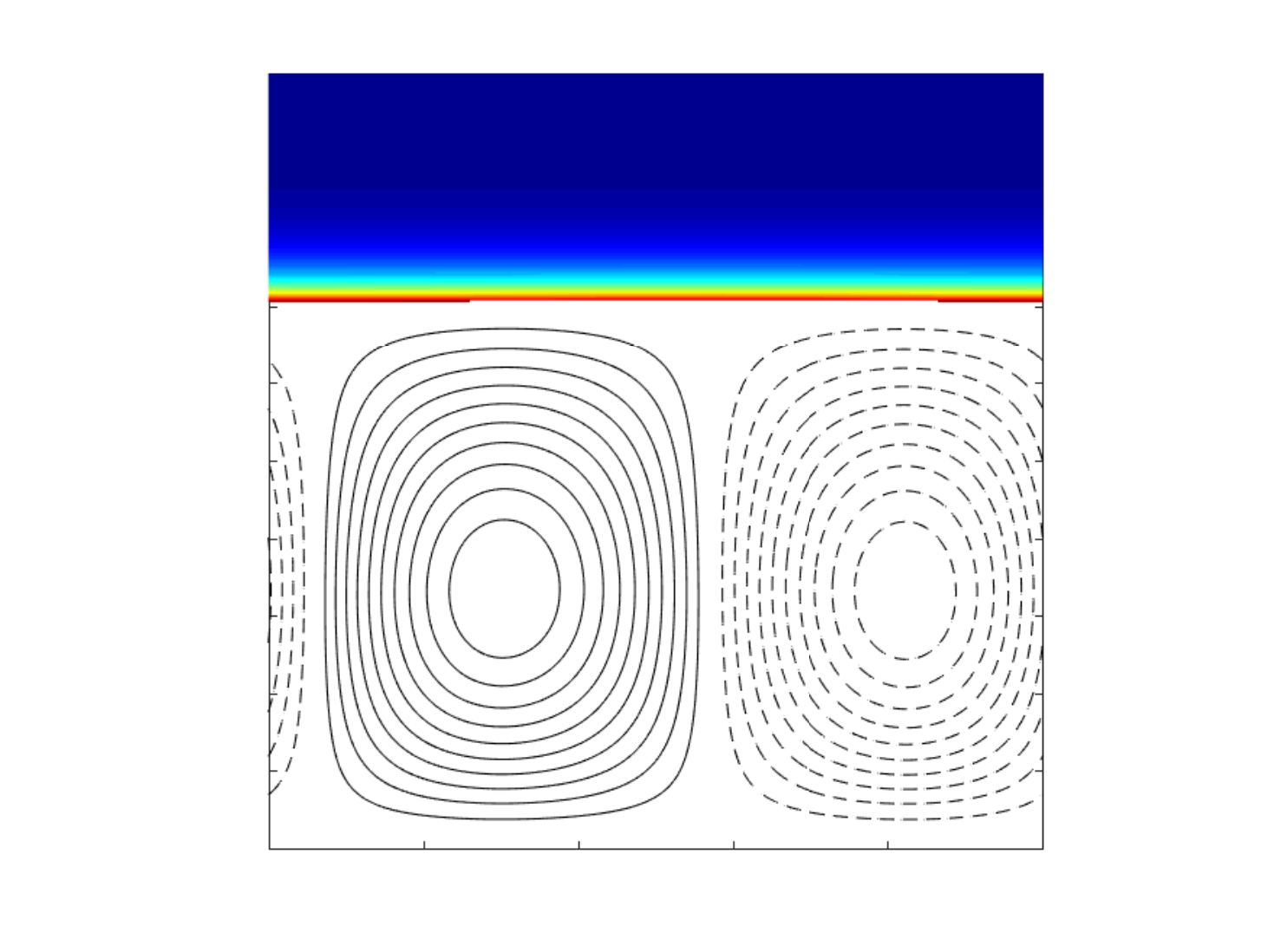}&
   \includegraphics[scale=0.25]{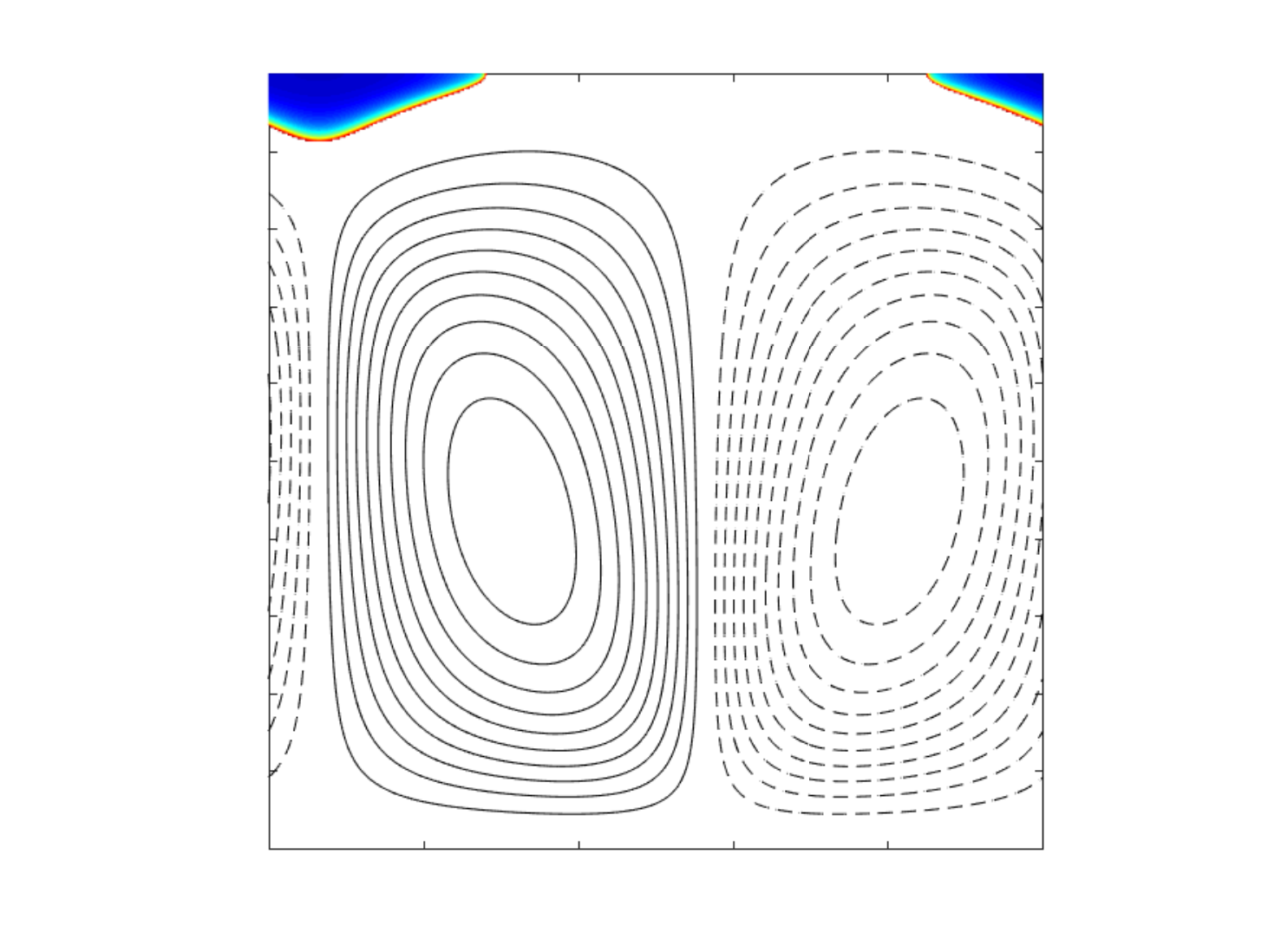}\\
  \rotatebox{90}{Temperature field}&
  \includegraphics[scale=0.25]{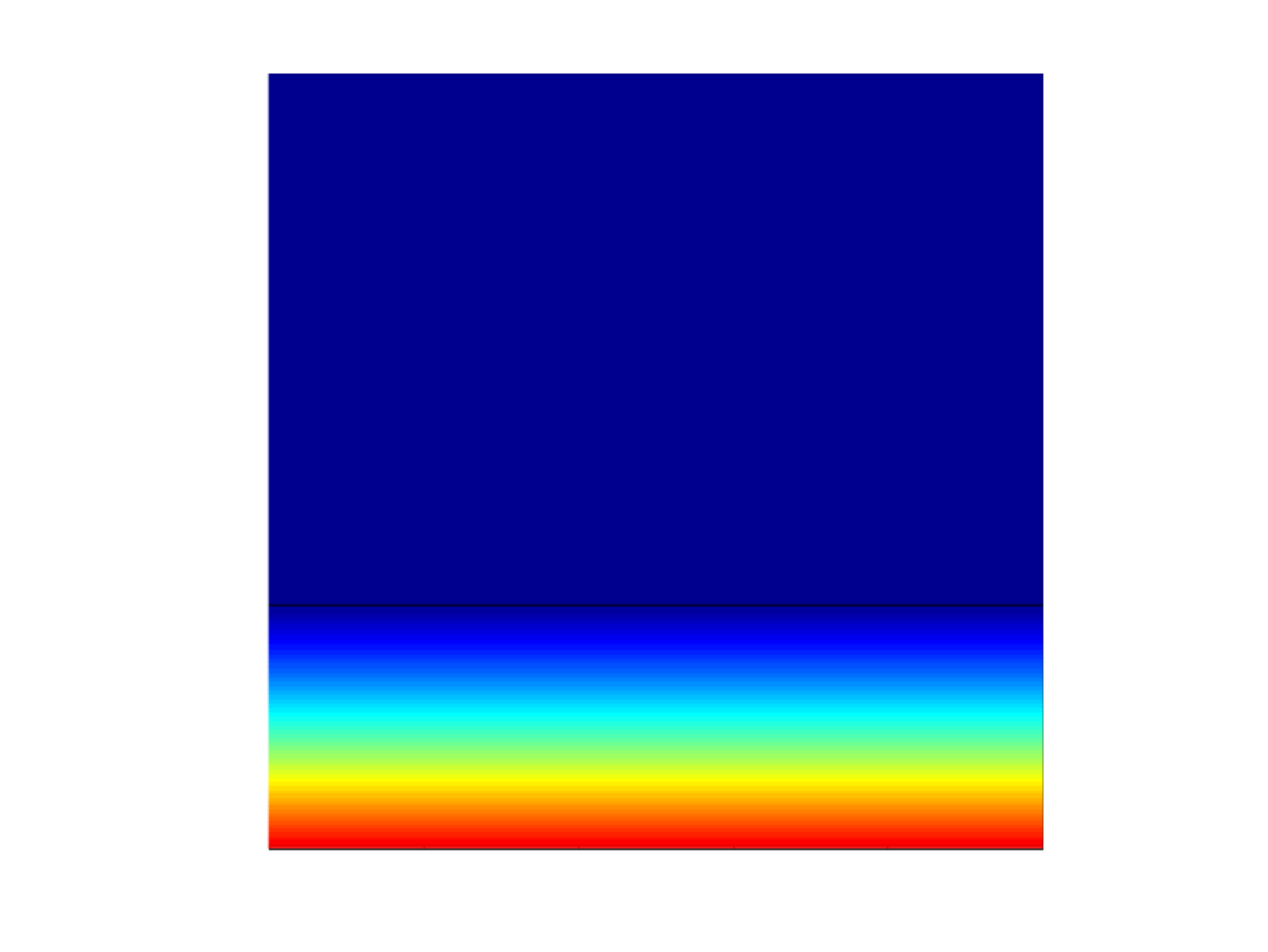}&
  \includegraphics[scale=0.25]{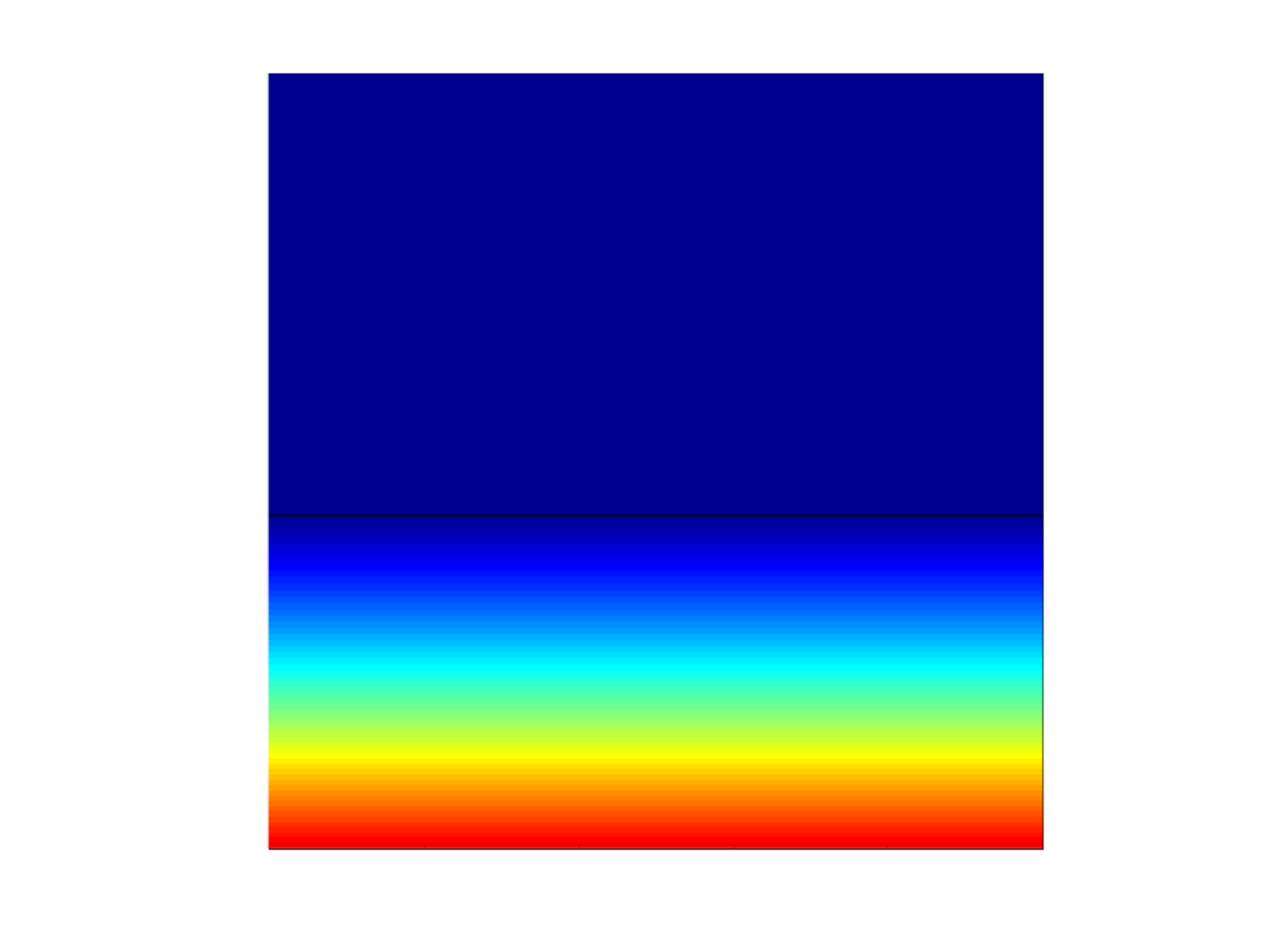}&
  \includegraphics[scale=0.25]{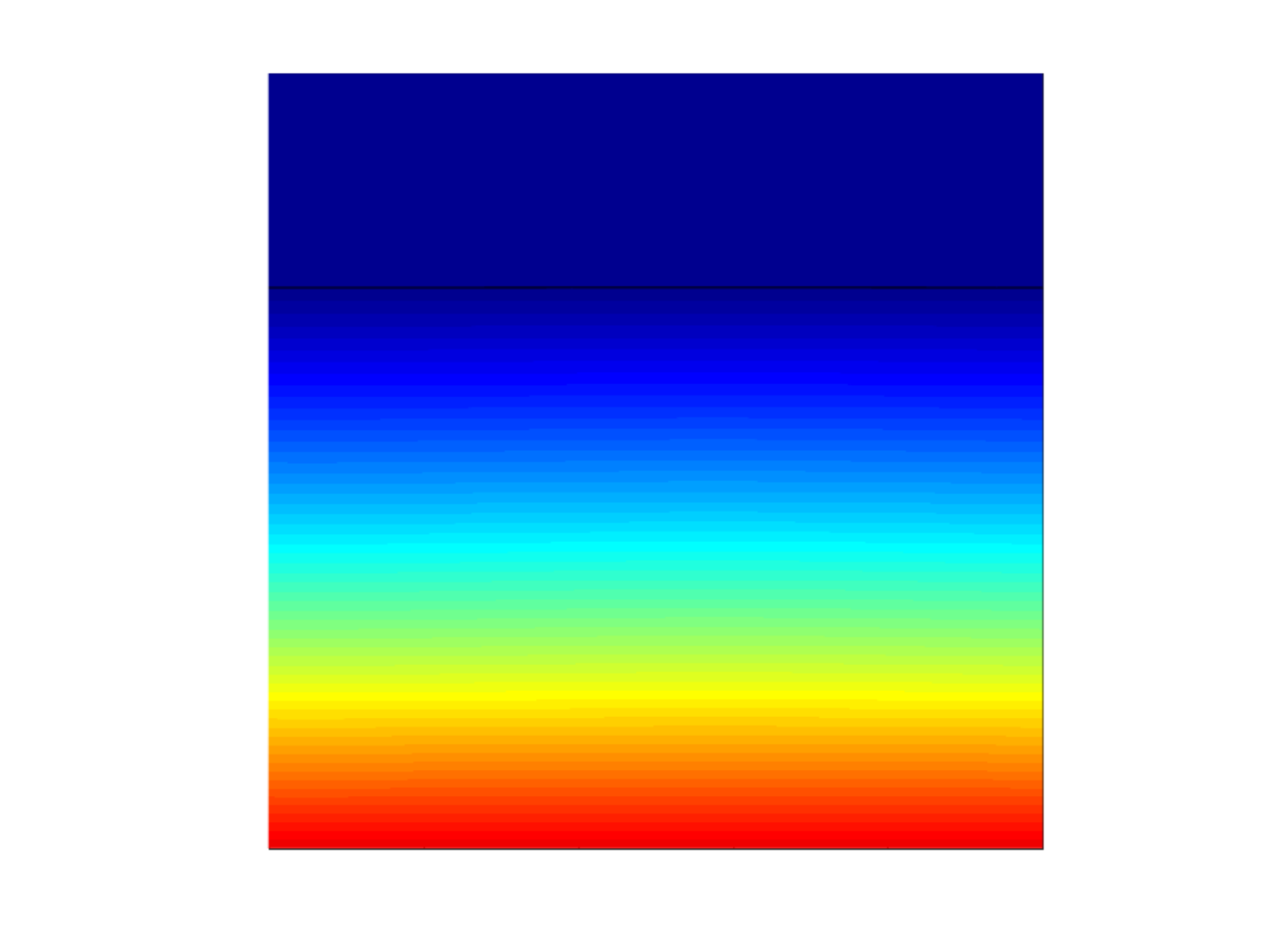}&
   \includegraphics[scale=0.25]{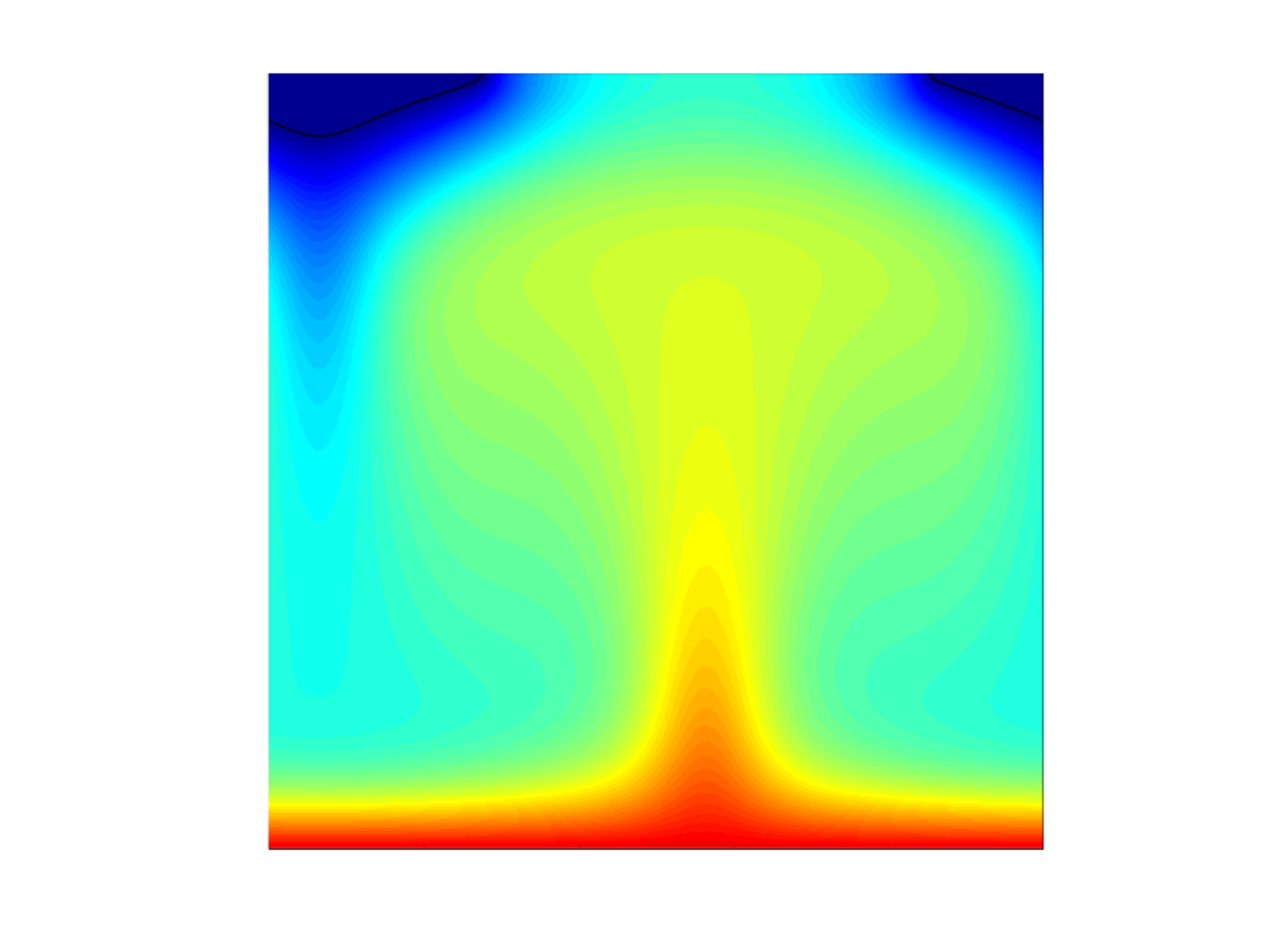}
\end{tabular}
\caption{Snapshots at different times $t=50,100,300,440s$ for the stream function (top row) and temperature field (bottom row) for a domain of side 
$L=0.005\,m.$}\label{F:Snapshot_0005}
\end{figure}
 The conductive regime starts at the beginning of melting and is governed by conductive heat transfer. The PCM is melted above the hot wall, there is not convective motion within the liquid phase, and a solid/liquid interface parallel to the hot wall develops moving upwards.

This regime is common to all domain sizes, and Fig. \ref{F:Snapshot_0005}(a) illustrates this state  for the  smallest domain of this study $L=0.005\,m$ at 
$t=50\,s$.  The height of the interface advances in time  following a power law $h\sim t^{0.49}$ % ok 
with a similar exponent for the rest of sizes $L$. The value of this exponent agrees remarkably well with that of the  position of the solid/liquid interface 
of the analytic two-phase Stefan problem in a semi-infinite slab  $\sim t^{1/2}$. In the Stefan problem,  the interface is supposed to be sharp, with zero 
thickness. It is worth noticing that in spite of adopting a model with a diffuse interface   in this work the value of the numerical exponent is very close to 
the  Stefan problem with a sharp  interface.

\subsubsection{Linear  regime}

 As time advances, a second regime  arises after the destabilization of the conductive layer through a linear Rayleigh-B\'enard instability and the ensuing convective motion.  Indeed, for  a horizontal layer of n-octadecane bounded by rigid plates, it suffices a thickness  $\sim 1.5\,mm$ to destabilize the conductive state.  The instability creates an array of counter-rotating convective cells and lengthening plumes.
 
\begin{figure}[!b]
\begin{tabular}{cccccc}
  &   a) $t=100s$ &   b) $t=200s$ &   c) $t=300s$ &  d) $t=400s$ \\
   \rotatebox{90}{ Stream Function}&
  \includegraphics[scale=0.25]{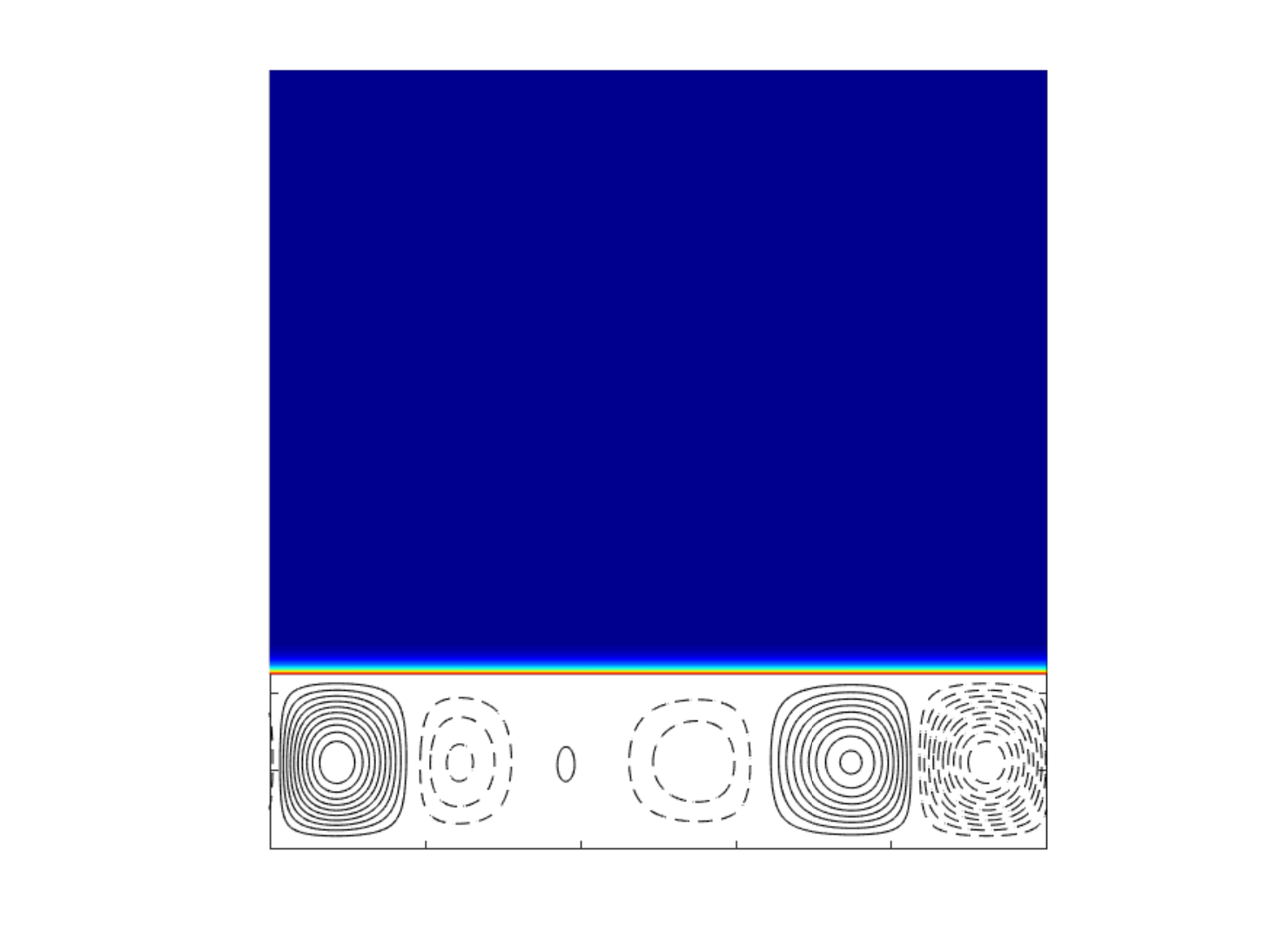}&
  \includegraphics[scale=0.25]{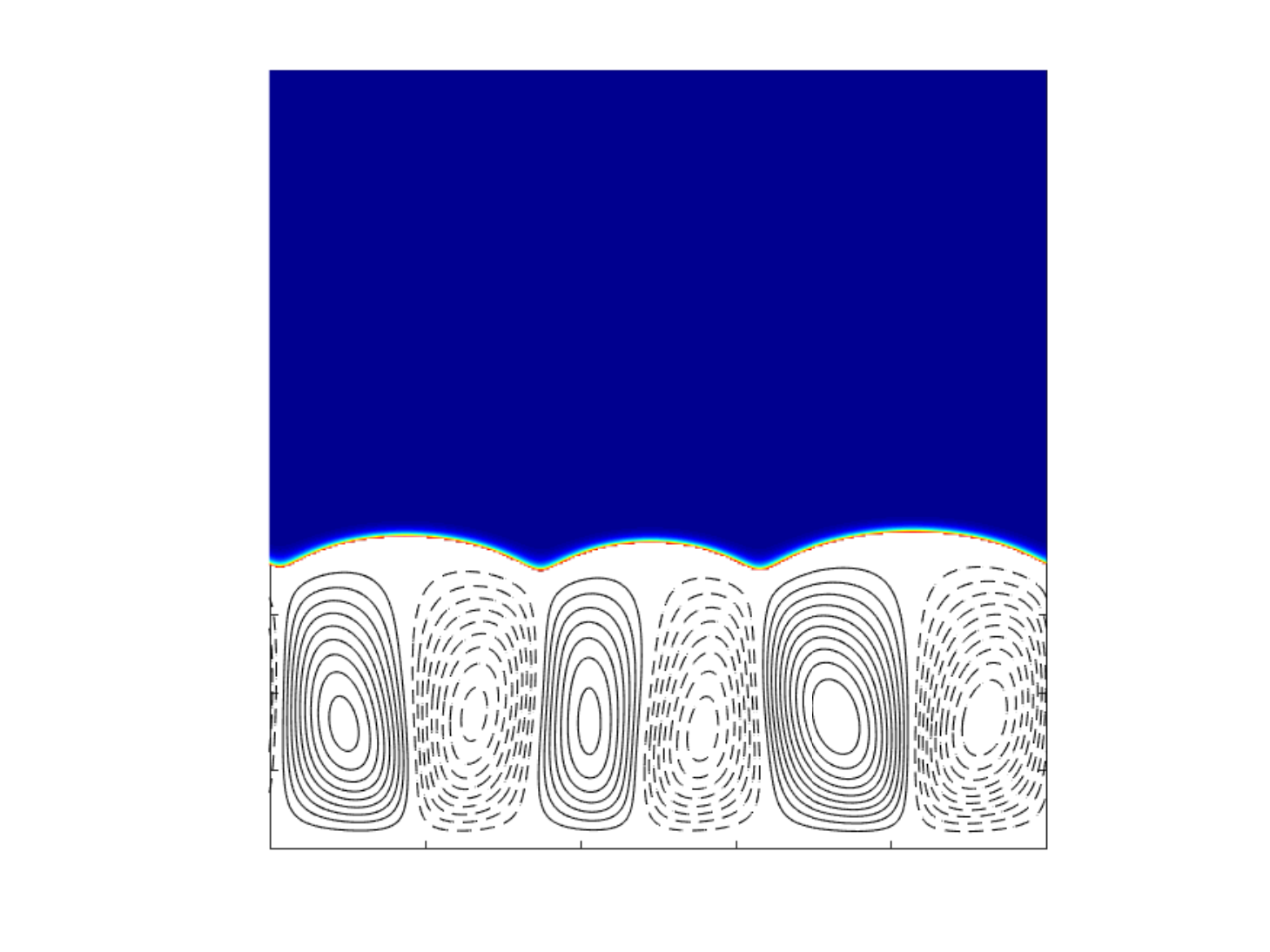}&
  \includegraphics[scale=0.25]{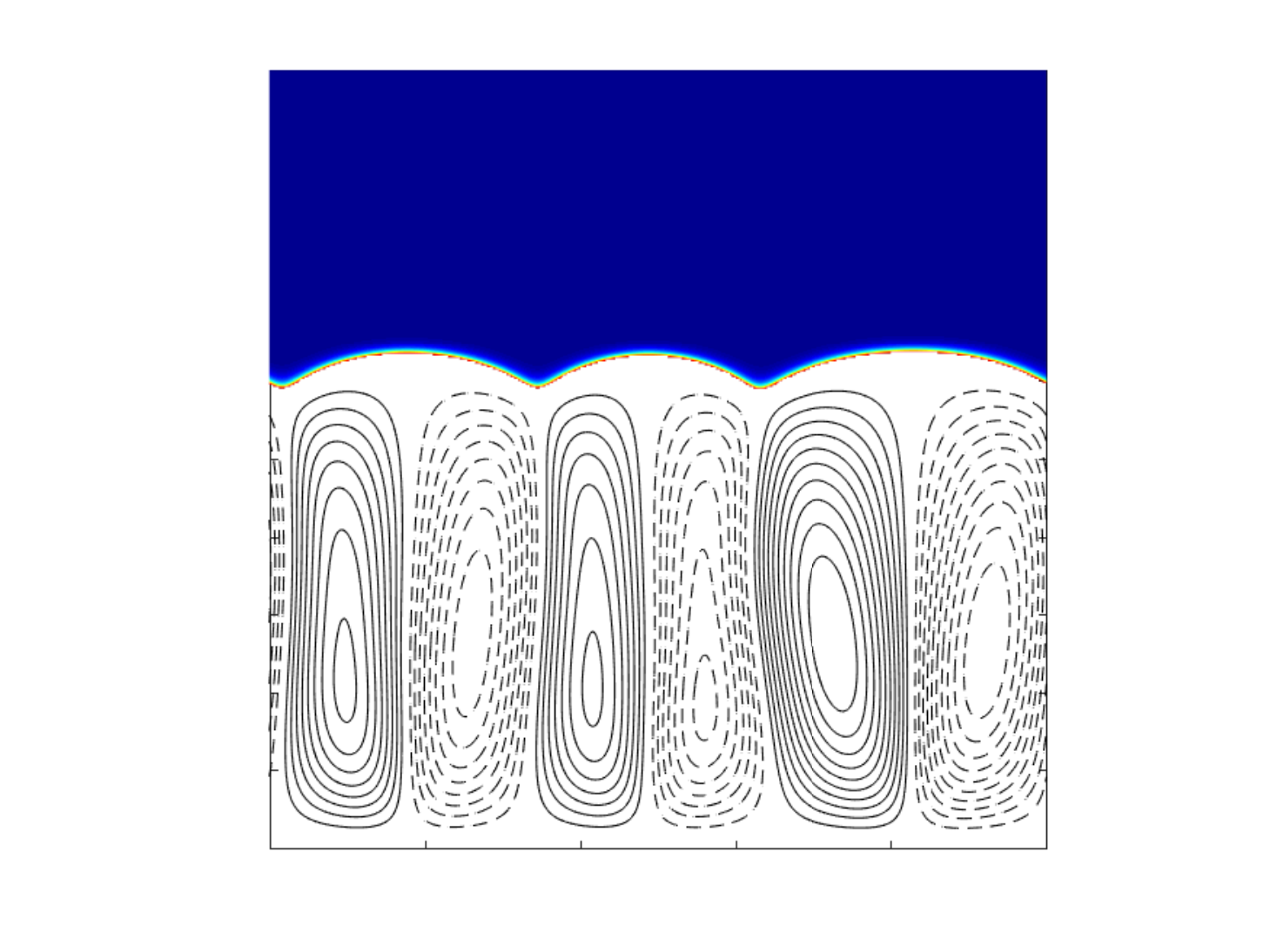}&
   \includegraphics[scale=0.25]{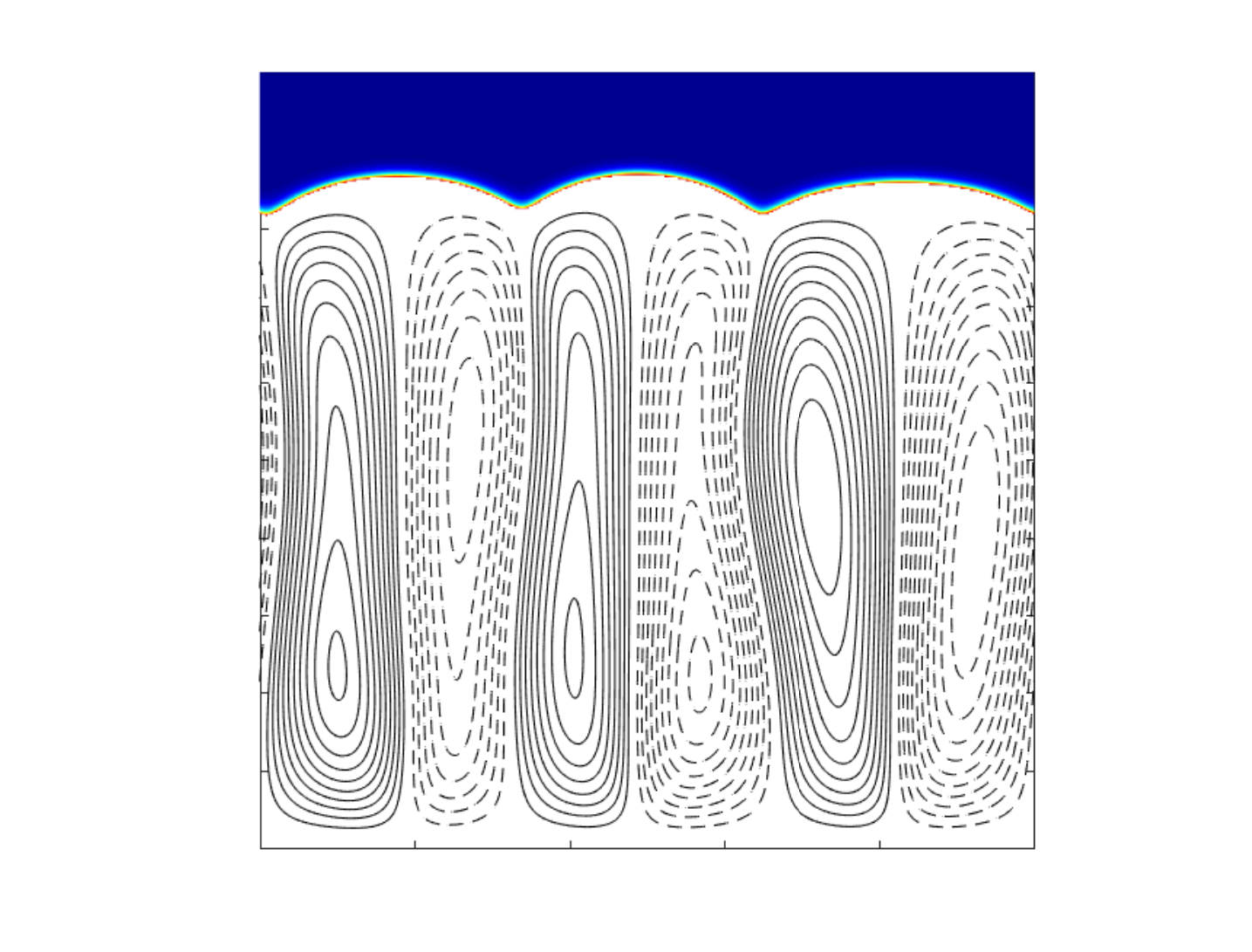}\\
  \rotatebox{90}{ Temperature field}&
  \includegraphics[scale=0.25]{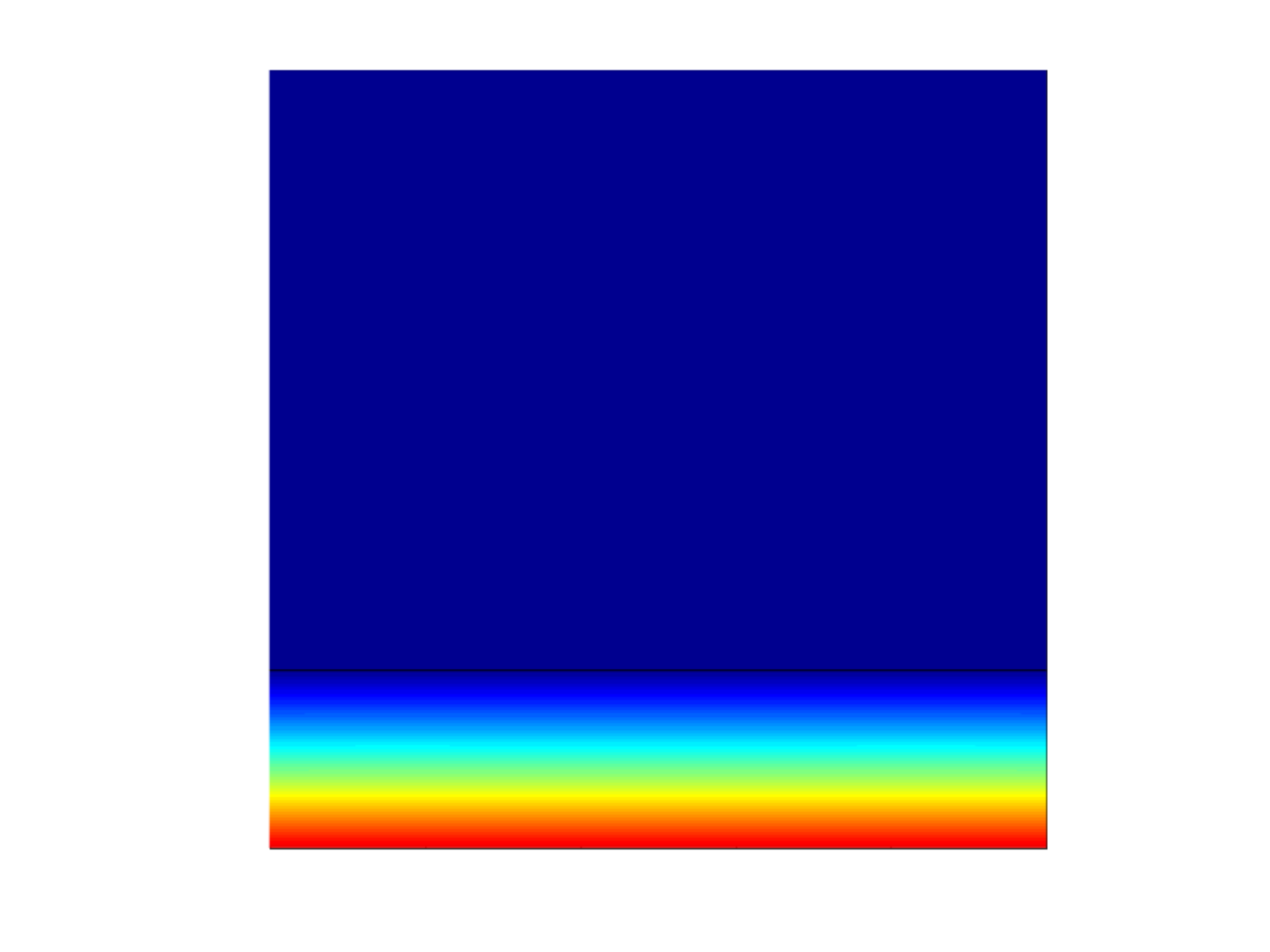}&
  \includegraphics[scale=0.25]{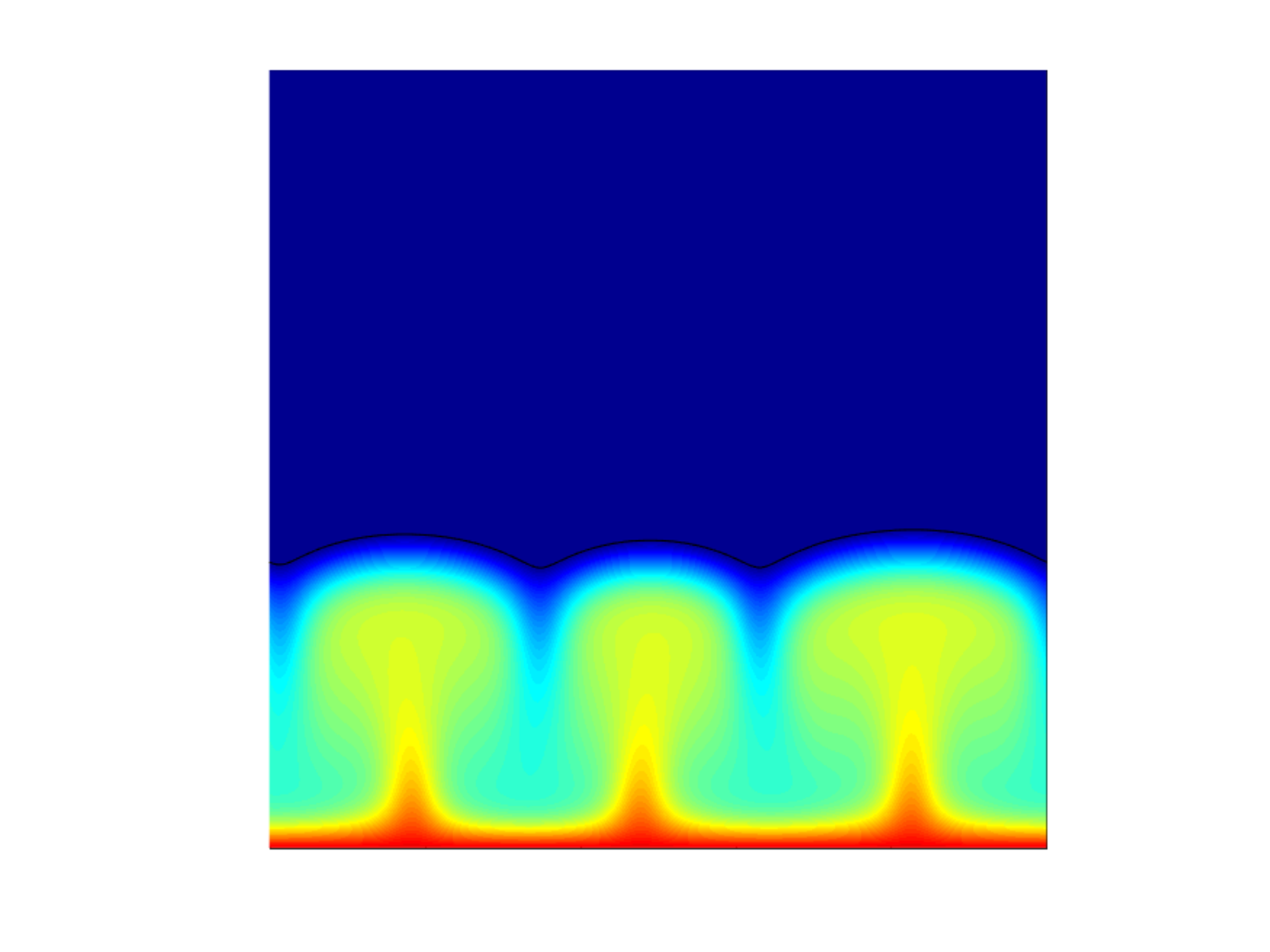}&
  \includegraphics[scale=0.25]{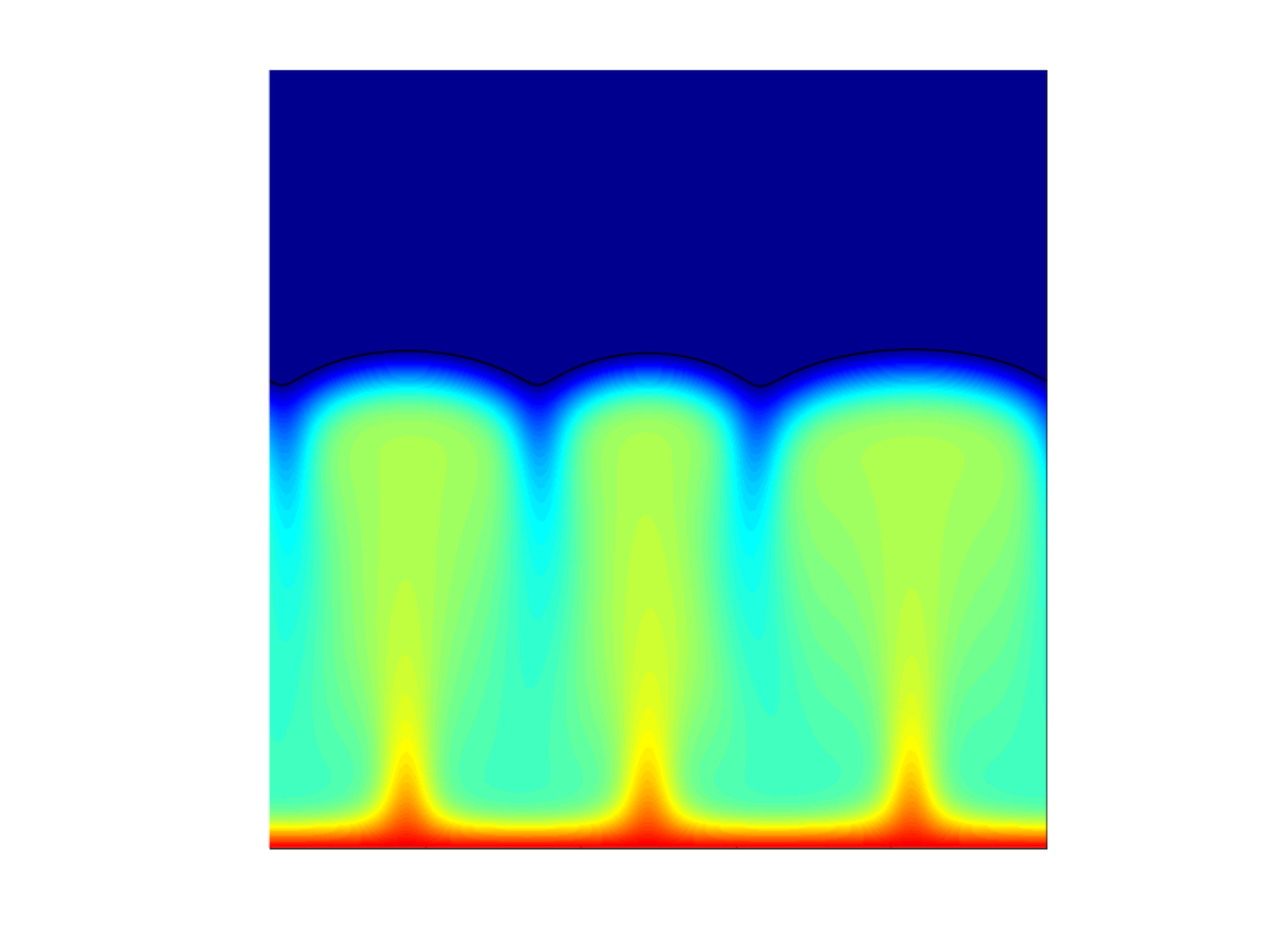}&
   \includegraphics[scale=0.25]{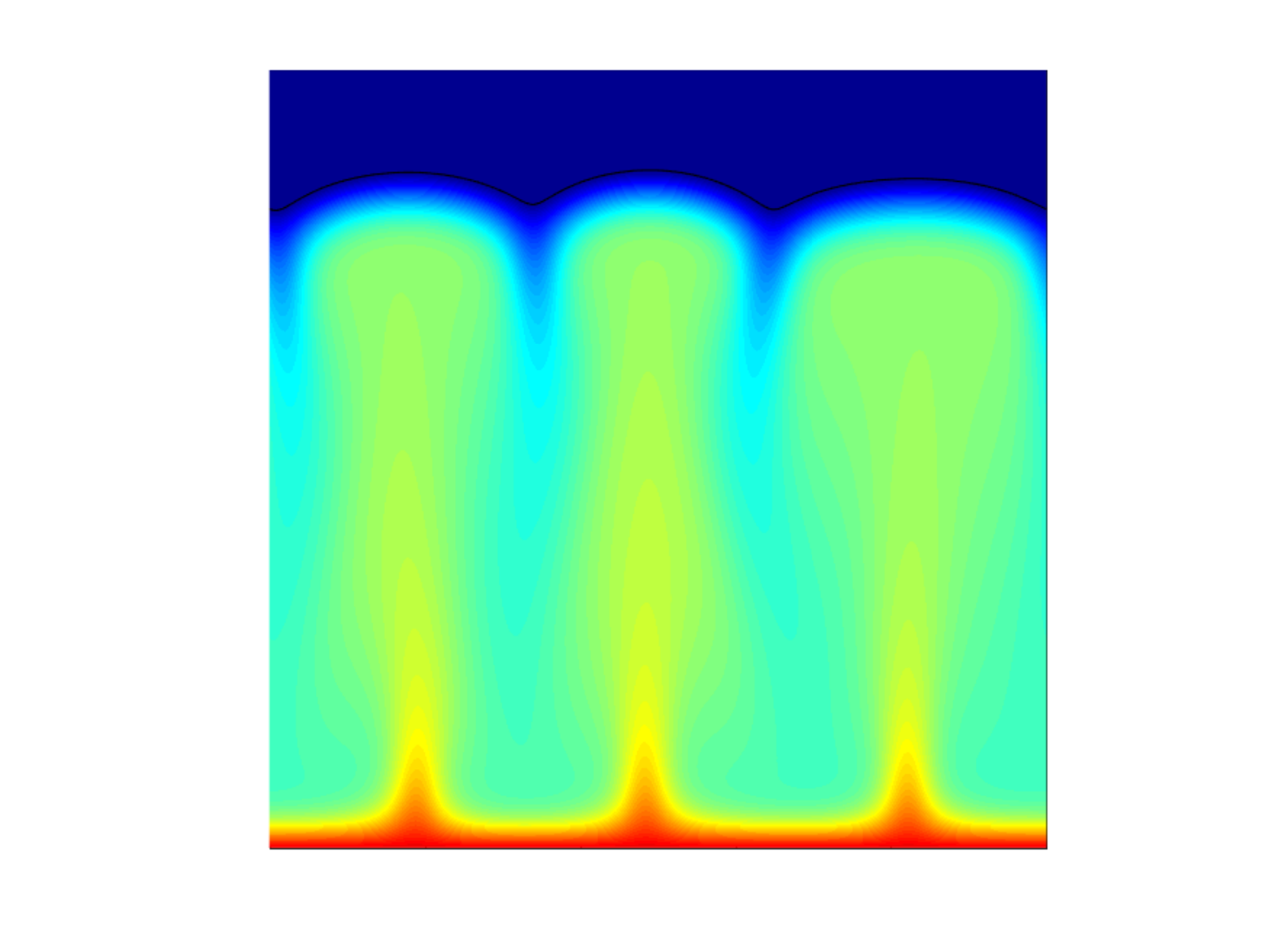}
\end{tabular}
\caption{Snapshots at different times $t=50,100,300,400s$ for the stream function (top row) and temperature field (bottom row) for a domain of side 
$L=0.01\,m.$}\label{F:Snapshot_001}
\end{figure}

Fig. \ref{F:Snapshot_0005} (b,c,d) show the evolution of this regime through a sequence of panels  for the temperature and 
velocity fields for  $L=0.005\,m$, and Fig. \ref{F:Snapshot_001} for  $L=0.01\,m$.
The wave number of the primary instability allows a single plume fitting into the domain for $L=0.005\,m$.  As time advances, the stem lengthens, and the head 
of the plume widens resulting in a plume with a form of {\em mushroom}. The widening and form of the head are responsible for a curved solid/liquid interface.  
During this process, there are two symmetric counter-rotating cells around the stem with preserved width.

For greater domain sizes, like $L=0.01\,m$ in Fig. \ref{F:Snapshot_001}, an array of plumes develop without interaction between them.  The  heads of this 
array of plumes generate a modulated melting front as appears in  experiments  of  melting of  n-octadecane by Gau and Viskanta (c.f. Fig 1a, 1b of Ref. 
\cite{Gau1983}). 

The separation between the plumes is determined by the critical wavenumber of the primary Rayleigh-B\'enard instability  and is conserved during this regime. 
As a consequence, this wavenumber determines the distance between peaks of the melting front. Given that the  main features of this regime come from the linear instability mode of the conductive regime, we refer to it as the {\em linear regime}.

This second melting regime exhibits relatively low Rayleigh numbers $<O(10^5)$. While the main dynamic features of this regime occur along the vertical direction we notice that when melting advances the stems of the plumes begin to bend, with roughly fixed  stem location on the thermal boundary layer.  Indeed, the thermal boundary layer extends along the hot wall with its thickness barely   changing. This  
is  opposed  to plumes created by point sources of heat \cite{Lappa2009}.   

\begin{figure}[!h]
\begin{tabular}{ccccc}
 &  a) $t=200s$  &  b) $t=500s$ &  c) $t=550s$ & d) $t=570s$\\
  \rotatebox{90}{\large Stream Function}&
  \includegraphics[scale=0.25]{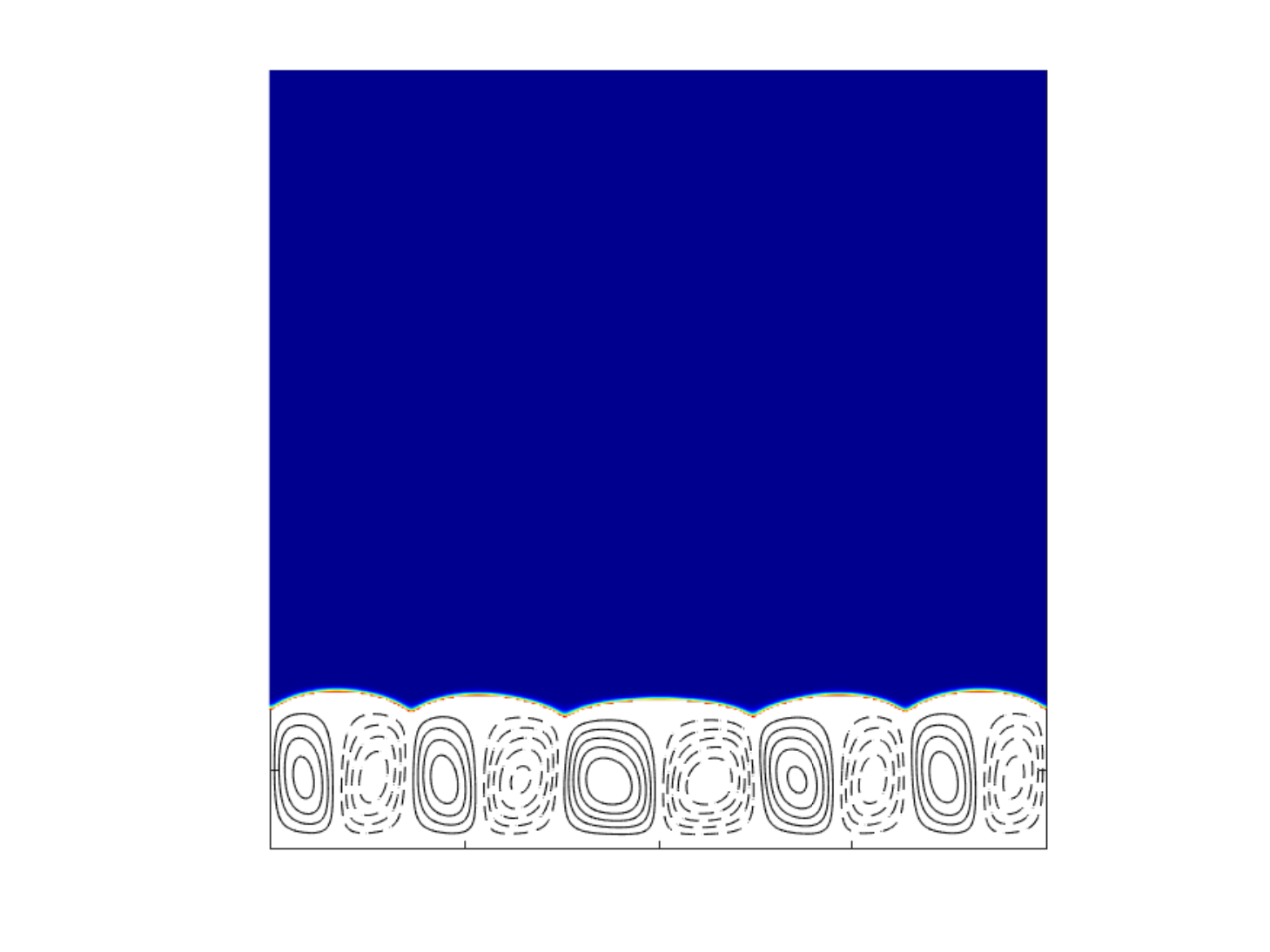}&  
  \includegraphics[scale=0.25]{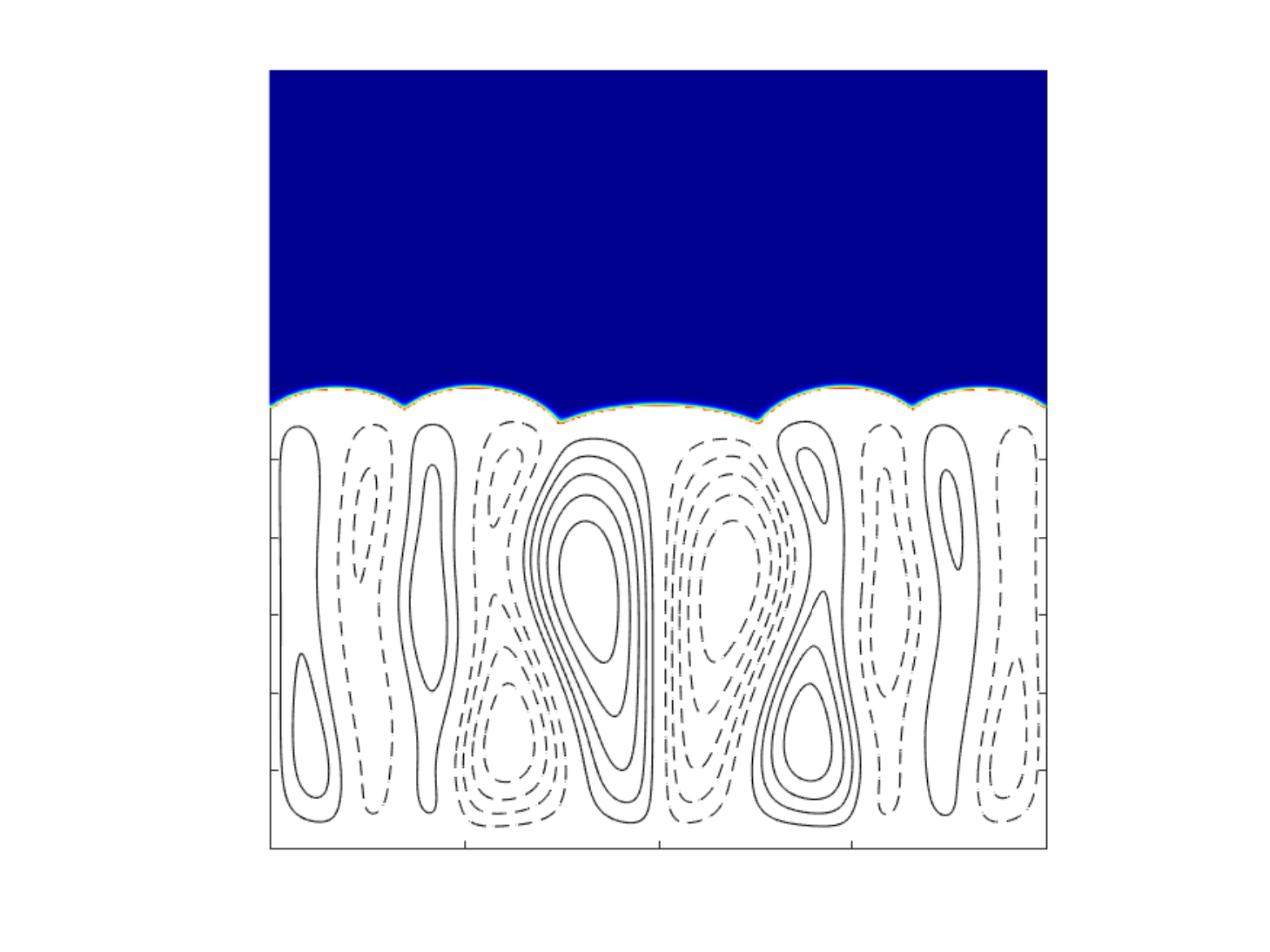}&
  \includegraphics[scale=0.25]{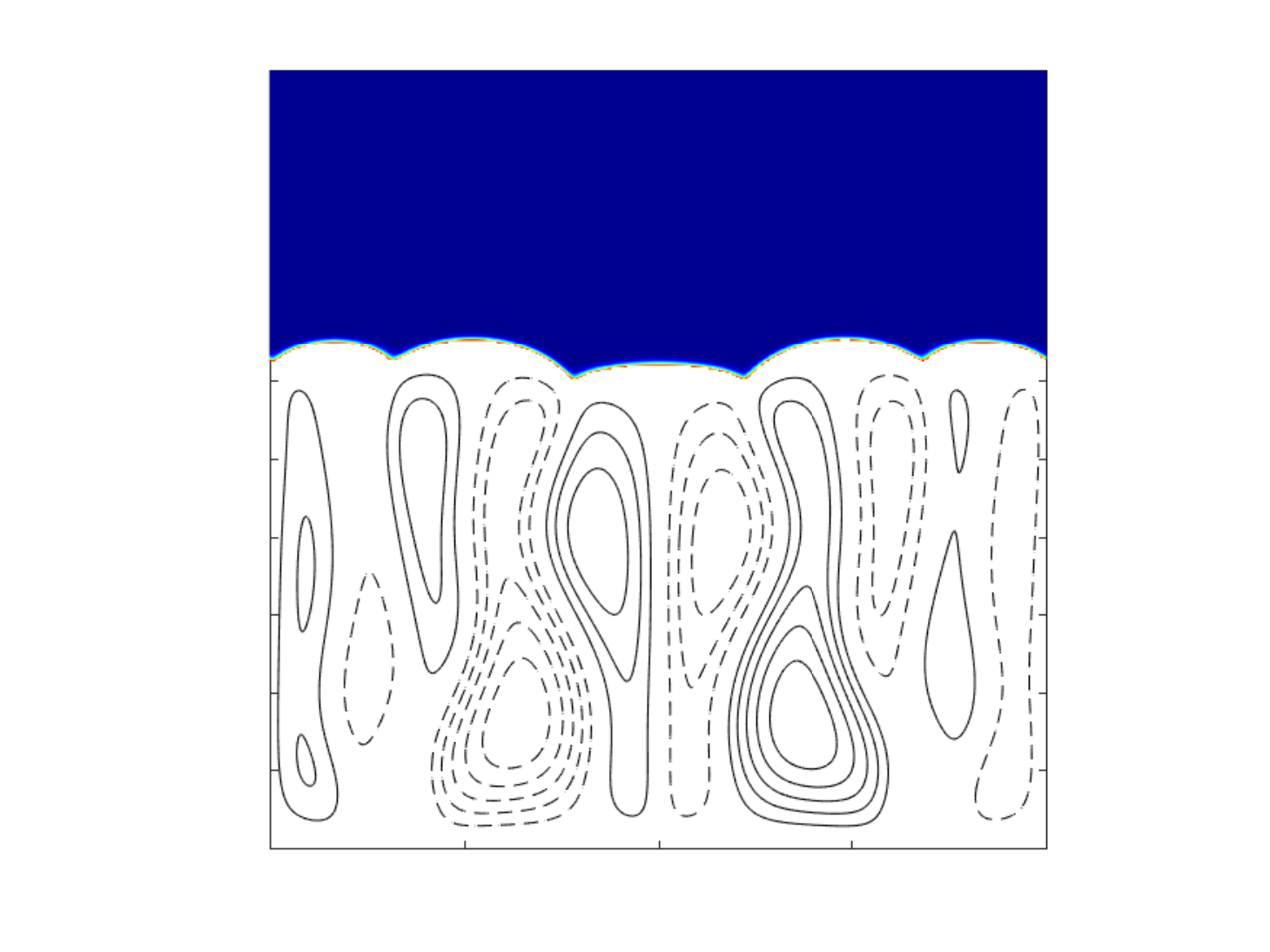}&
  \includegraphics[scale=0.25]{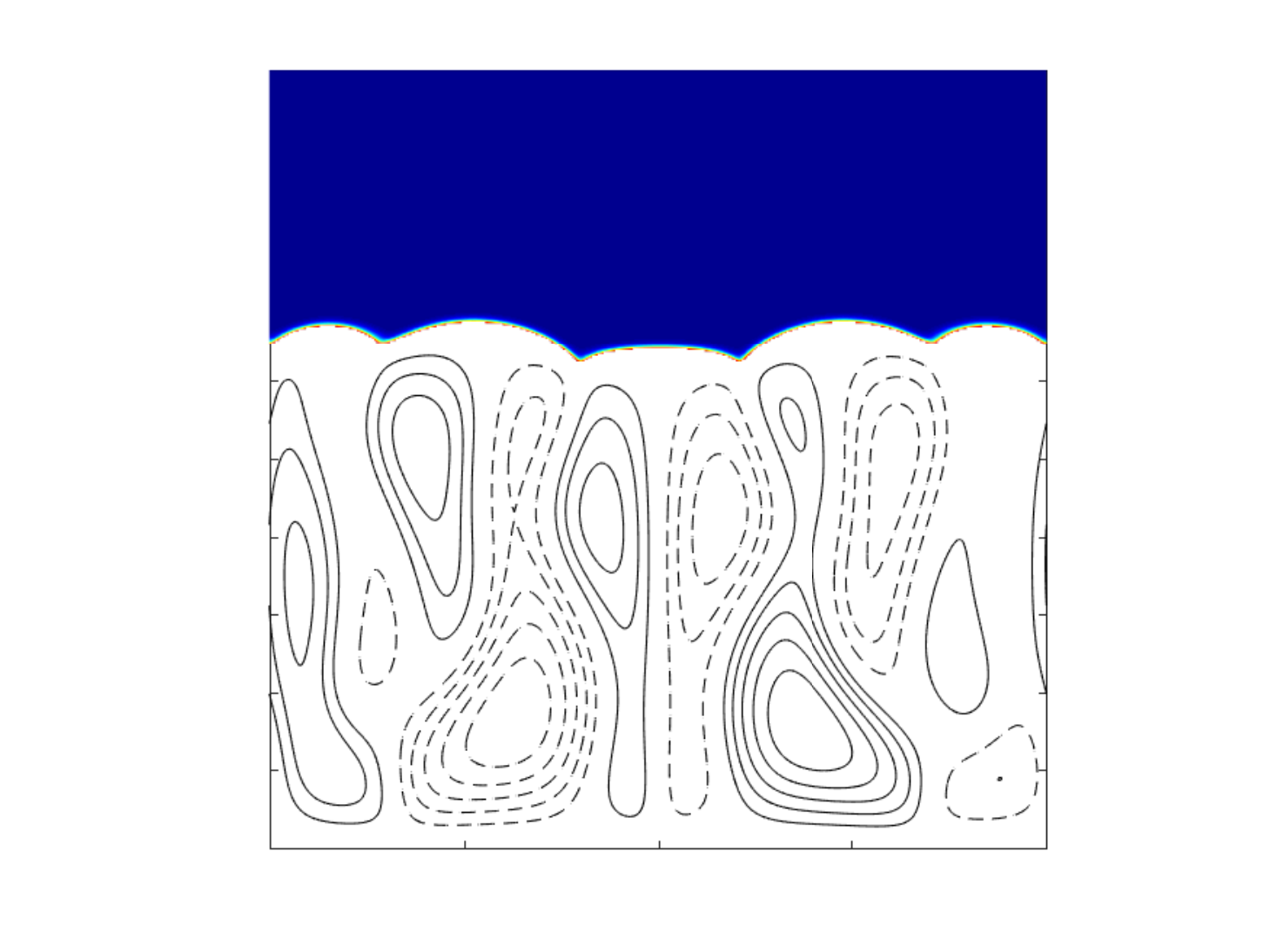}\\
   \rotatebox{90}{\large Temperature field}&
   \includegraphics[scale=0.25]{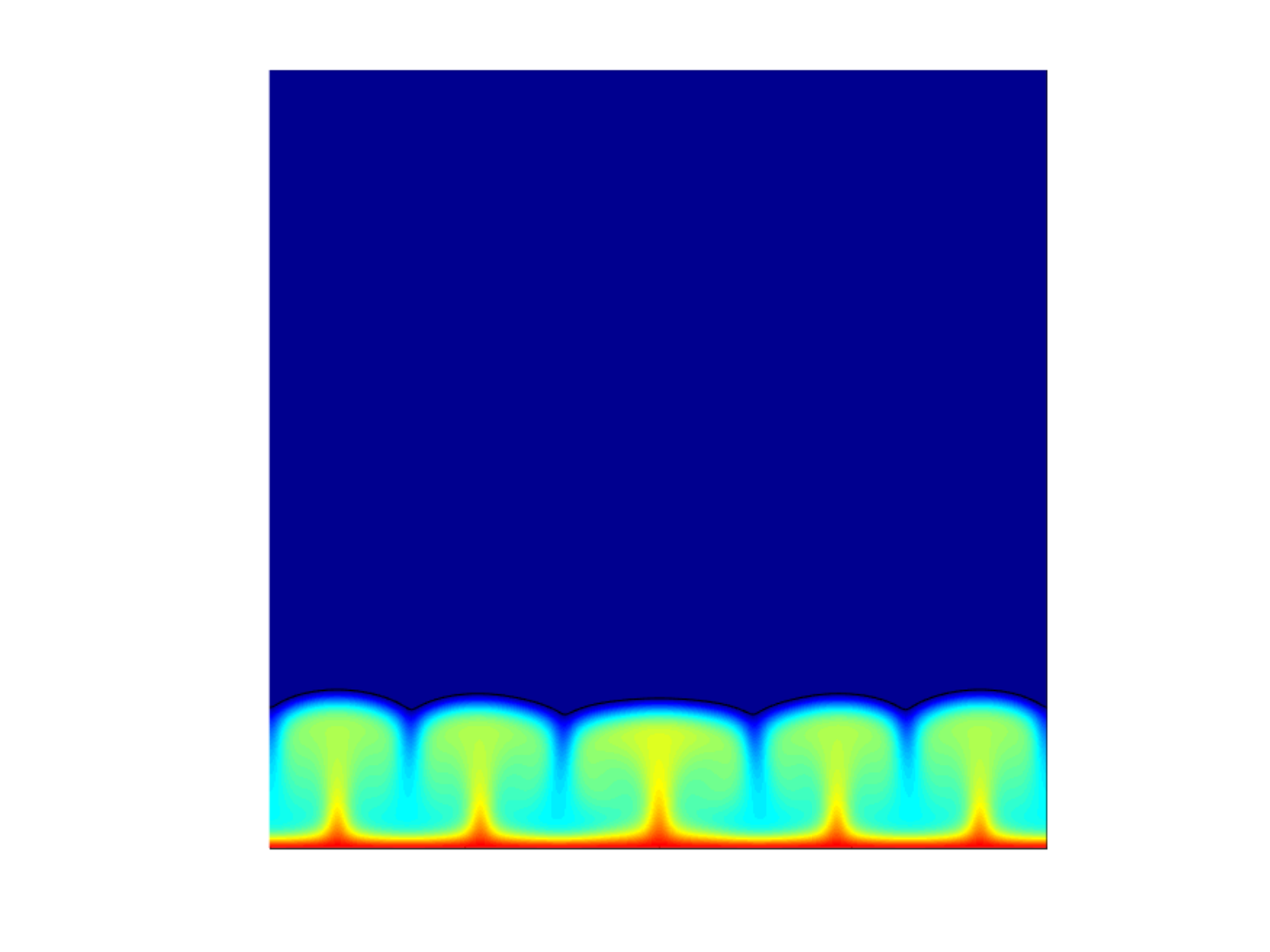}&
  \includegraphics[scale=0.25]{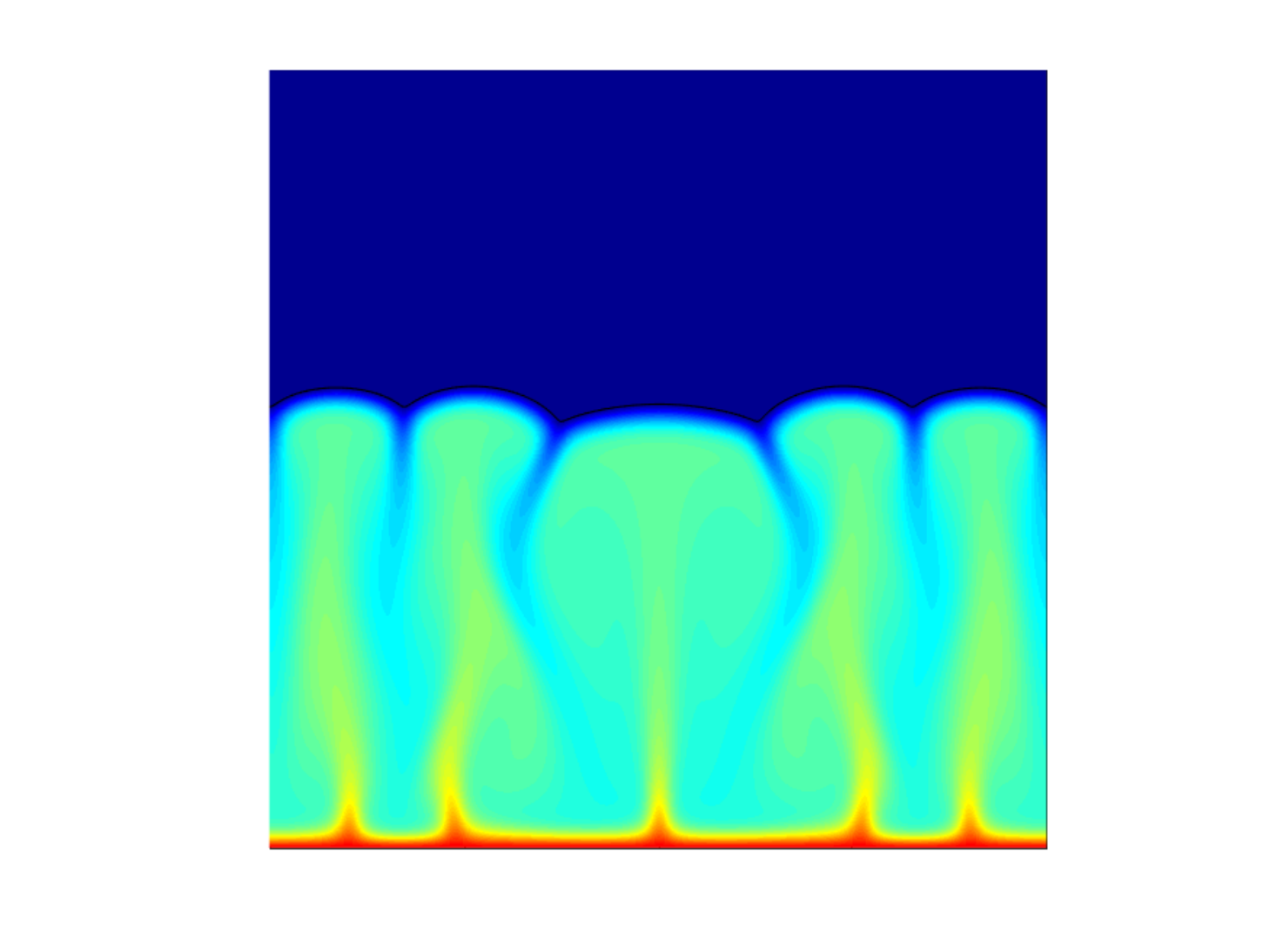}&
  \includegraphics[scale=0.25]{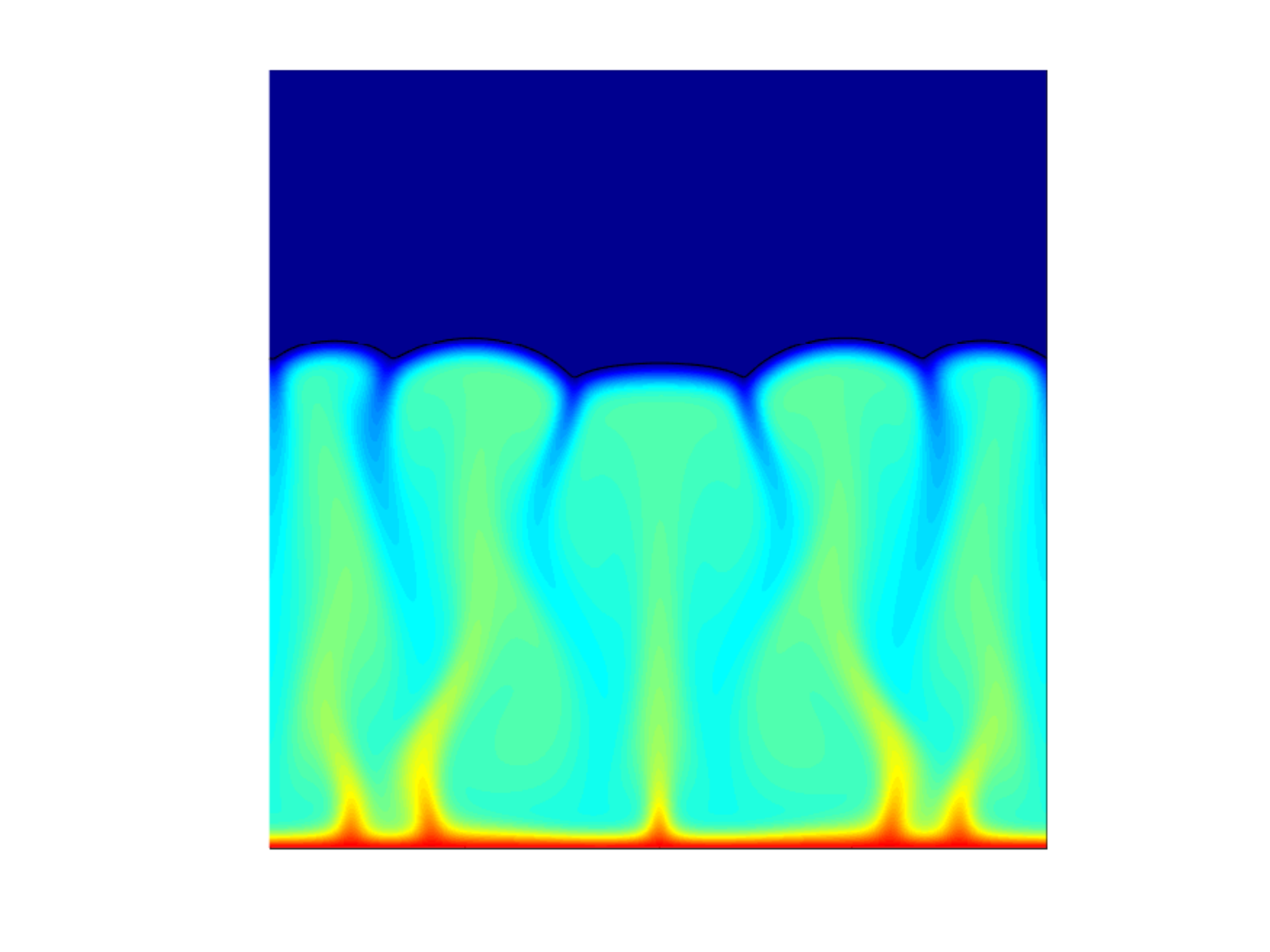}&
  \includegraphics[scale=0.25]{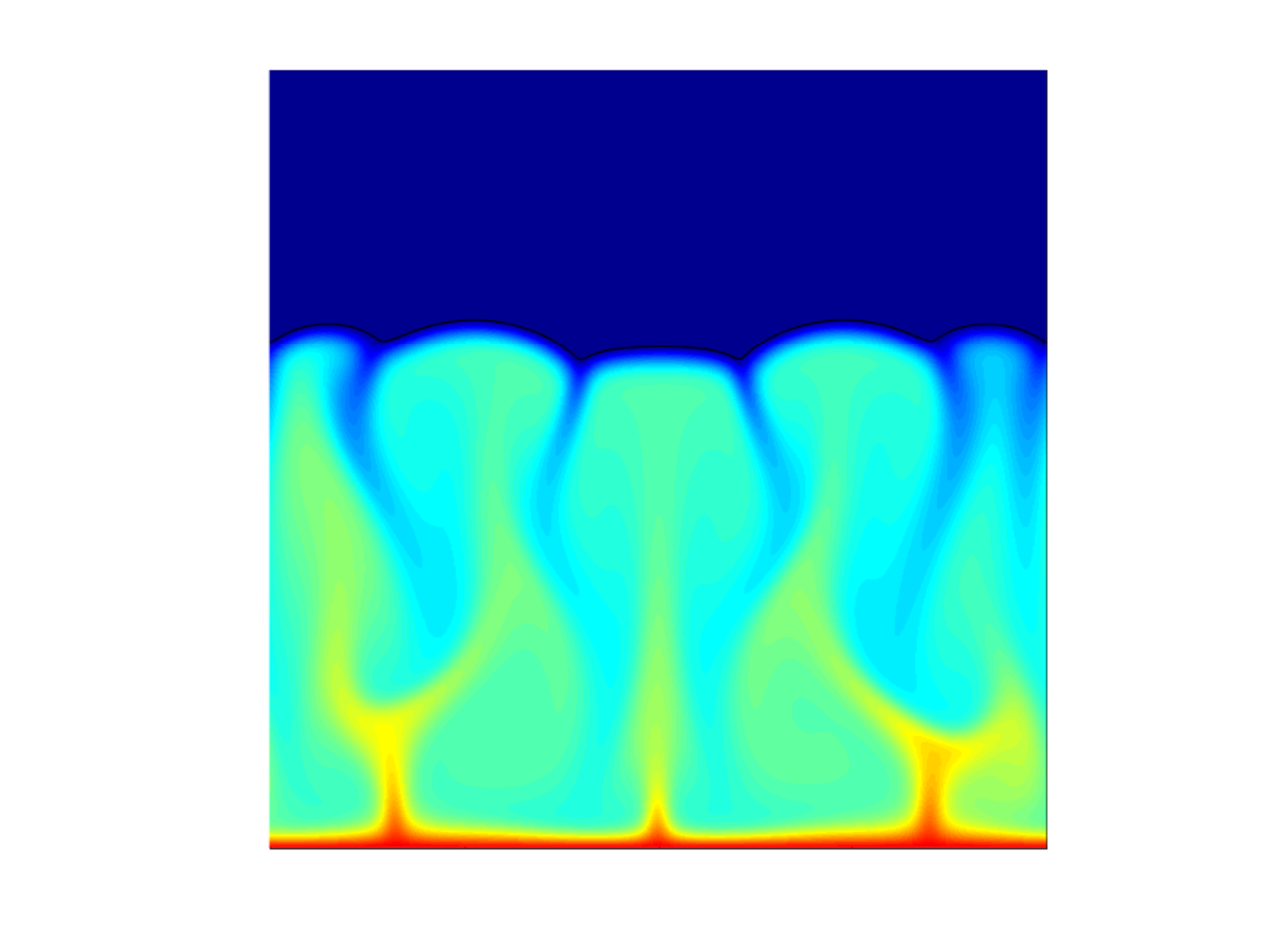}\\
  \hline \\
  & e) $t=590s$ &  f) $t=600s$ &  c) $t=620s$ &  d) $t=670s$ \\
   \rotatebox{90}{\large Stream Function}&
   \includegraphics[scale=0.25]{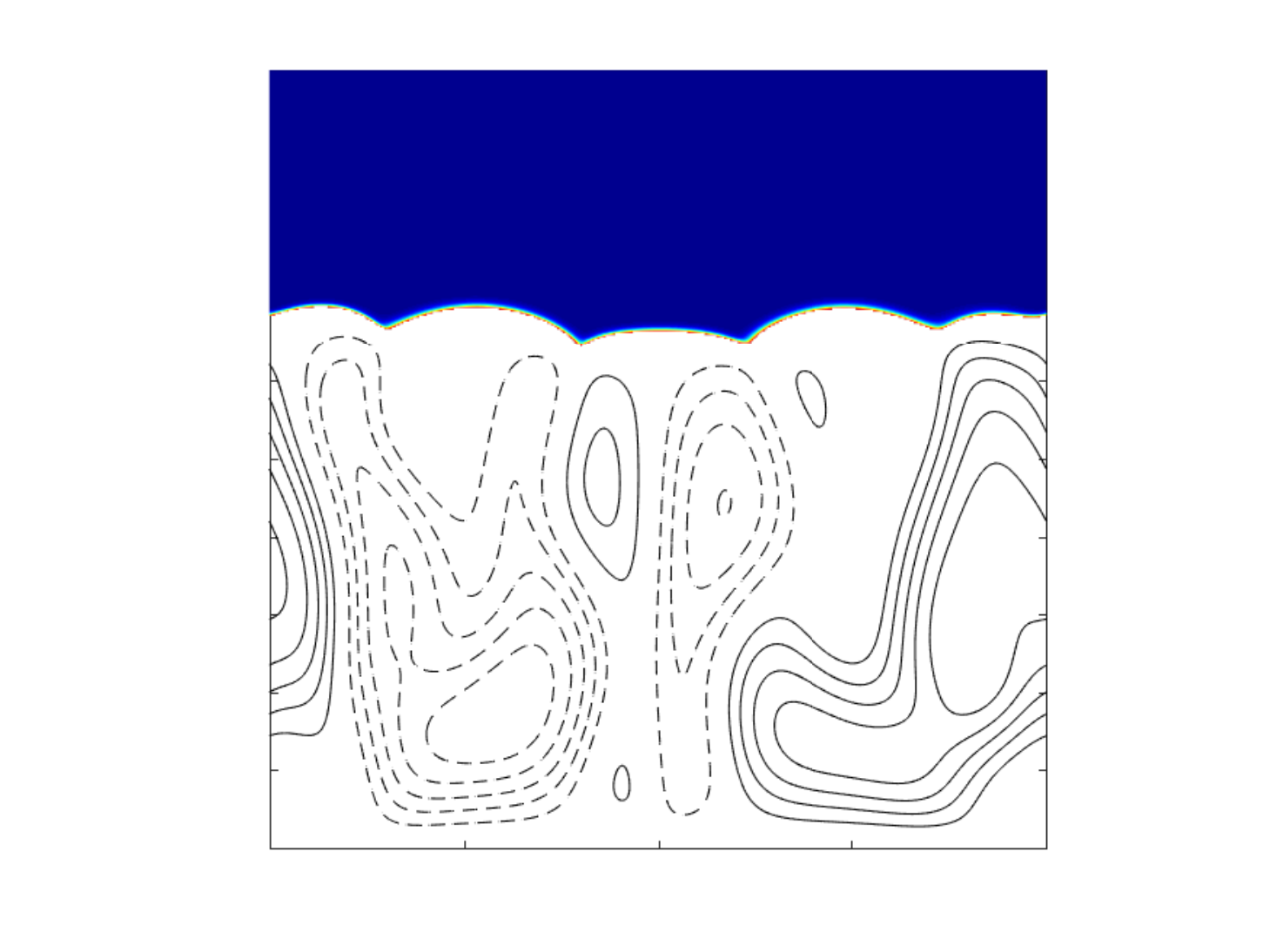}&
   \includegraphics[scale=0.25]{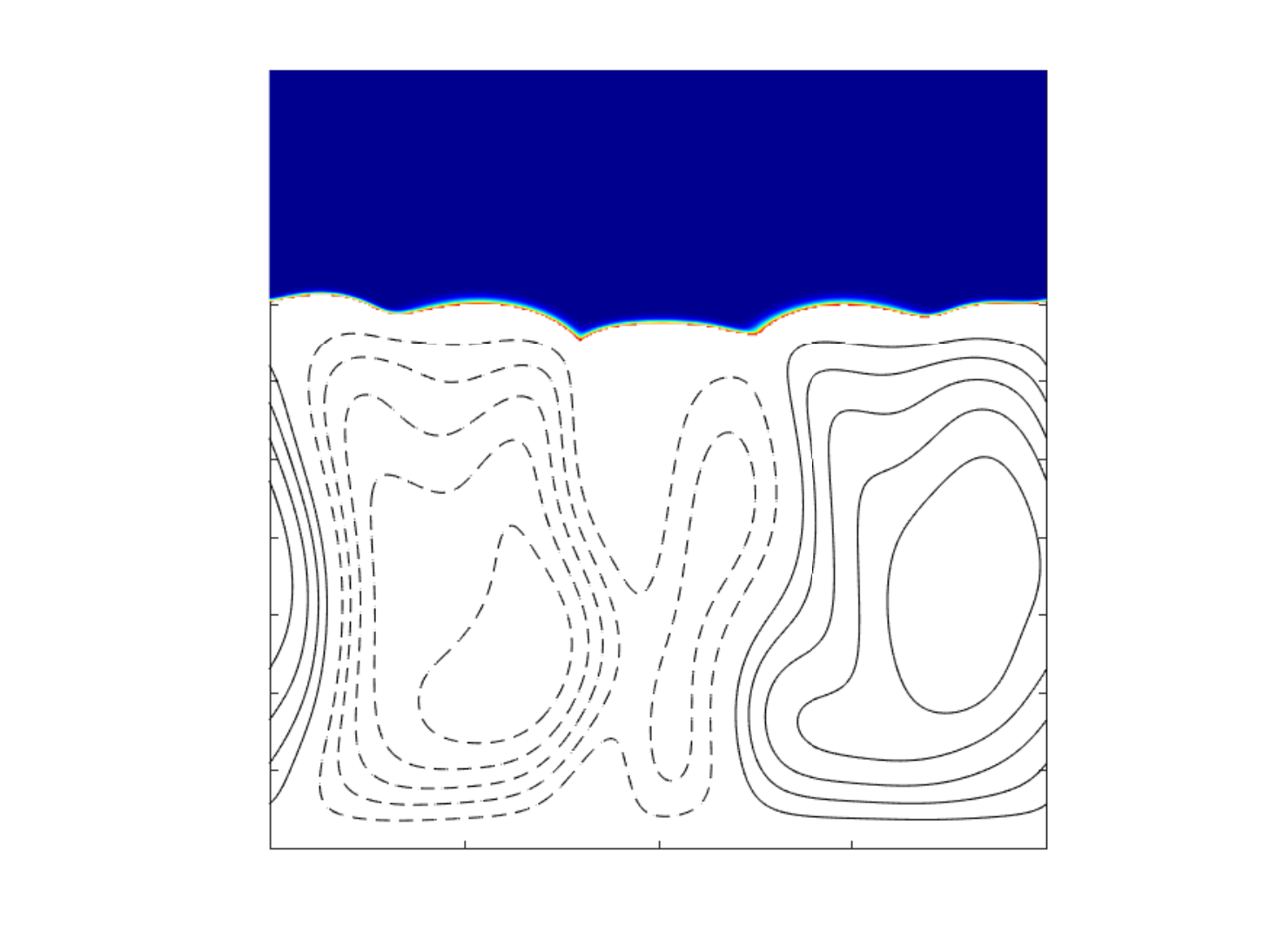}&
   \includegraphics[scale=0.25]{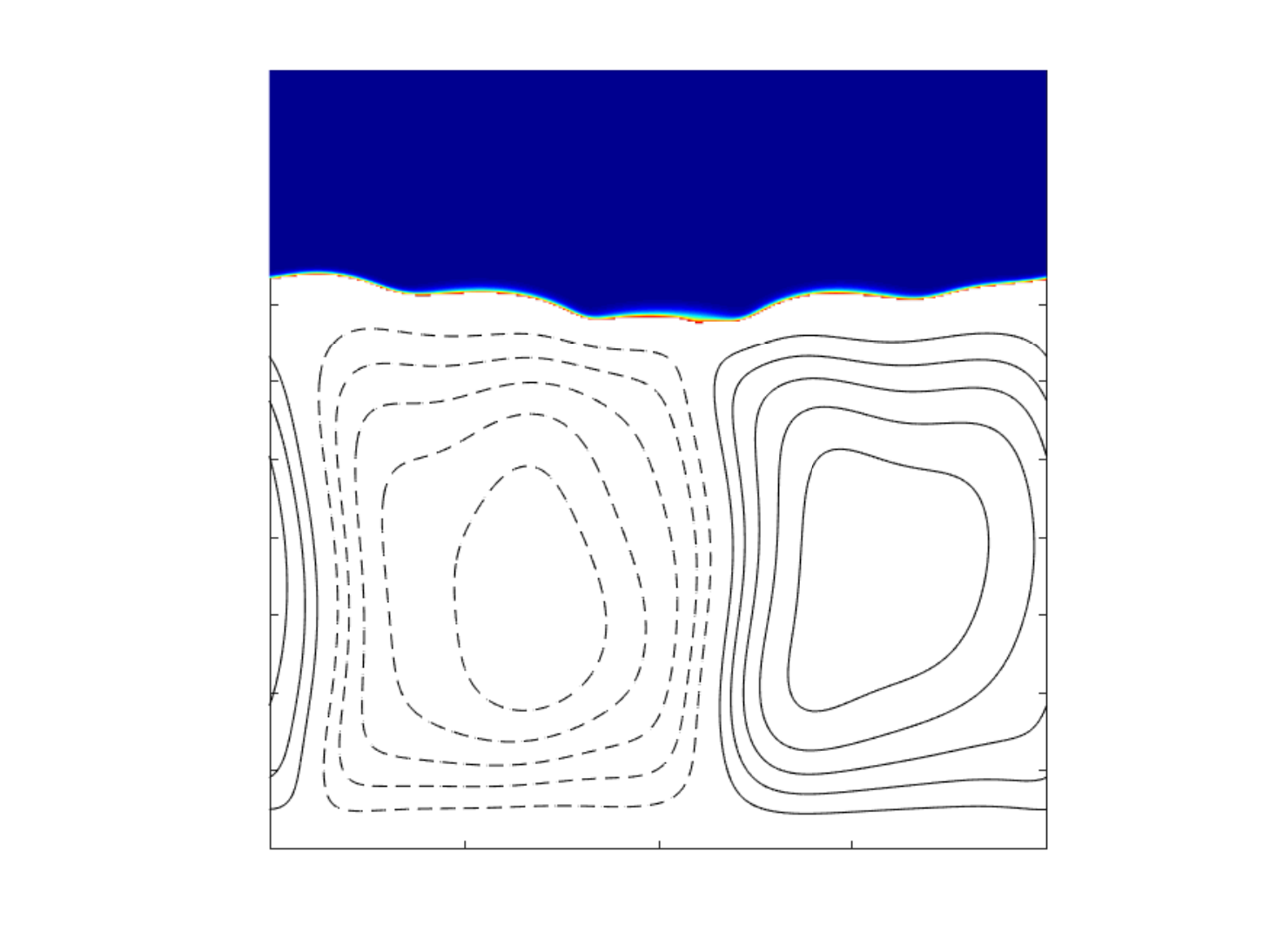}&
   \includegraphics[scale=0.25]{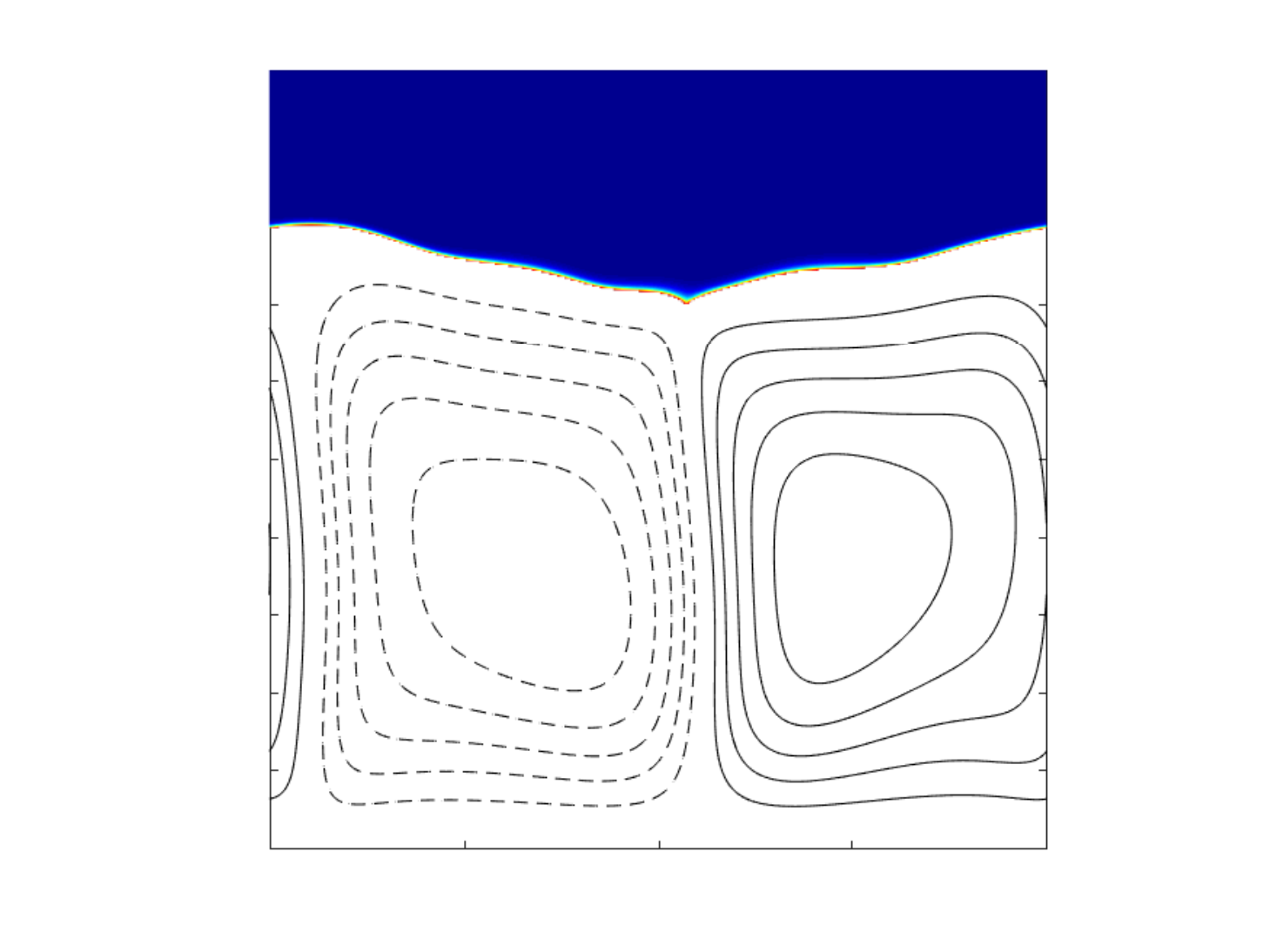}\\
  \rotatebox{90}{\large Temperature field}&  
   \includegraphics[scale=0.25]{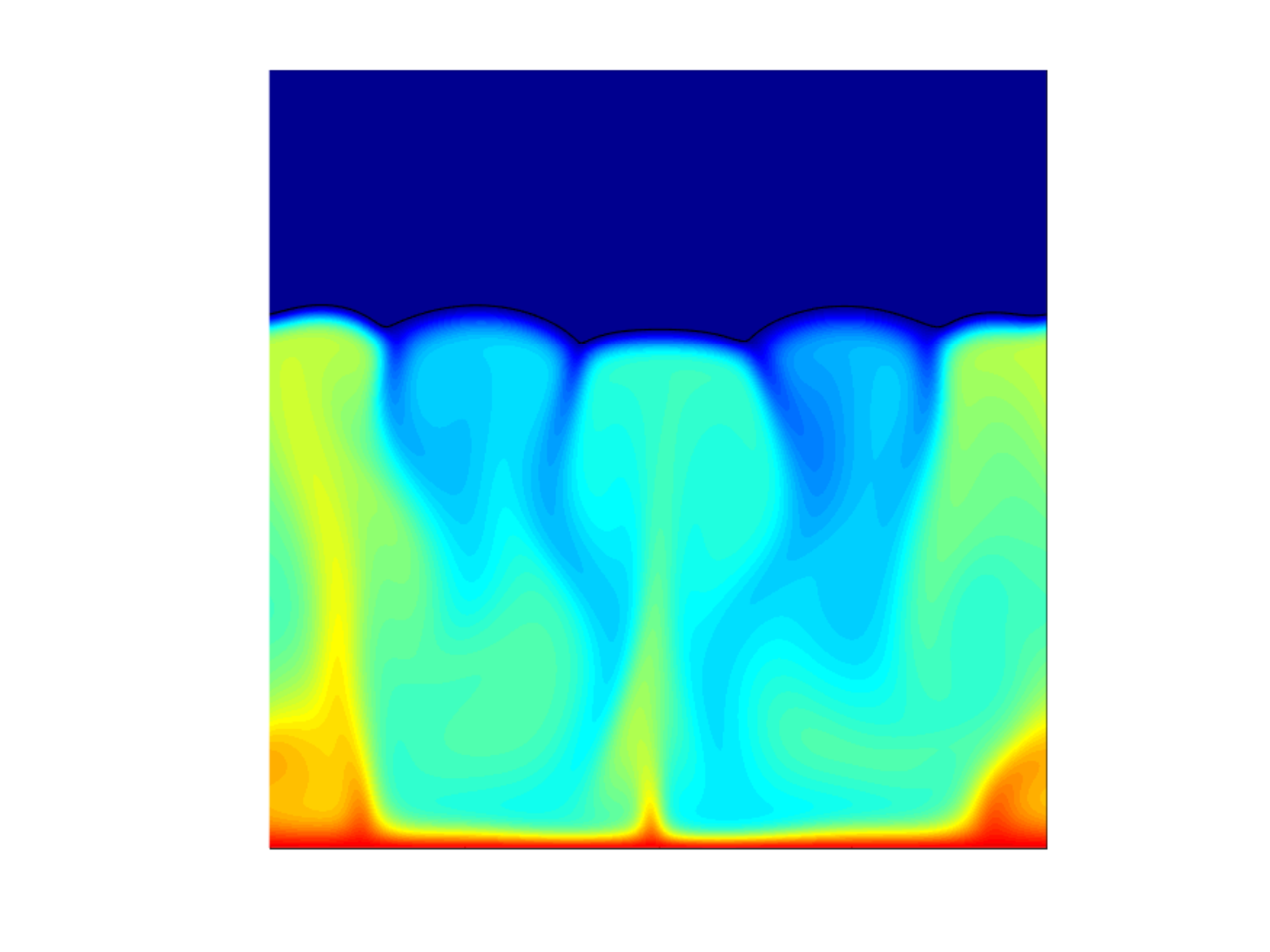}&
  \includegraphics[scale=0.25]{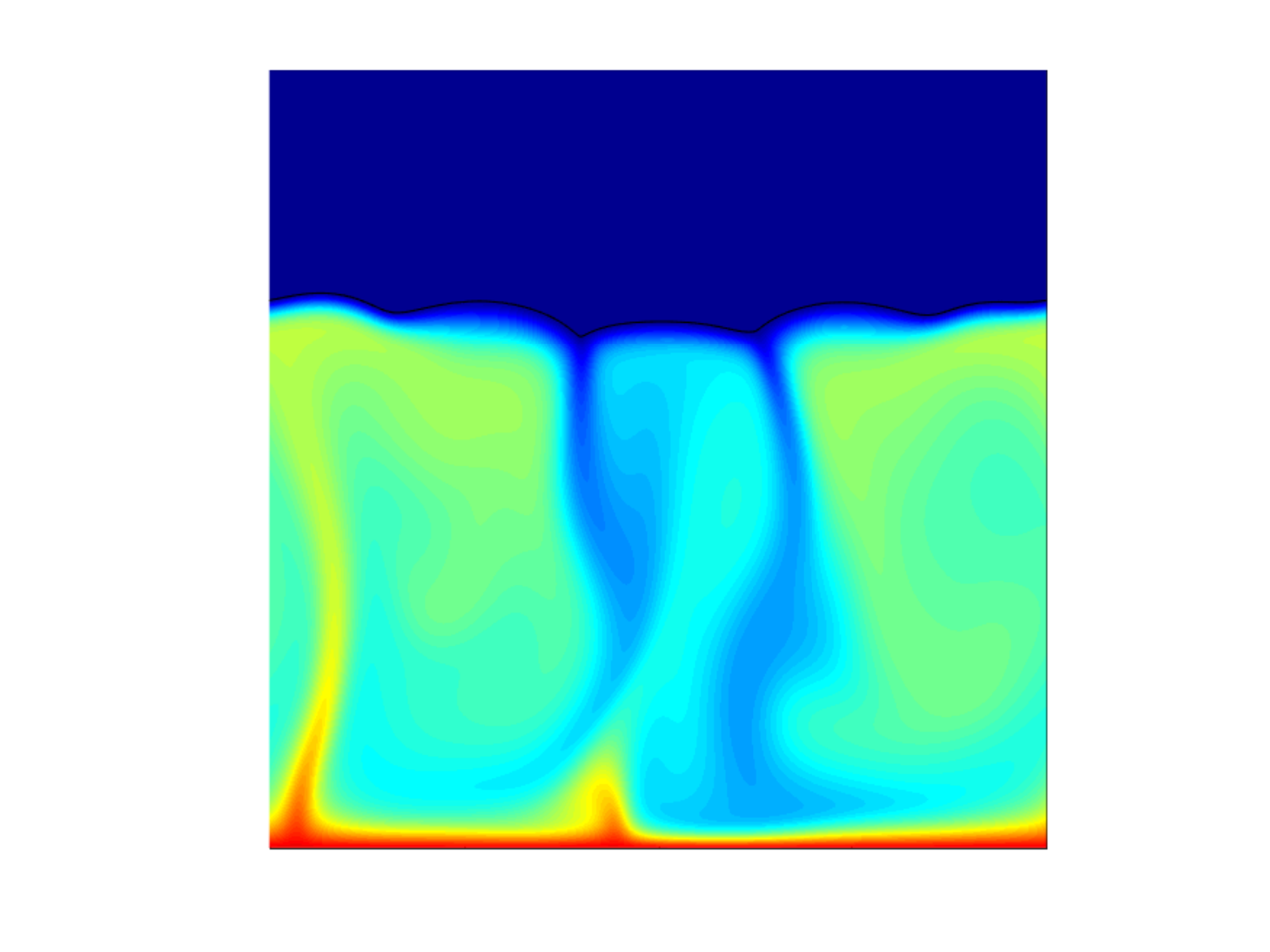}&
   \includegraphics[scale=0.25]{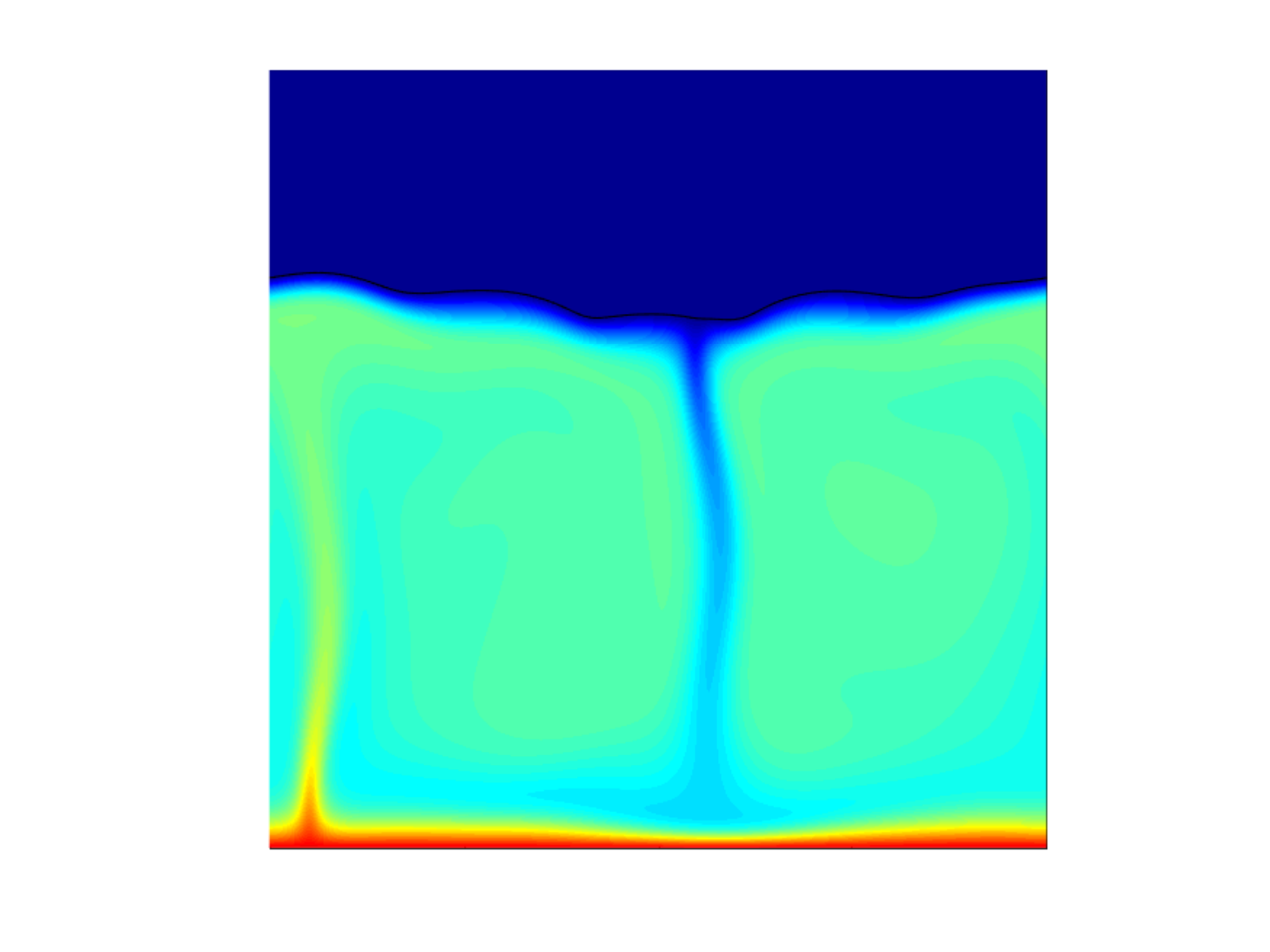}&
    \includegraphics[scale=0.25]{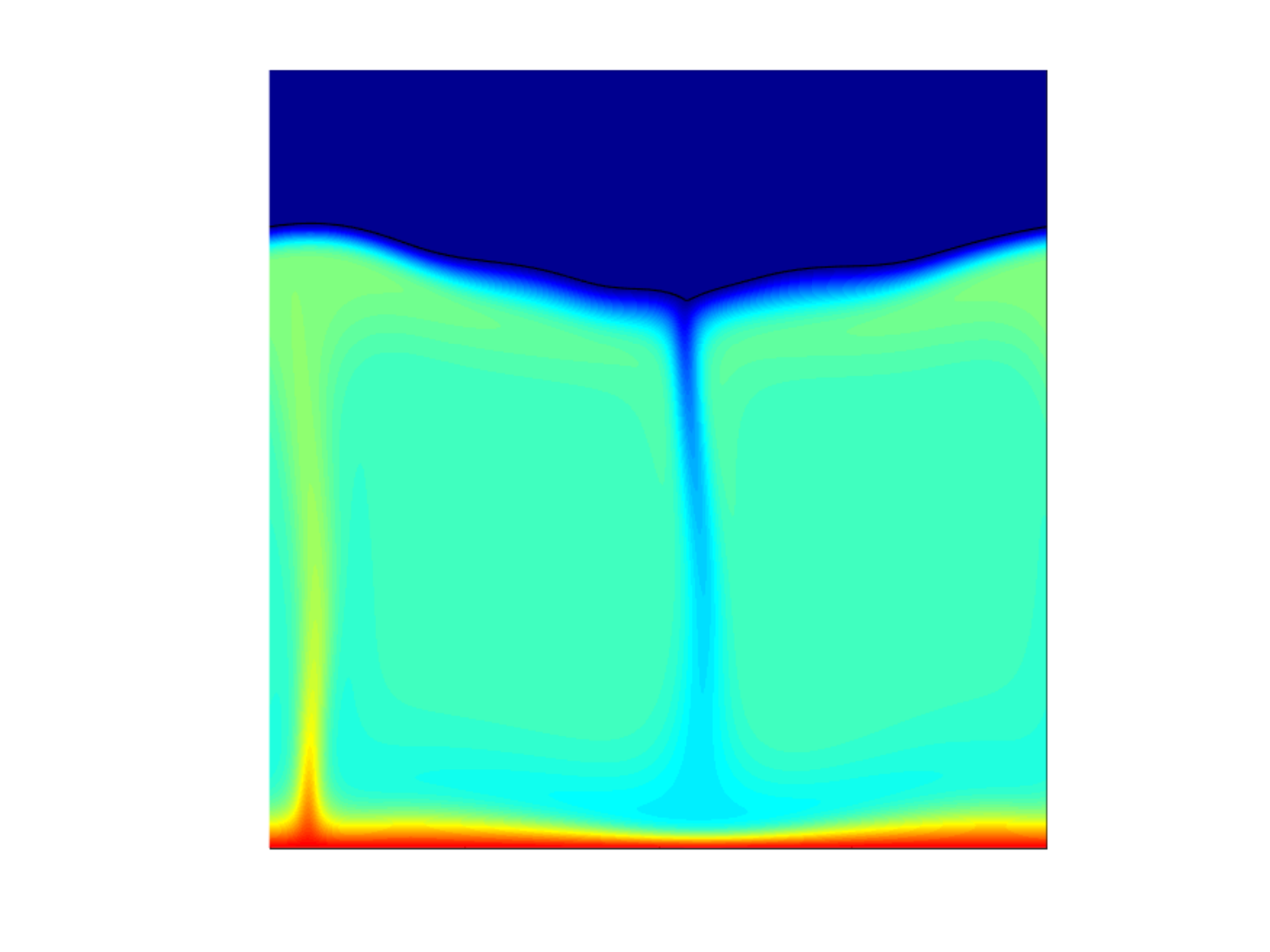}
\end{tabular}
\caption{Sequence of snapshots  for the stream function (top row) and temperature field (bottom row) at different times for a domain of side 
$L=0.02\,m$. The sequence shows the coarsening of five plumes in the first panel into a single plume in the last panel after suffering a secondary instability. 
  \label{fig:coarsening_regime_002}}
%\caption{Stream function and Temperature field for a square grid of $L=2cm$ at different times.}\label{F:Snapshot_002}
\end{figure}
\subsubsection{Coarsening regime}

 With the further advance in melting appears a regime where the convective cells of the linear regime destabilize and merge  up to 
recover roughly an aspect ratio close to the cells at the onset of the Rayleigh-B\'enard instability. {\em Notice that this aspect ratio is the fraction of the average depth of the convective cells with respect to the domain size $L$.}  This merging takes places  with the coarsening of the  thermal plumes of the linear regime.

\begin{figure}[!t]
\begin{tabular}{ccc|ccc}
& \multicolumn{2}{c}{First coarsening} && \multicolumn{2}{c}{Second coarsening}\\
\hline
& Stream Function & Temperature field & & Stream Function & Temperature field \\
 \rotatebox{90}{\quad \quad $t=400s$}&
\includegraphics[scale=0.23]{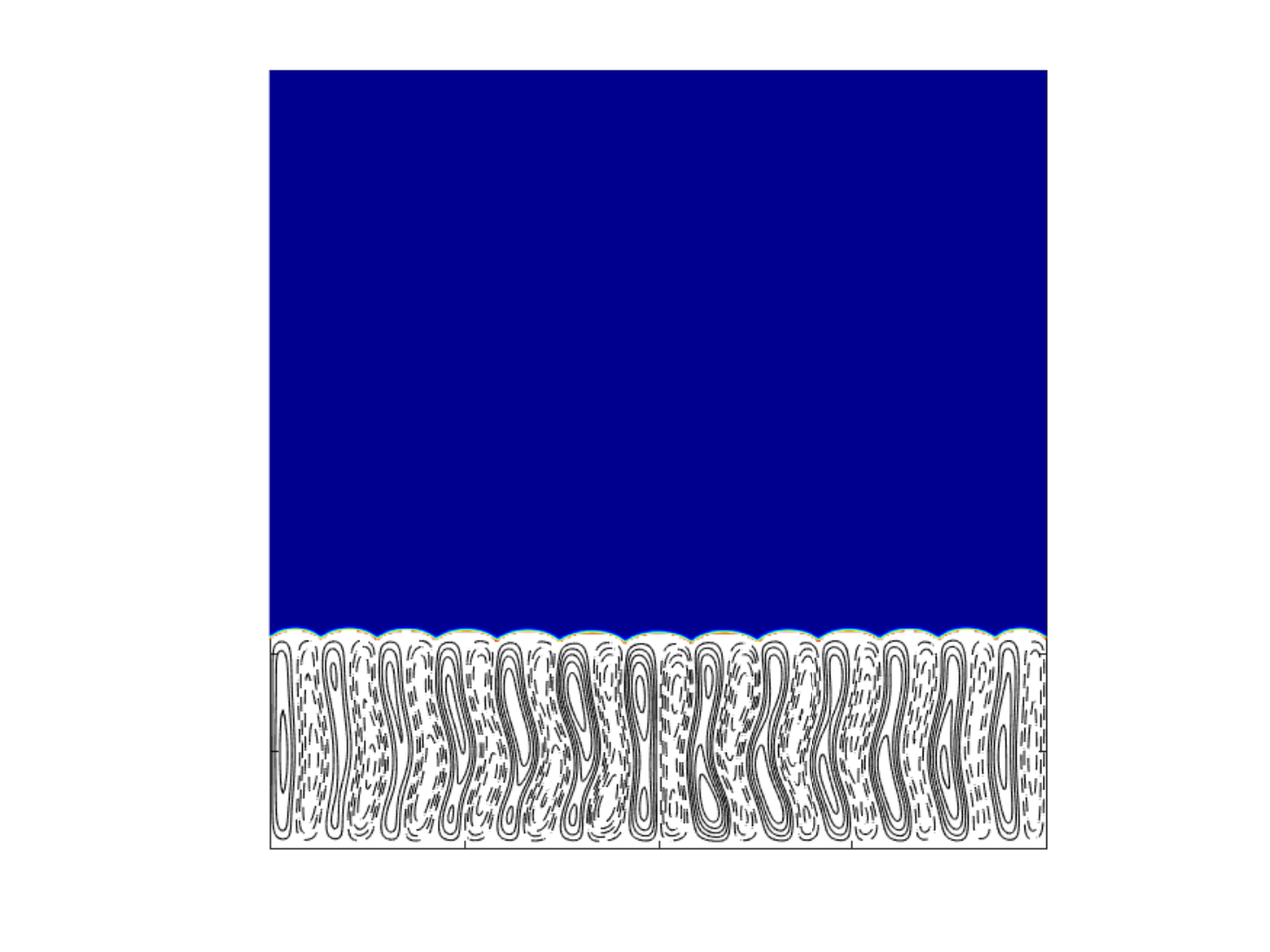}&
\includegraphics[scale=0.23]{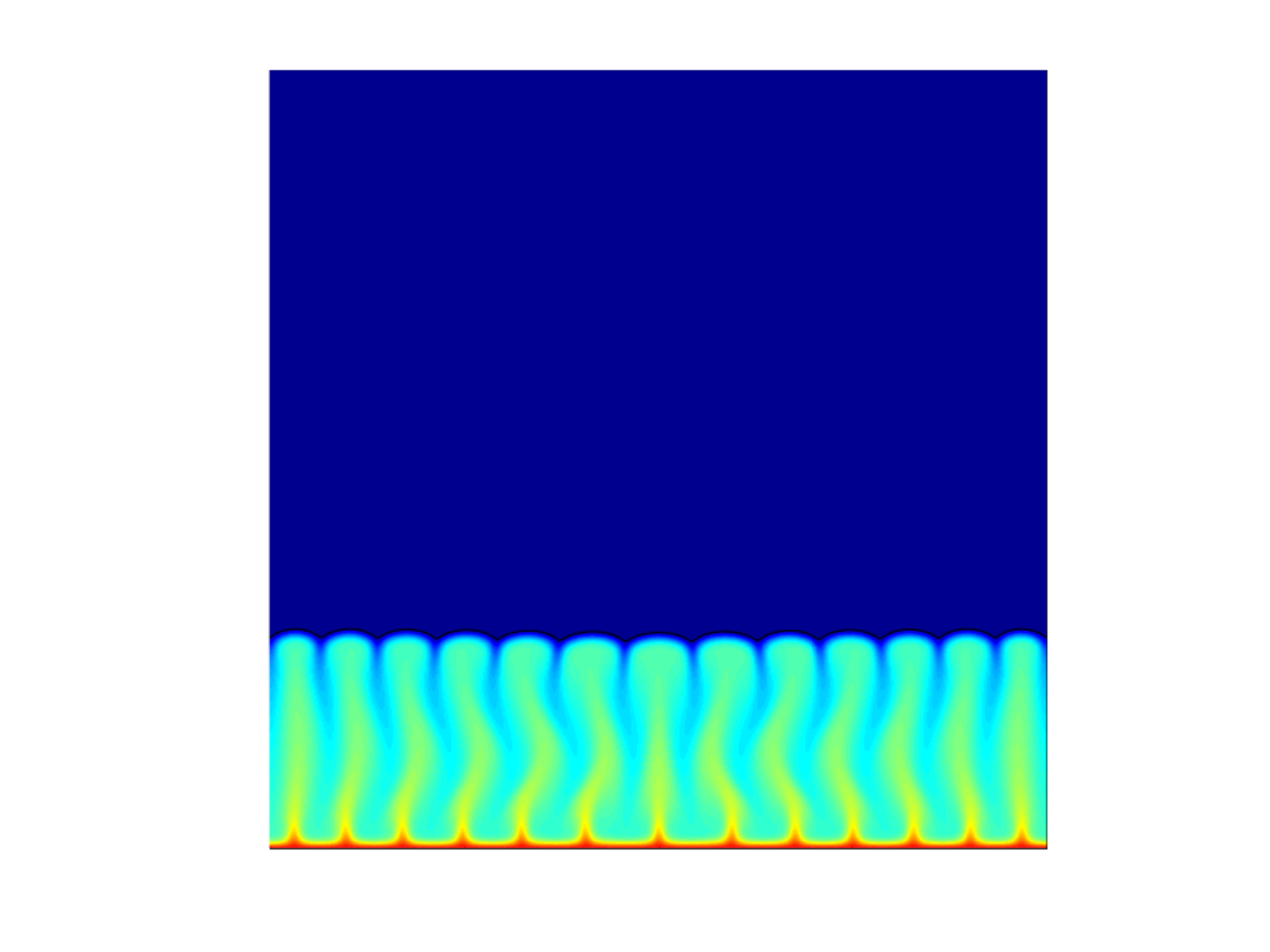}&
\rotatebox{90}{ \quad \quad $t=600s$}&
\includegraphics[scale=0.23]{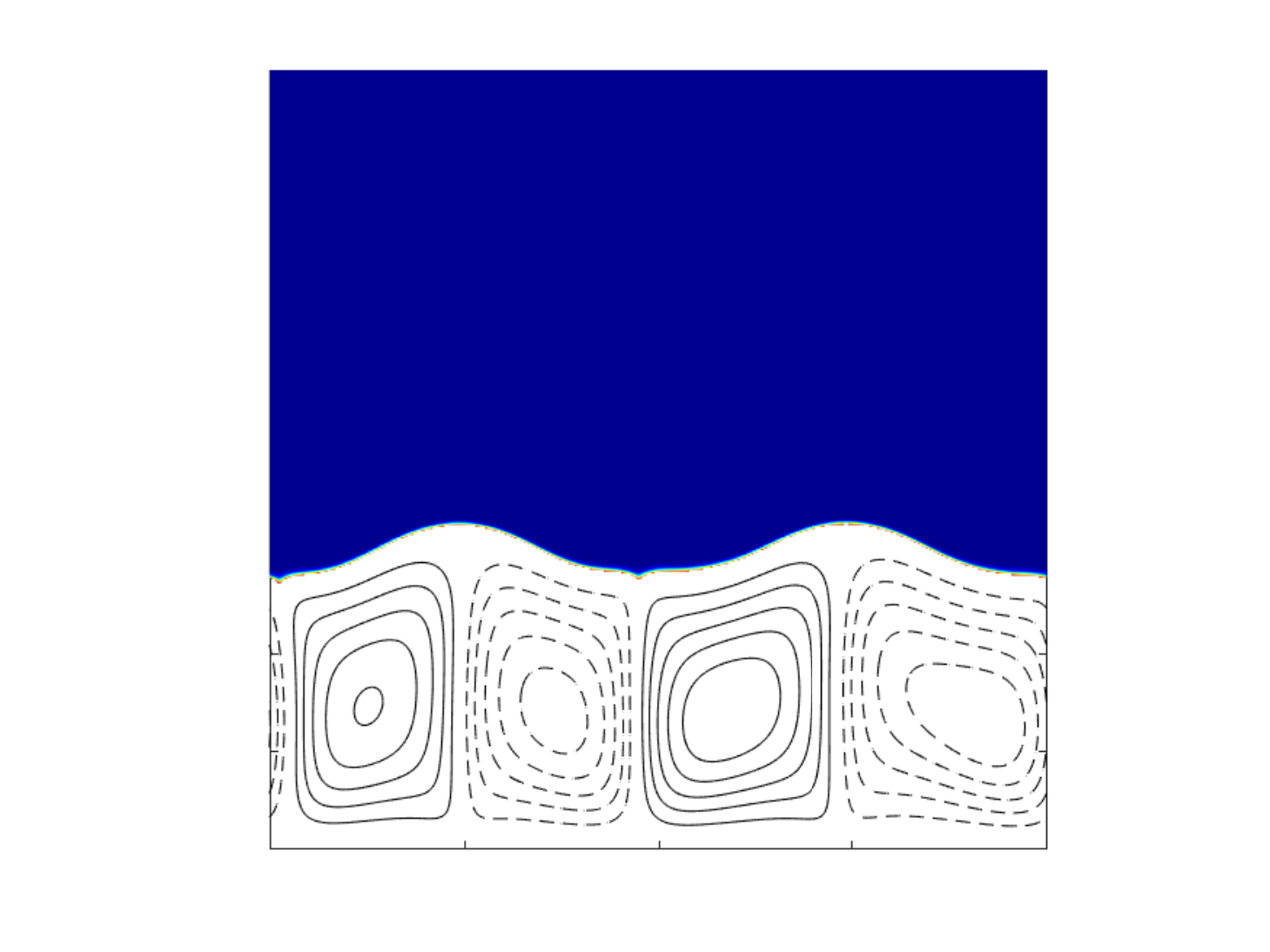}&
\includegraphics[scale=0.23]{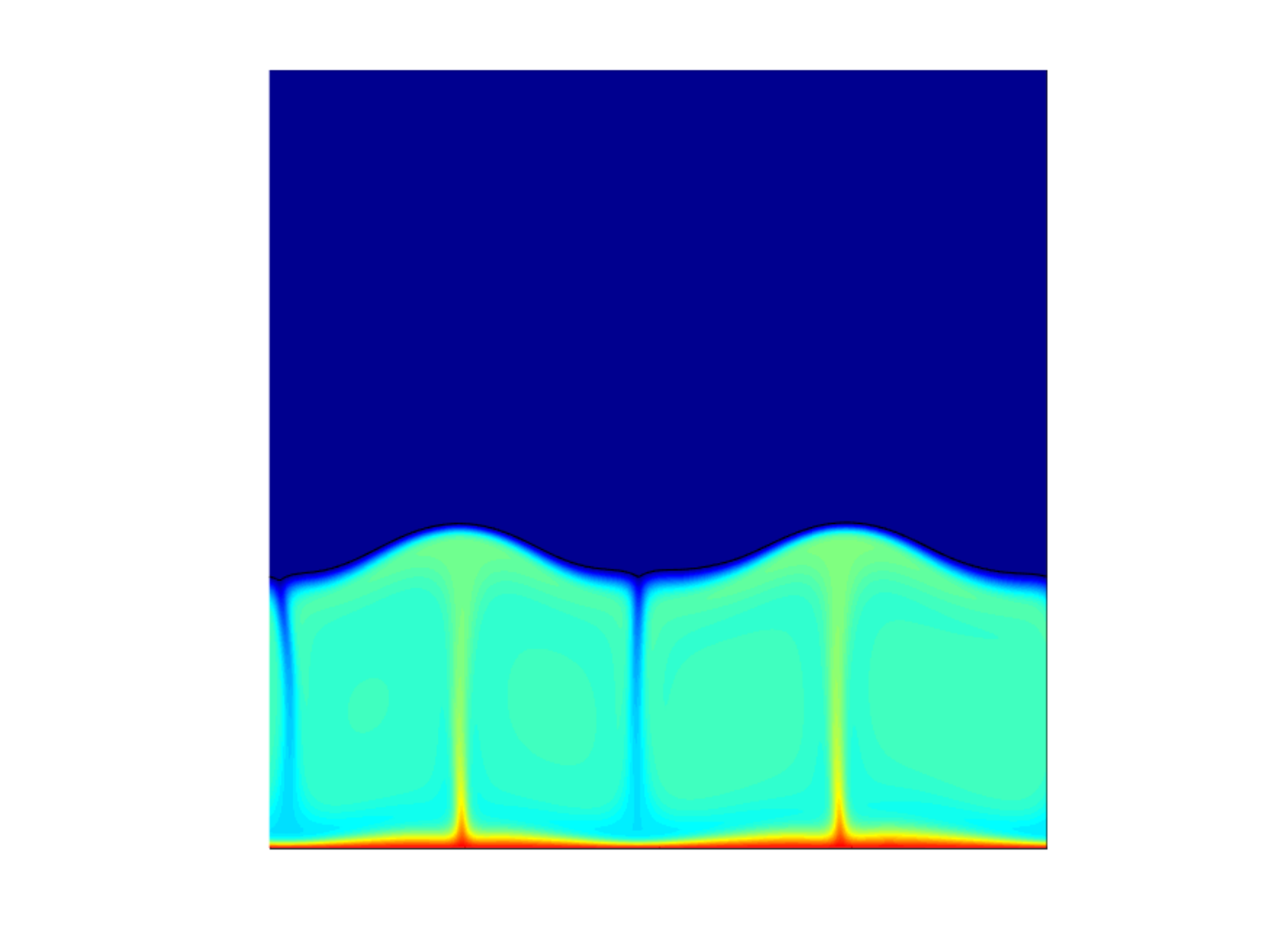}  \\
 \rotatebox{90}{\quad \quad $t=420s$}&
\includegraphics[scale=0.23]{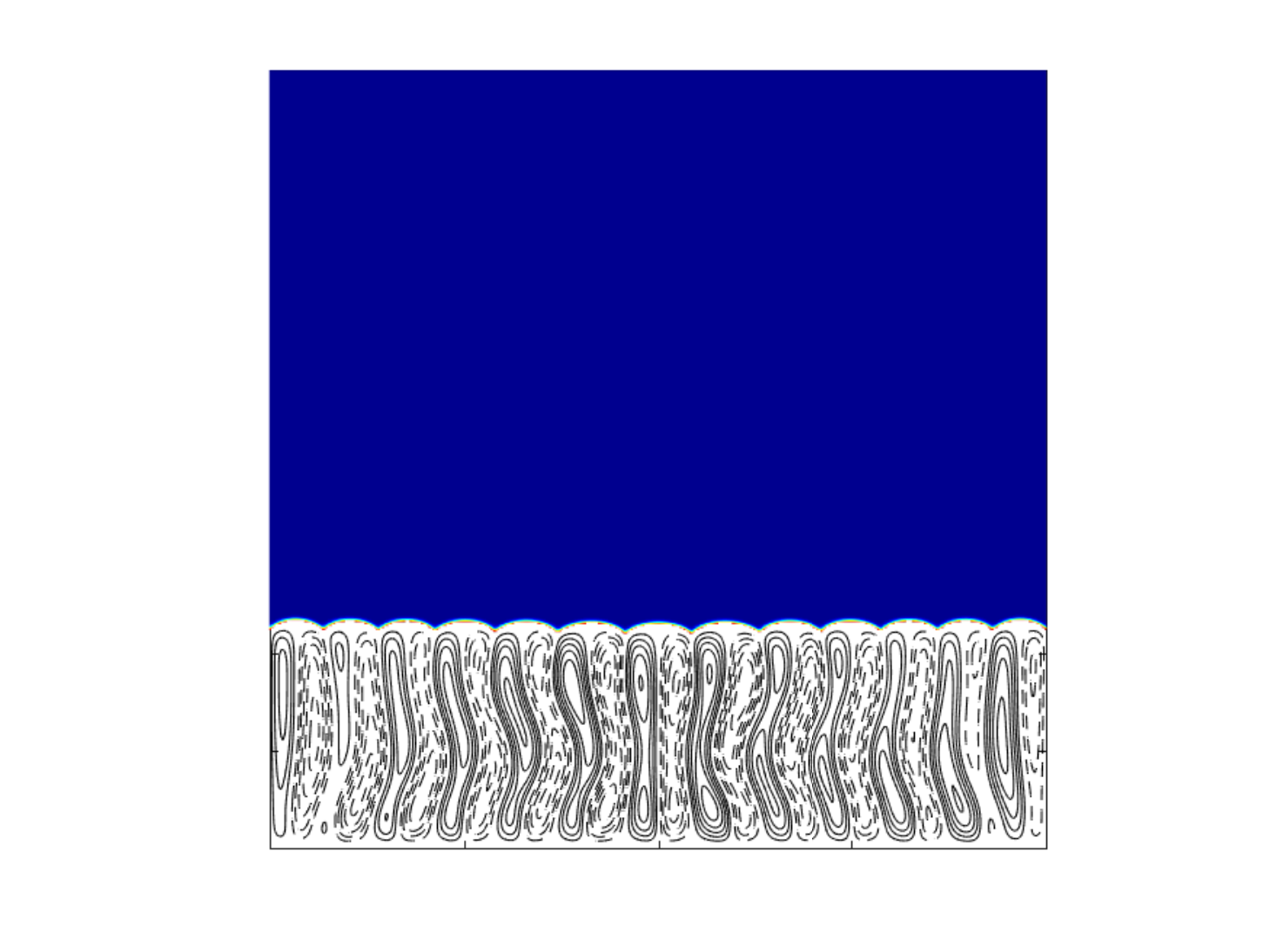}&
\includegraphics[scale=0.23]{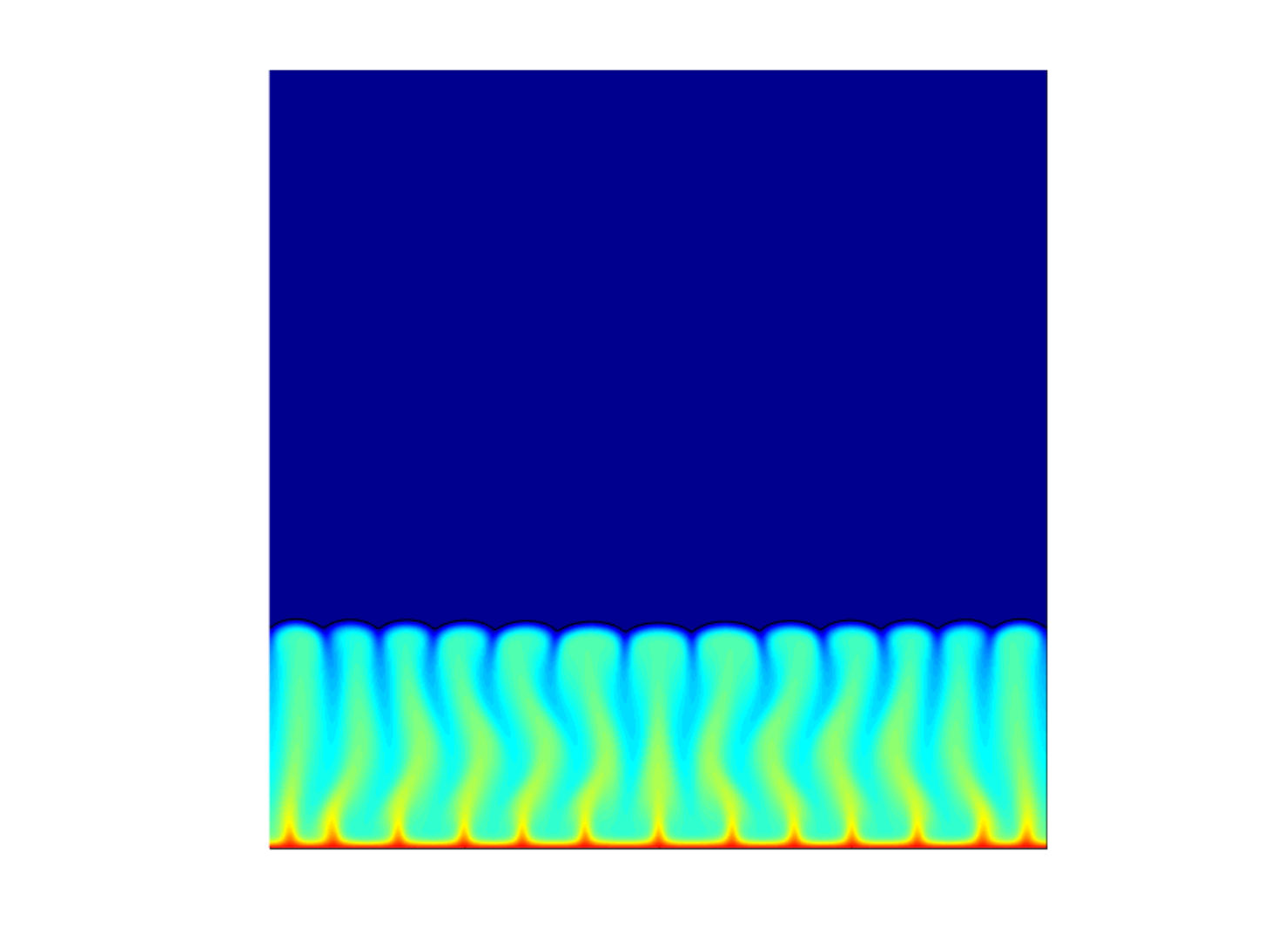}& 
\rotatebox{90}{ \quad \quad $t=750s$}&
\includegraphics[scale=0.23]{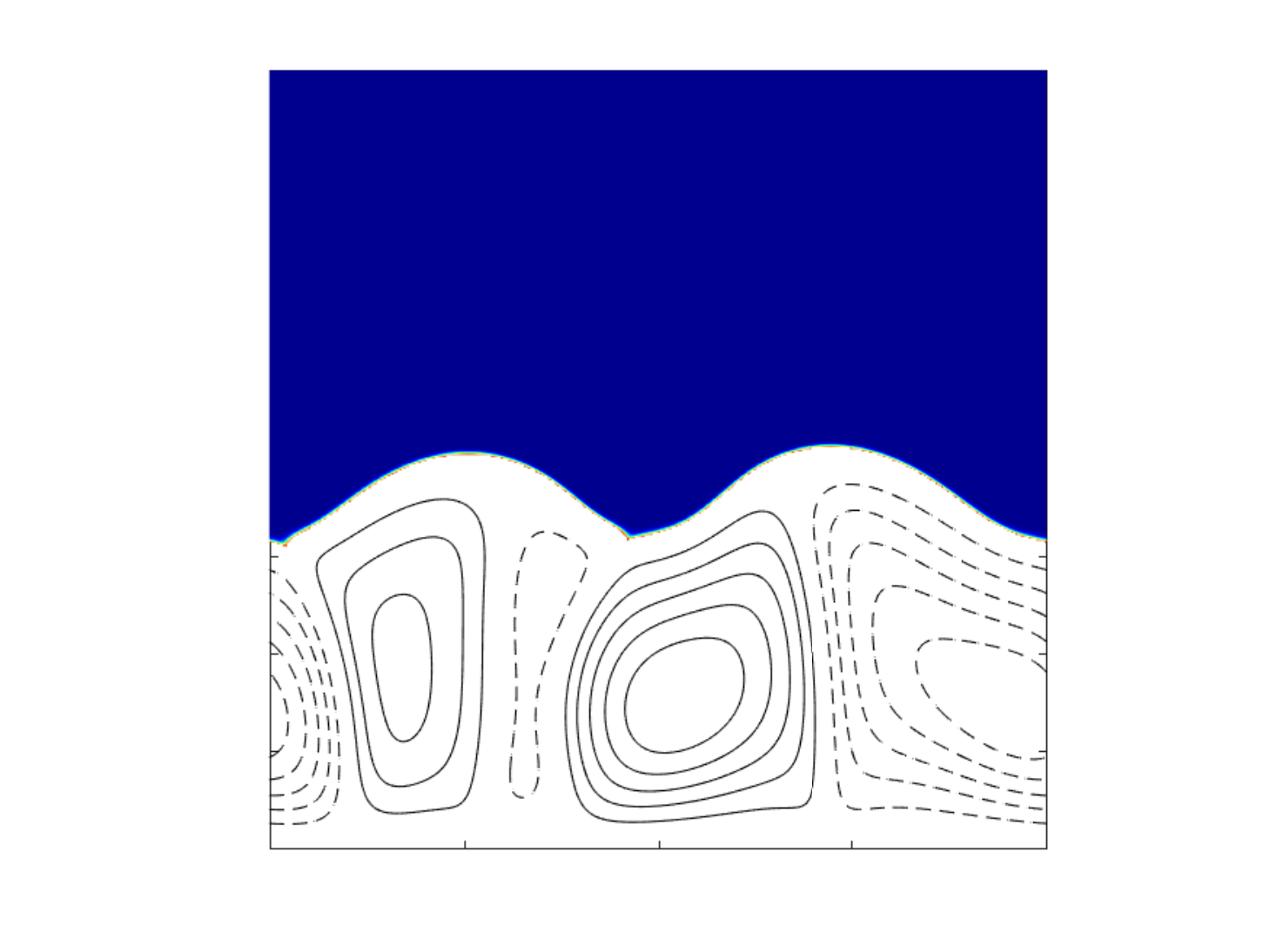}&
\includegraphics[scale=0.23]{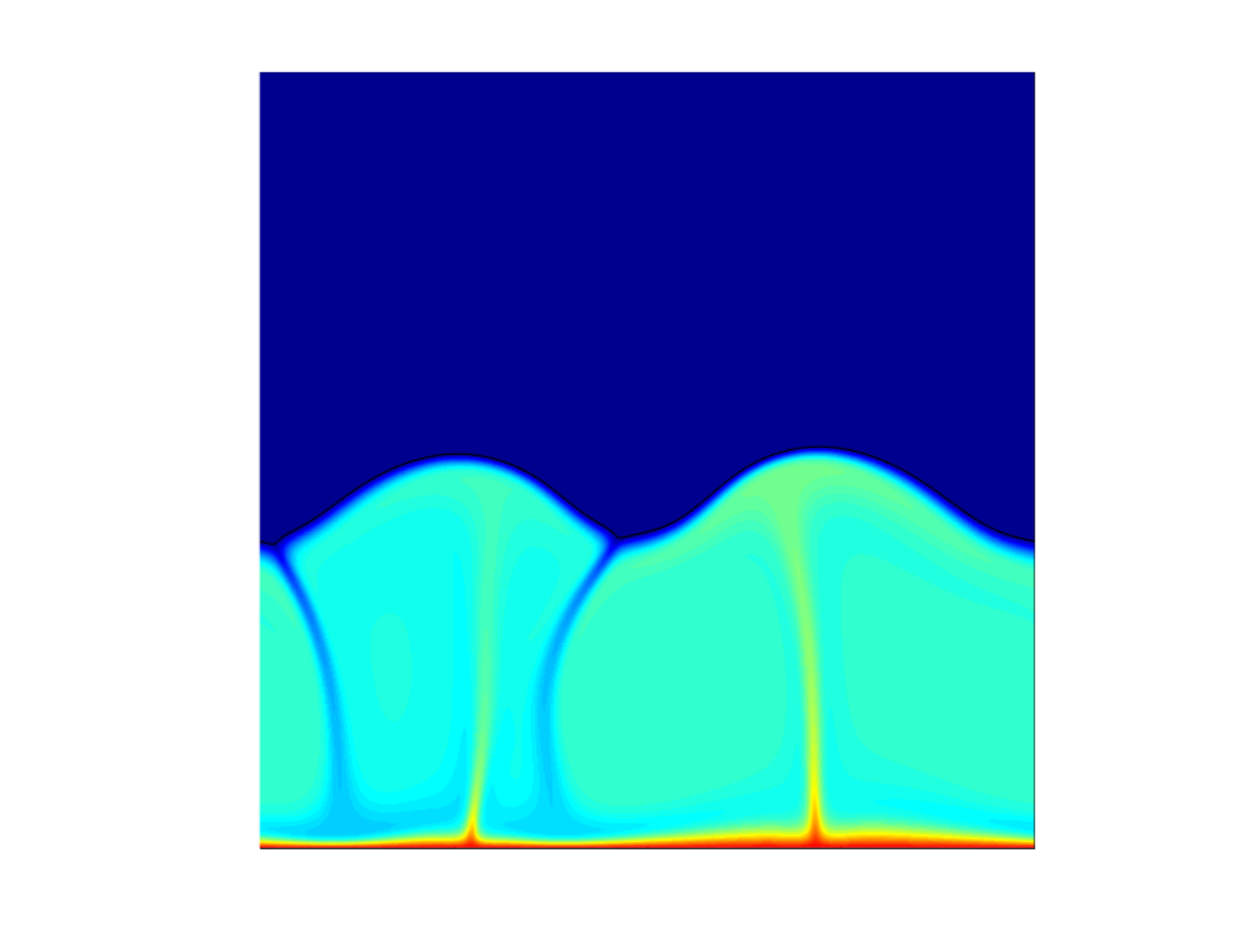} \\
\rotatebox{90}{\quad \quad $t=440s$}&
\includegraphics[scale=0.23]{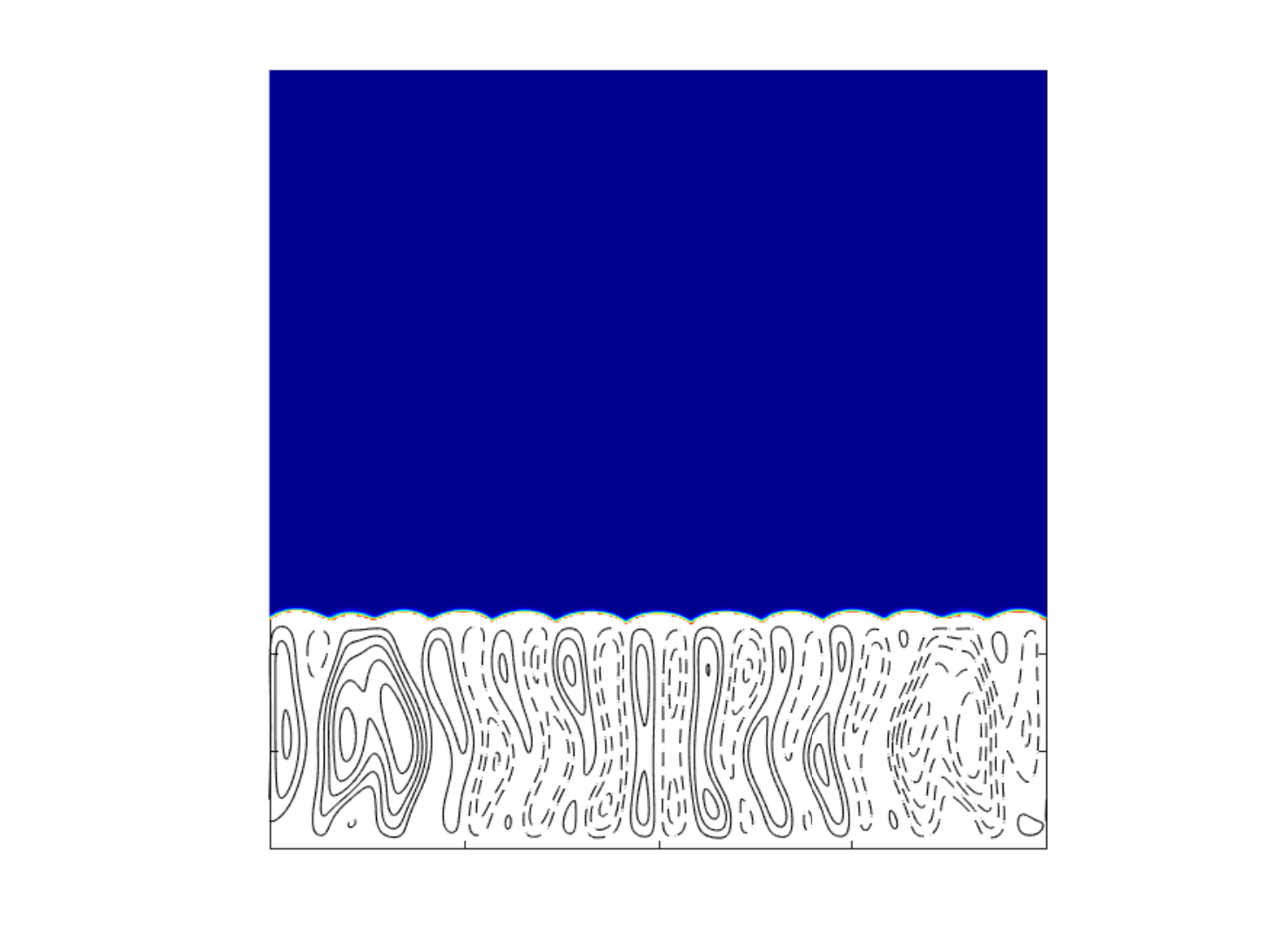}&
\includegraphics[scale=0.23]{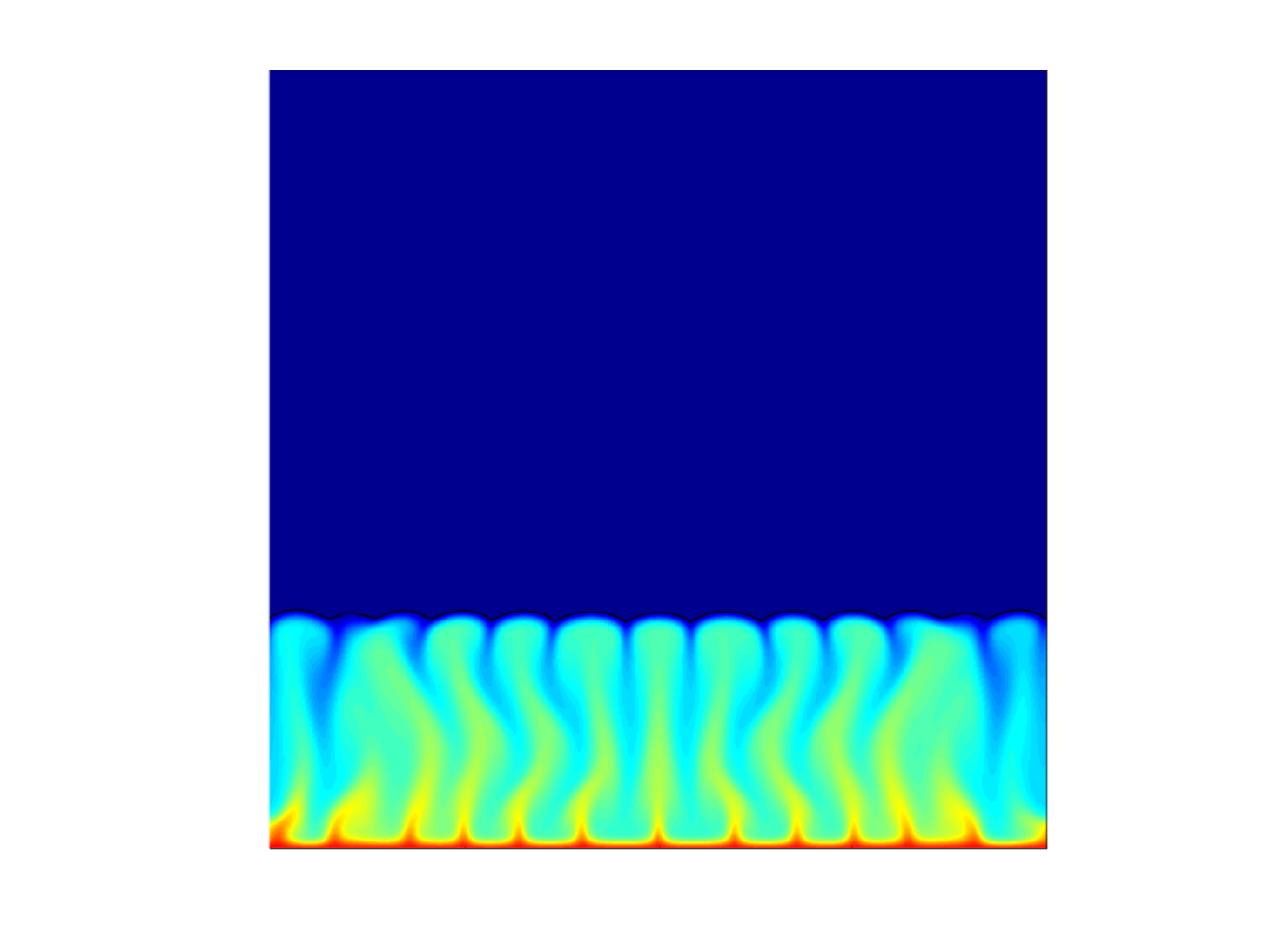}& 
\rotatebox{90}{ \quad \quad $t=765s$}&
\includegraphics[scale=0.23]{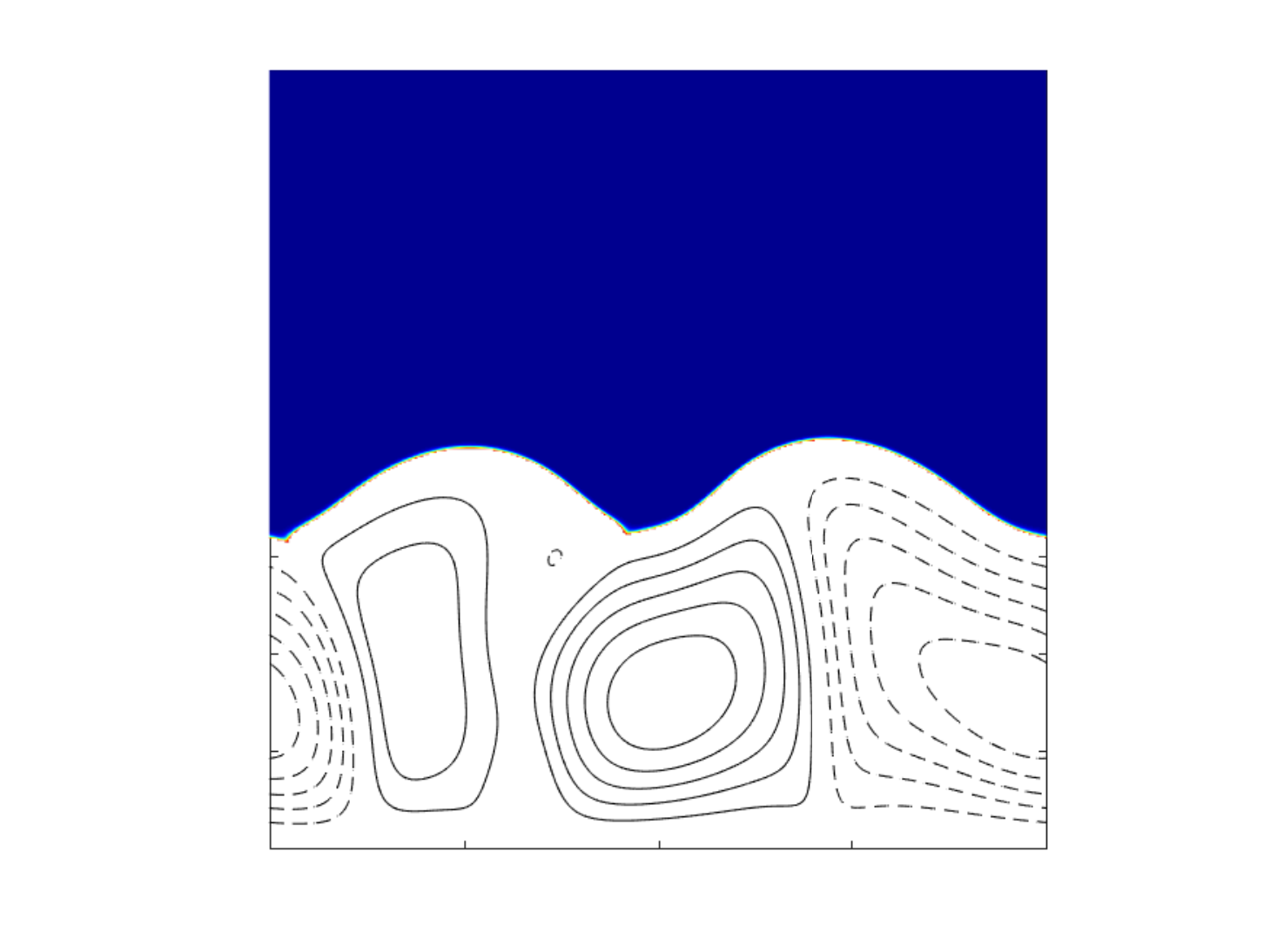}&
\includegraphics[scale=0.23]{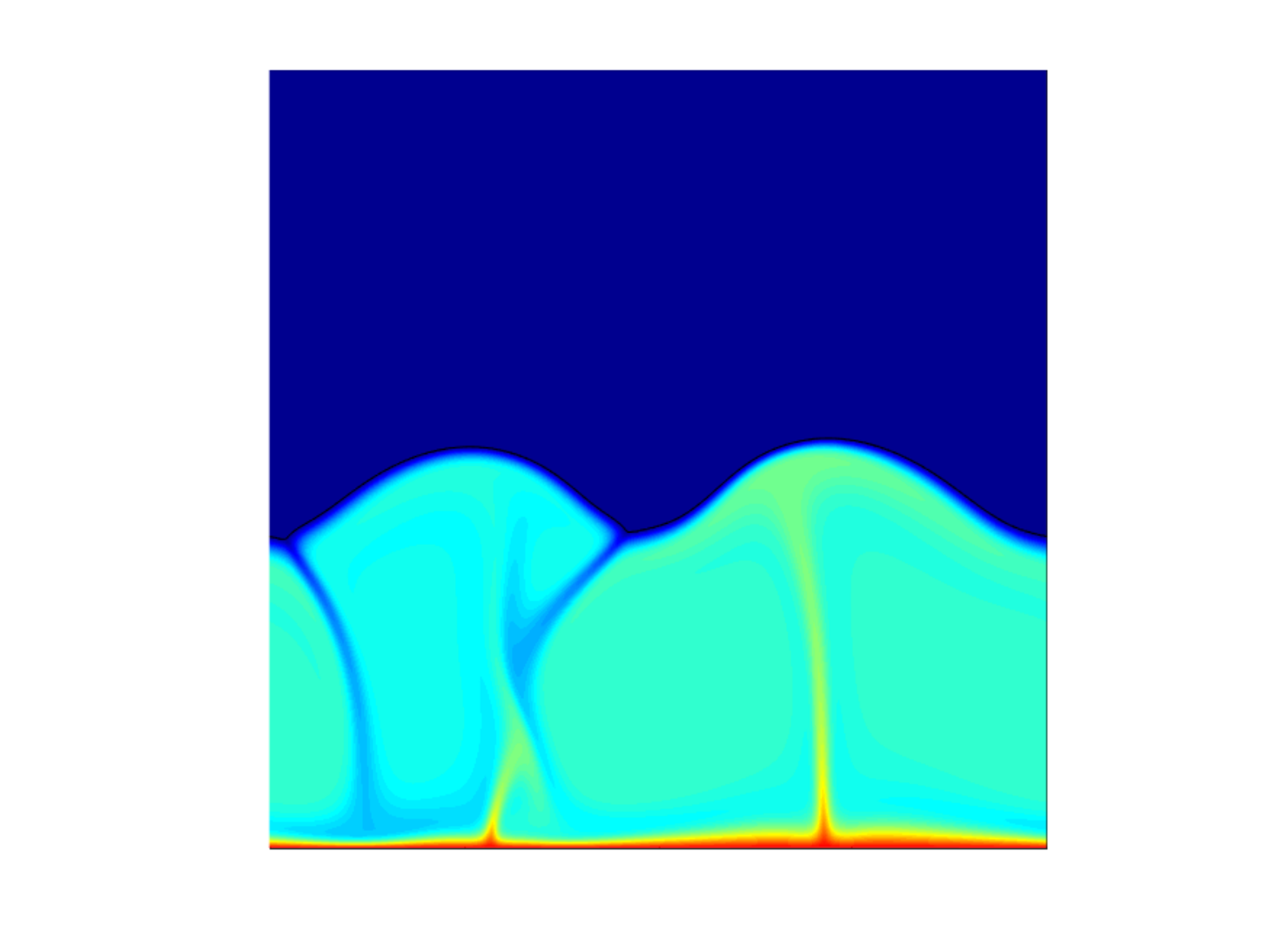} \\
\rotatebox{90}{\quad \quad $t=460s$}&
\includegraphics[scale=0.23]{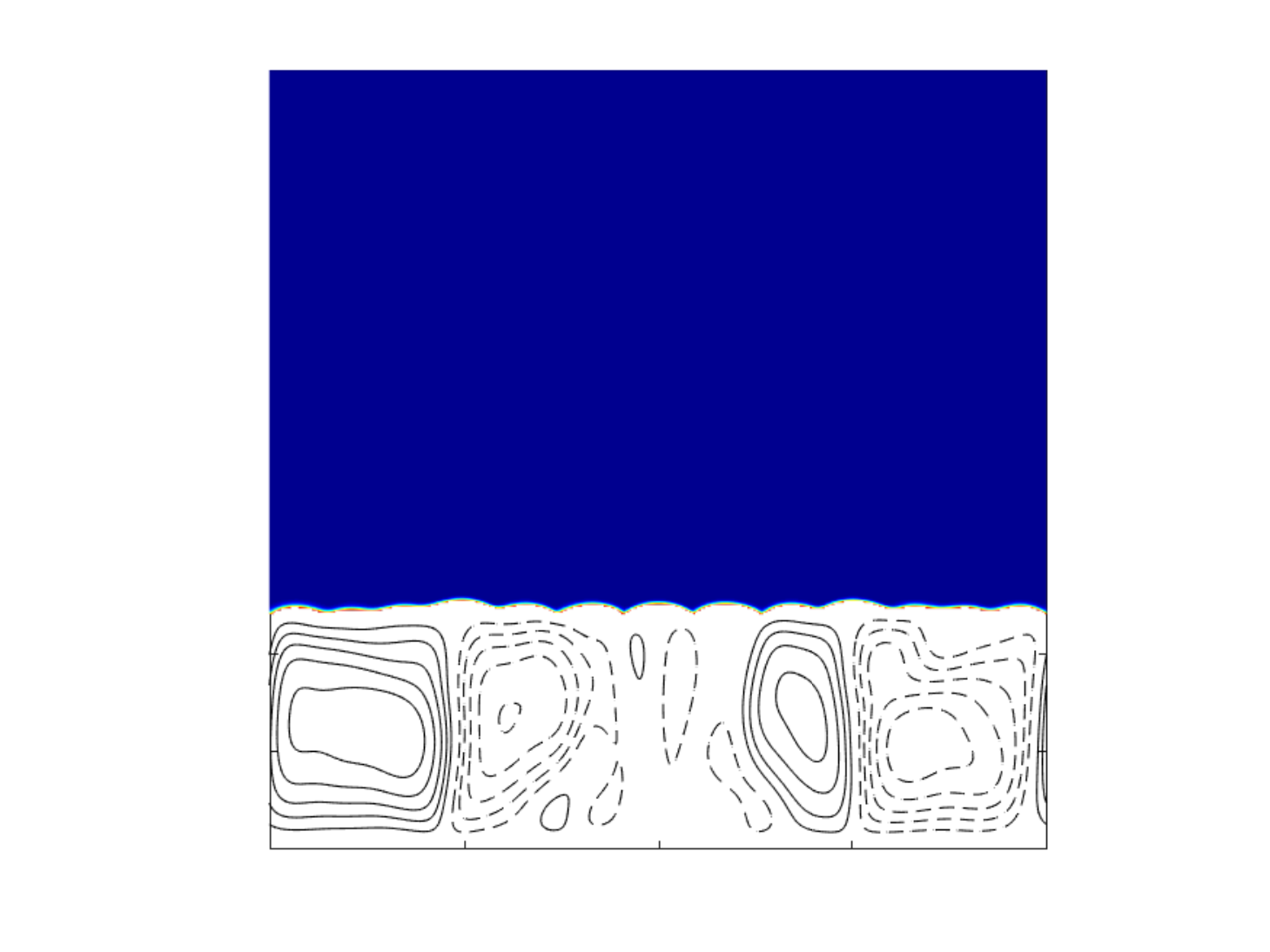}&
\includegraphics[scale=0.23]{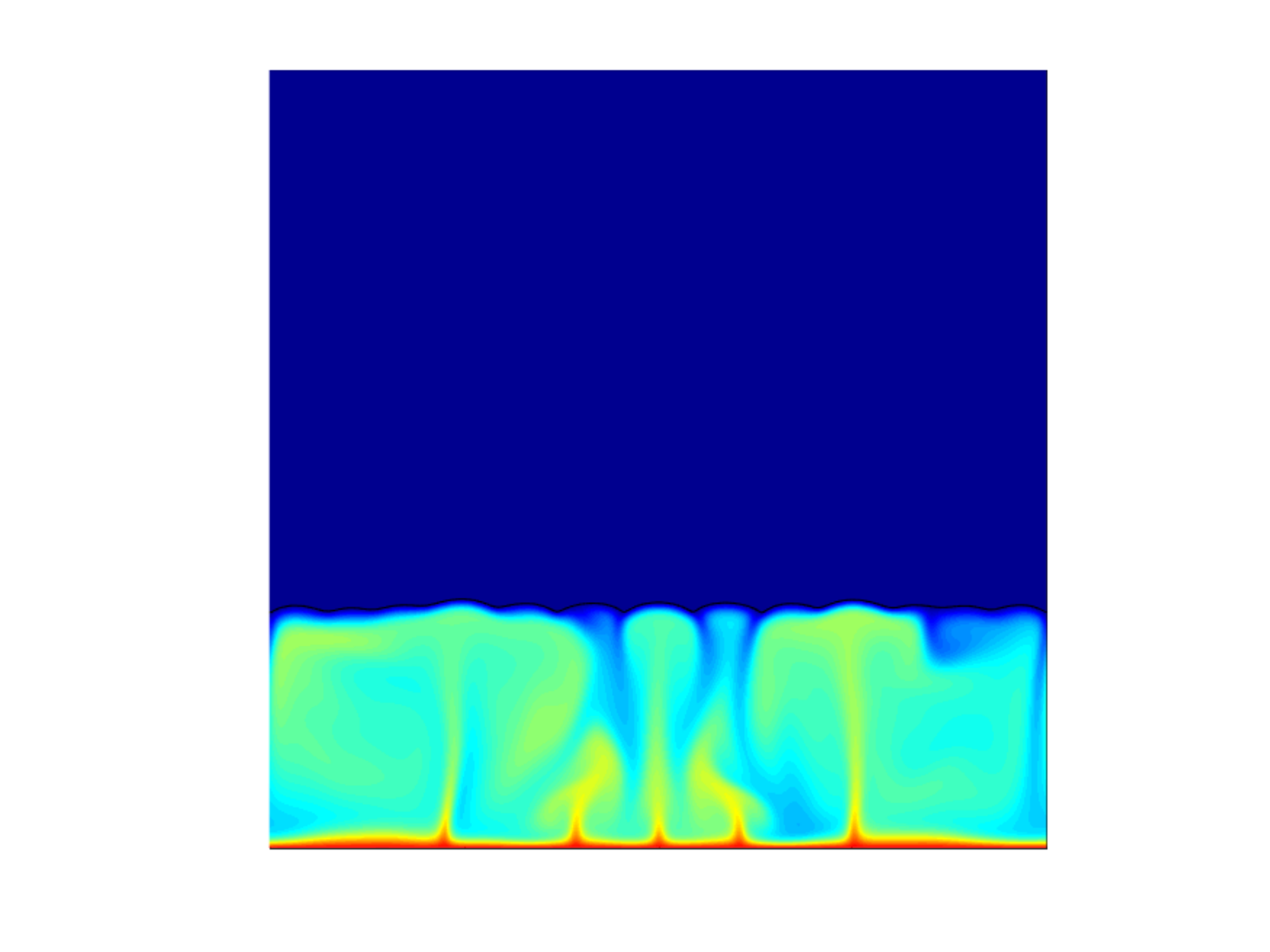}& 
\rotatebox{90}{ \quad \quad $t=770s$}&
\includegraphics[scale=0.23]{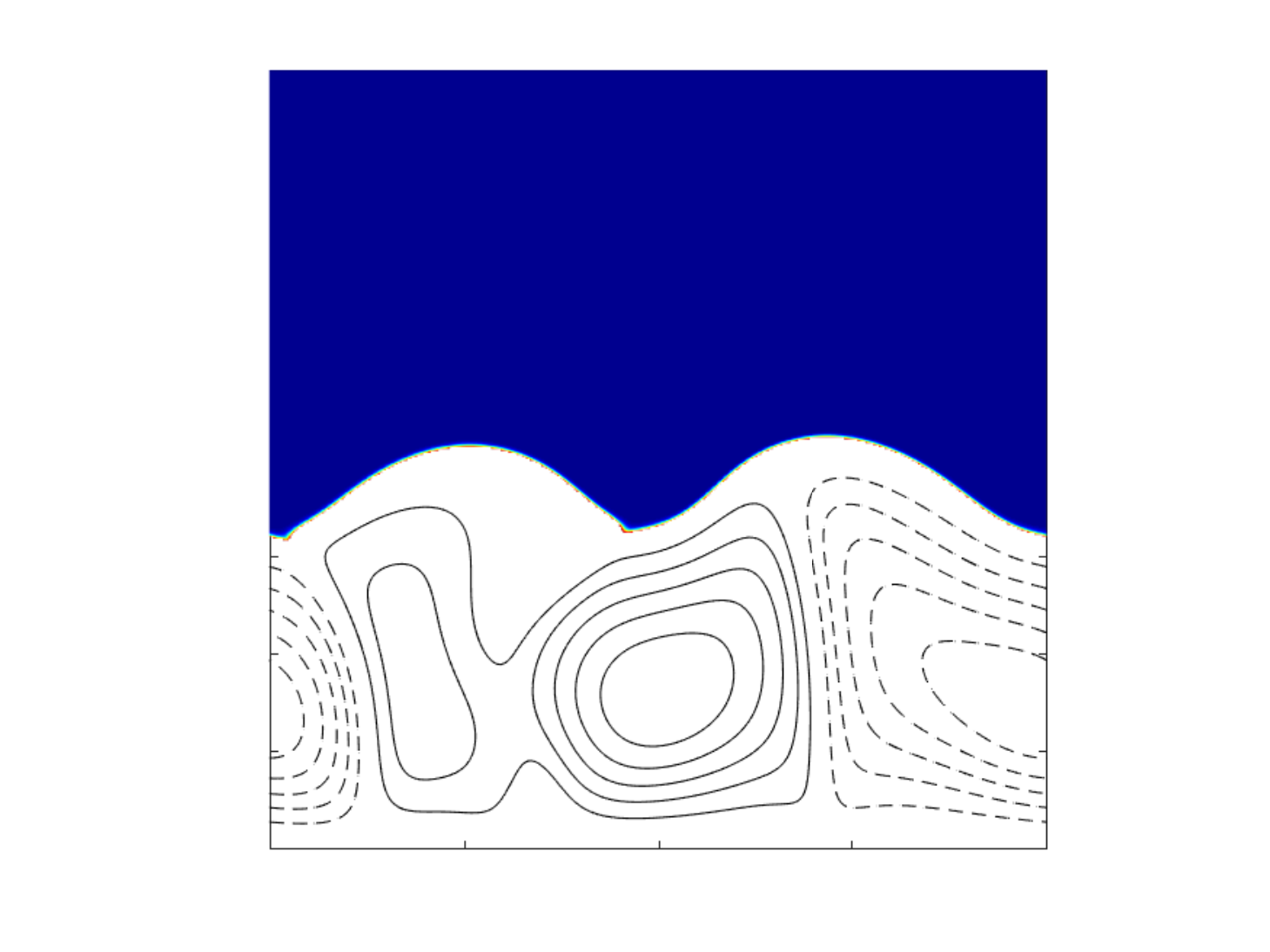}&
\includegraphics[scale=0.23]{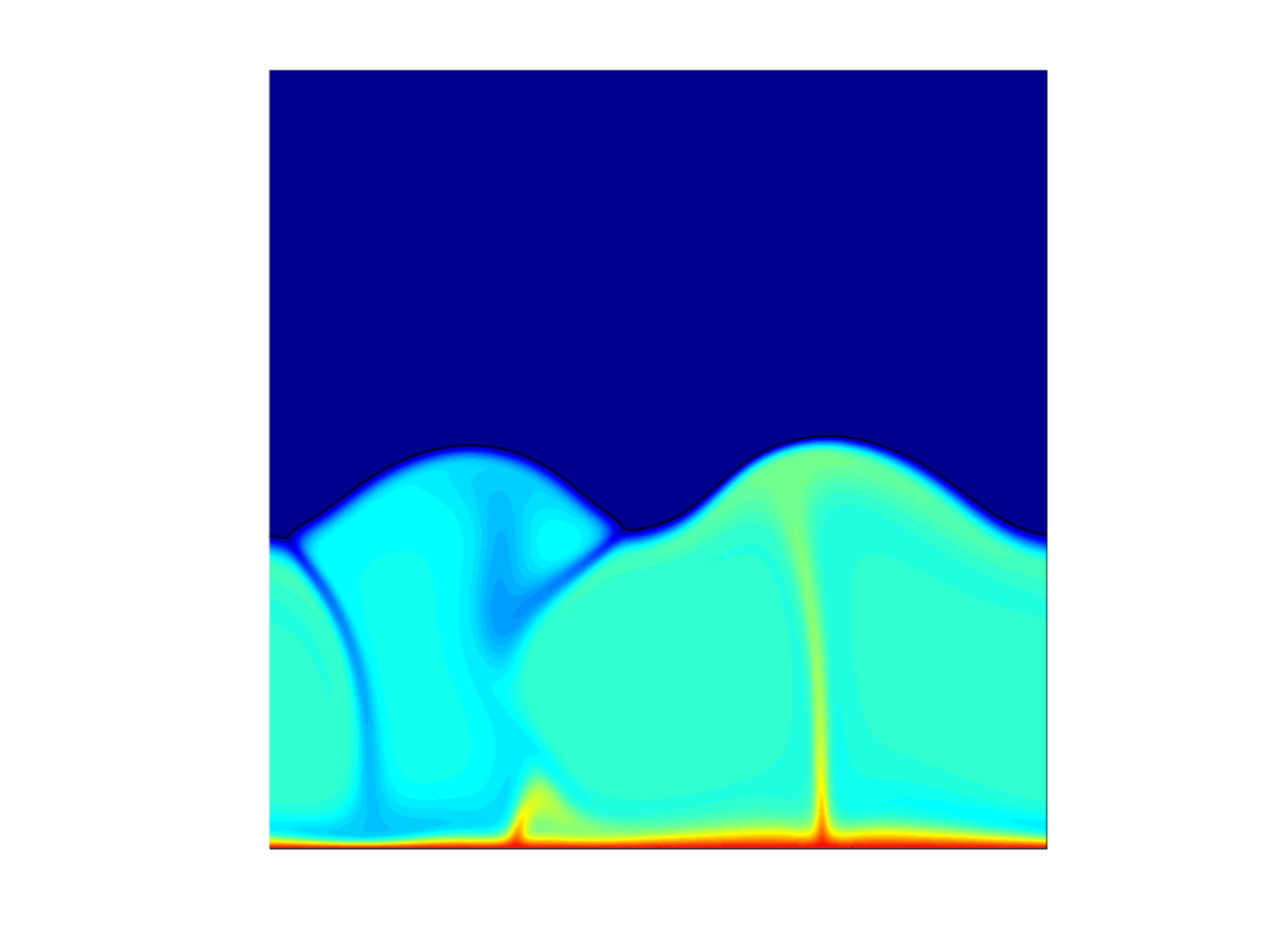}\\
\rotatebox{90}{\quad \quad $t=480s$}&
\includegraphics[scale=0.23]{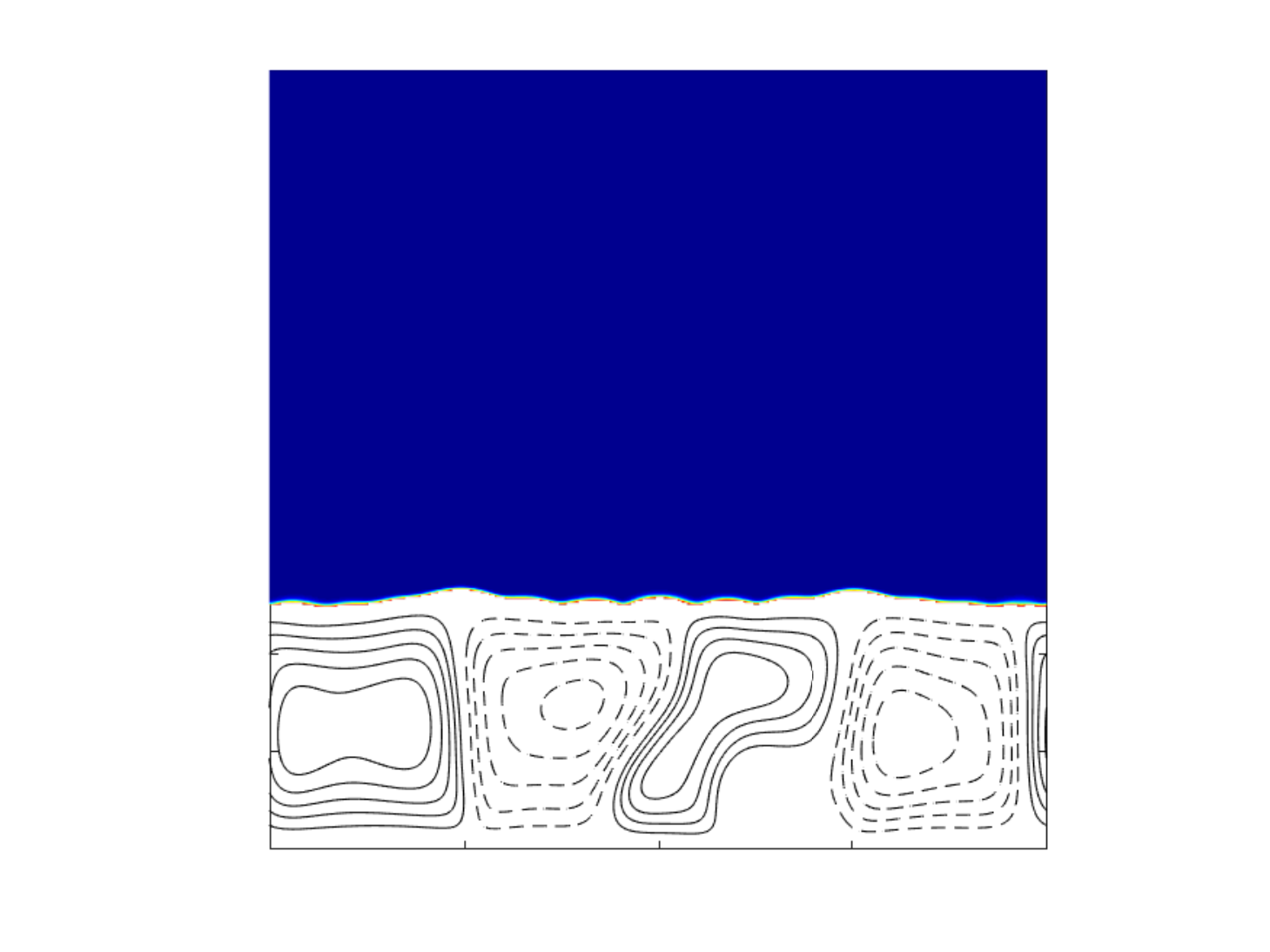}&
\includegraphics[scale=0.23]{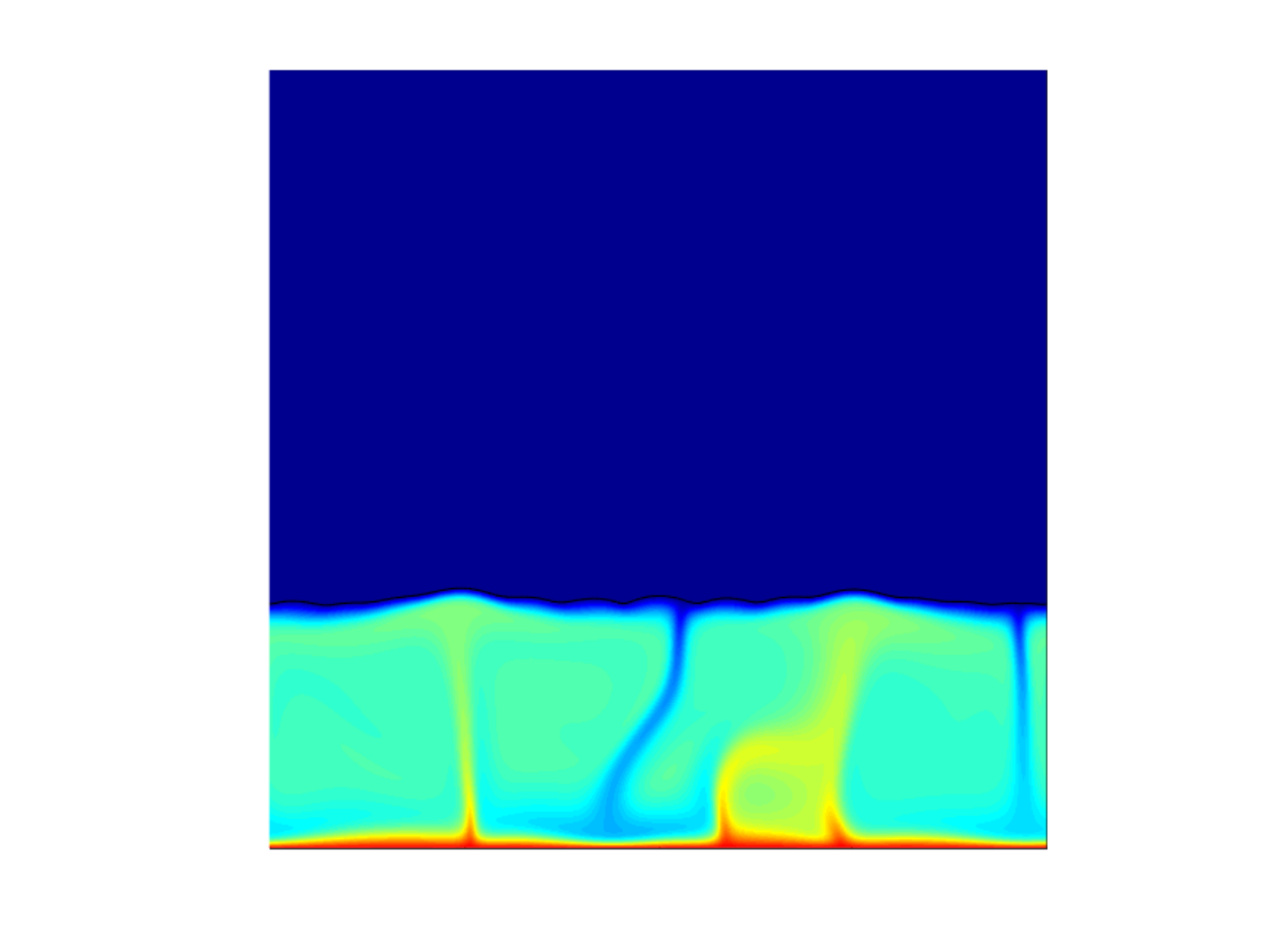}& 
\rotatebox{90}{ \quad \quad $t=780s$}&
\includegraphics[scale=0.23]{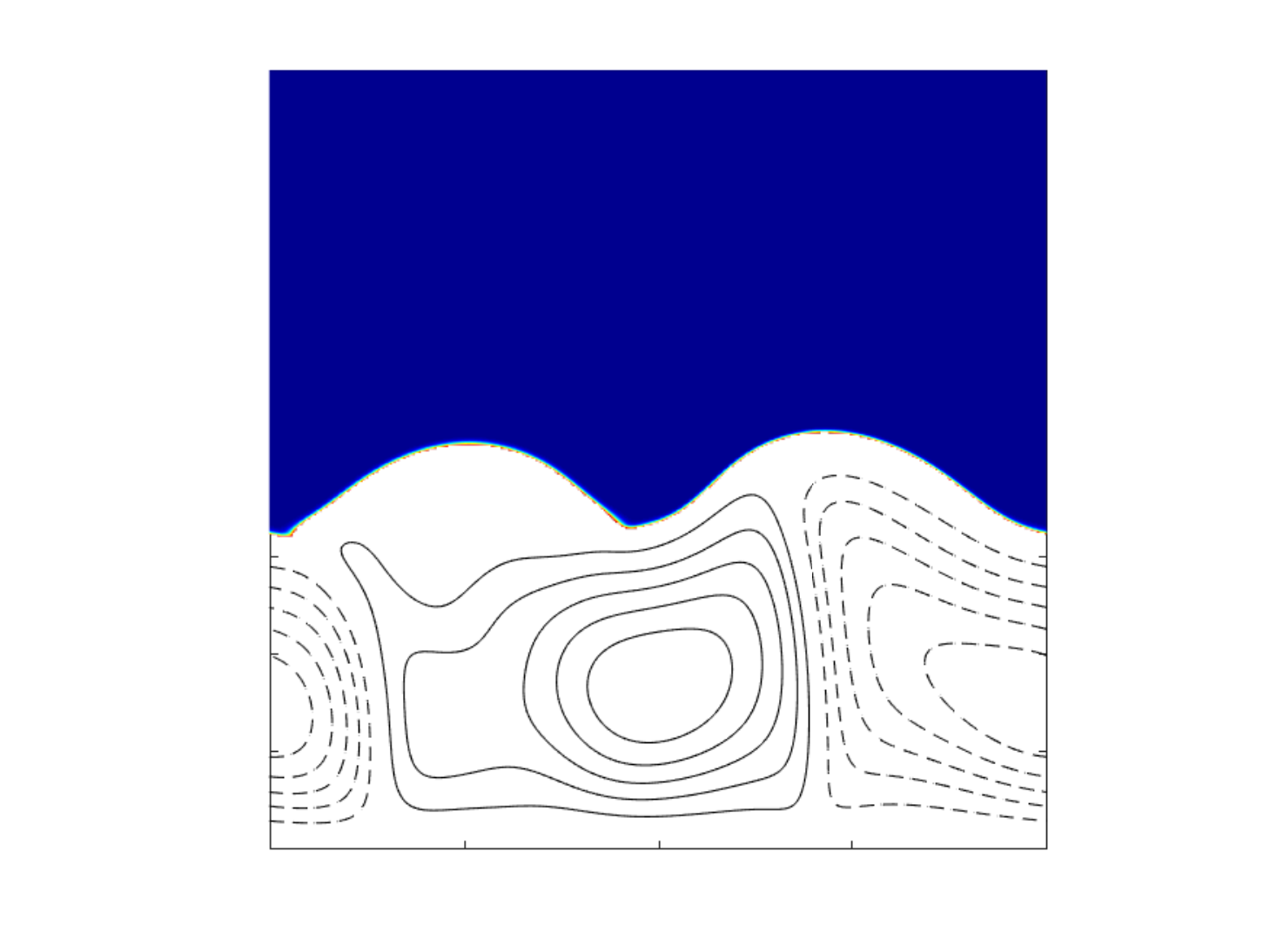}&
\includegraphics[scale=0.23]{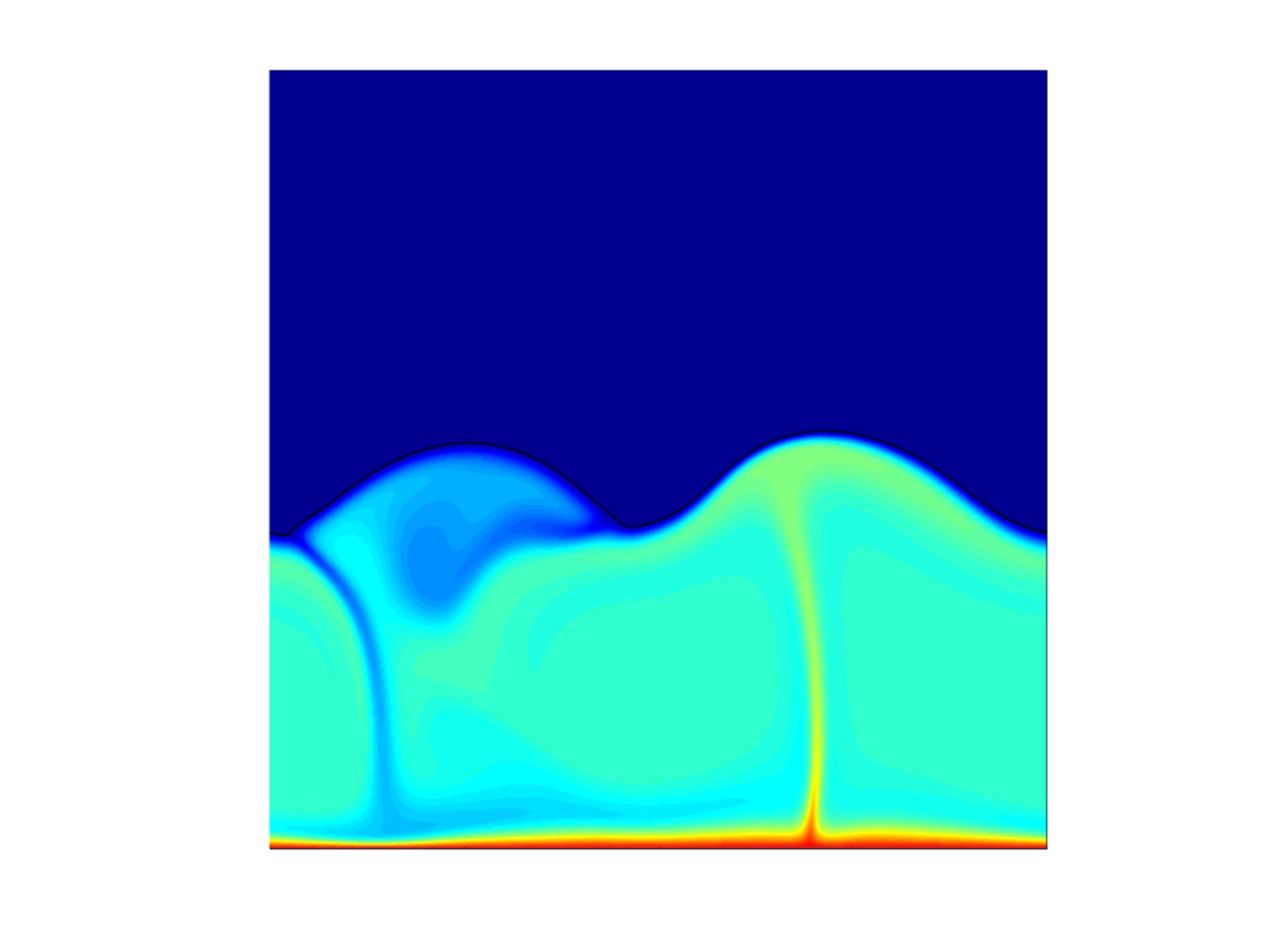}\\
\rotatebox{90}{\quad \quad $t=500s$}&
\includegraphics[scale=0.23]{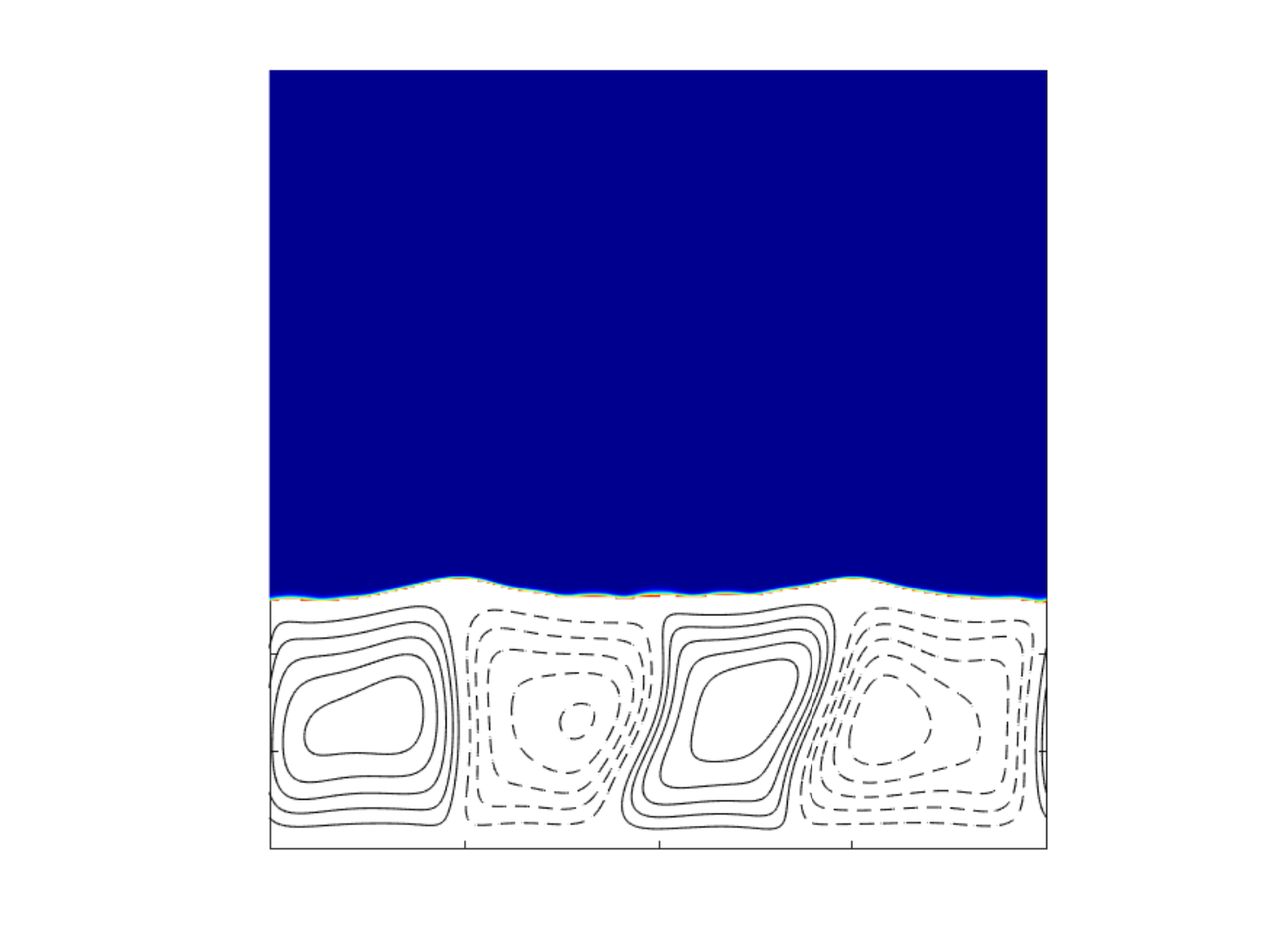}&
\includegraphics[scale=0.250]{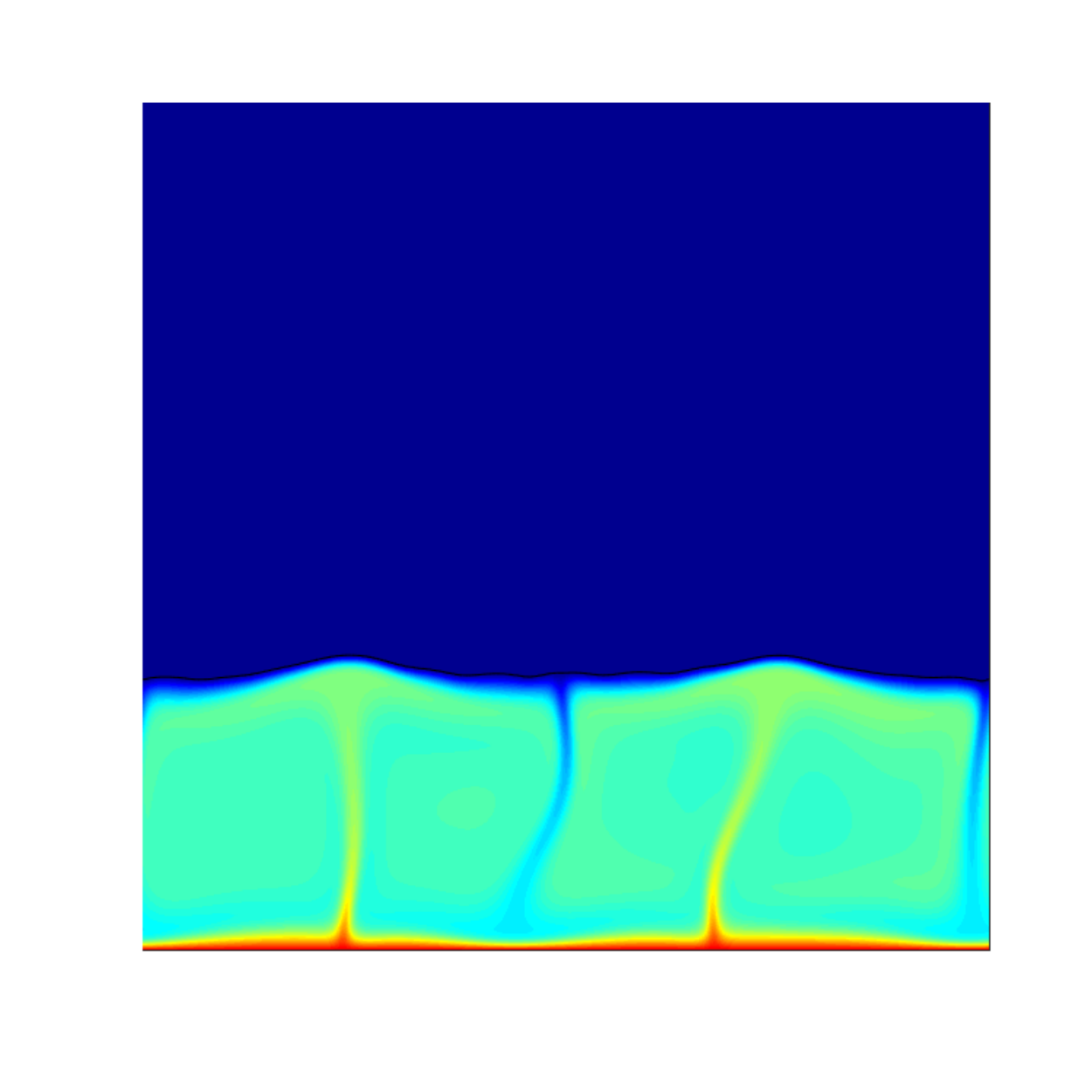}& 
\rotatebox{90}{ \quad \quad $t=800s$}&
\includegraphics[scale=0.25]{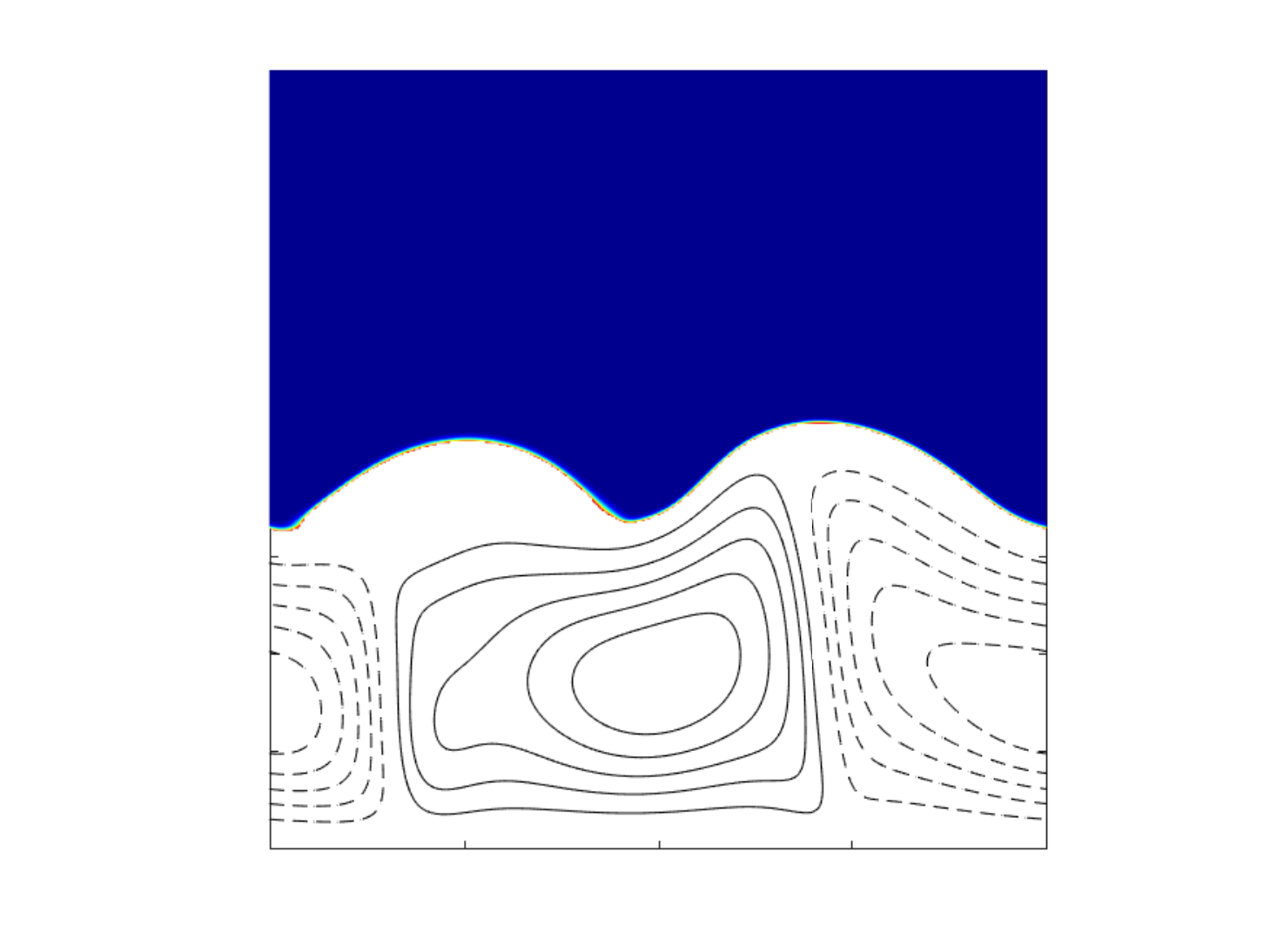}&
\includegraphics[scale=0.25]{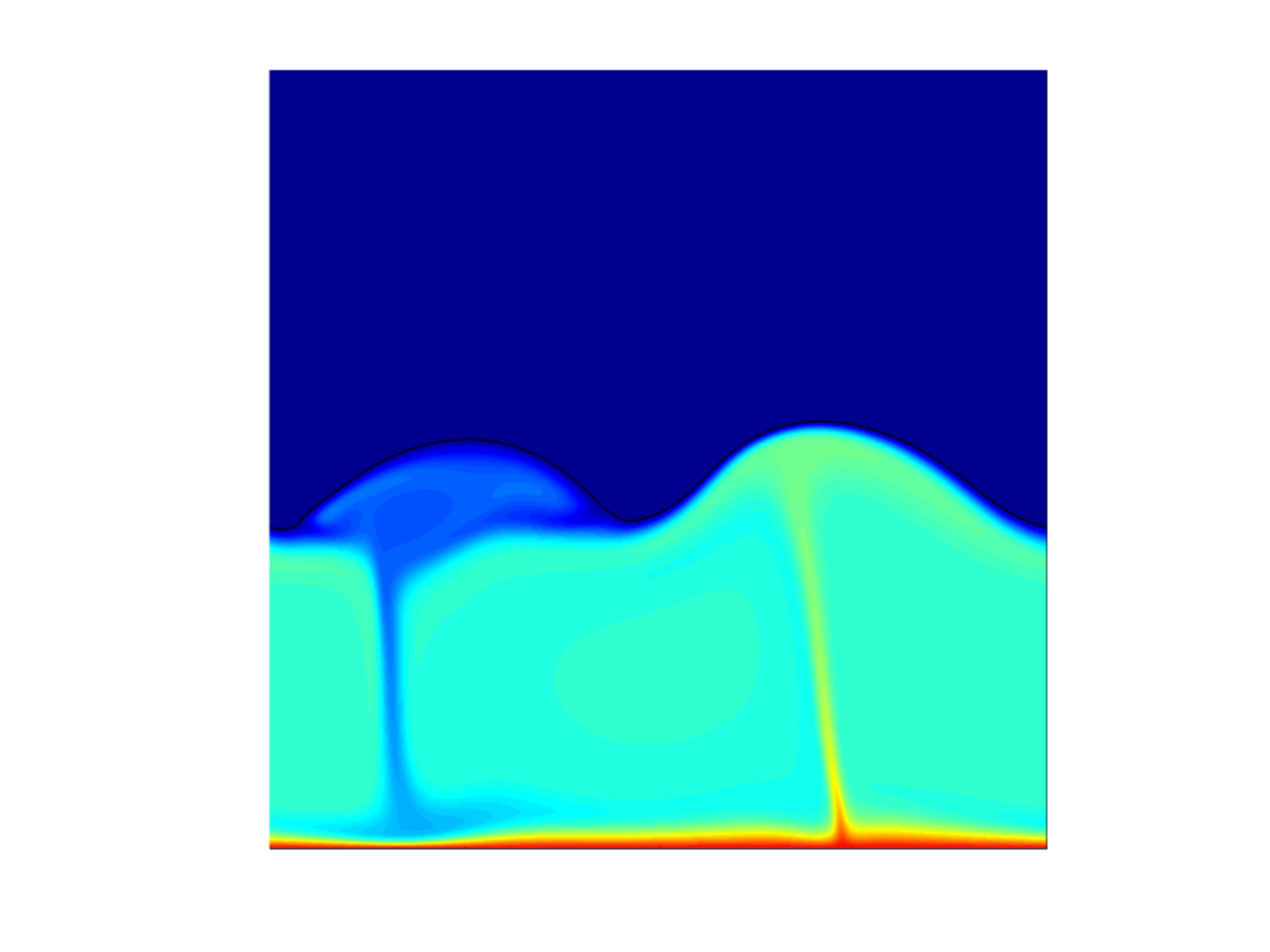}\\
\end{tabular}
\caption{Stream function (left) and temperature field (right) for a domain of side $L=0.04\,m$ at different times. The sequence shows a first coarsening of 
the thermal plumes between $400\,s$ and $500\,s$, and a second coarsening after $600\,s$. }\label{fig:coarsening_regime_004}
\end{figure}

 Figs. \ref{fig:coarsening_regime_002} and \ref{fig:coarsening_regime_004}(left) show the coarsening and development of this secondary instability in time through a sequence of snapshots
 of the temperature and velocity fields for $L=0.02\,m$ and $L=0.04\,m$, respectively.

A closer look at the early stage of the linear regime ($t=500\,s$ in Fig. \ref{fig:coarsening_regime_002}, or $t=400\,s$ in Fig.  
\ref{fig:coarsening_regime_004}) shows that the stem of some plumes becomes very 
 deflected in their central section. Next, the stem and cells begin to move erratically  up to the coarsening of the plumes. This occurs with a central plume holding roughly a  non-deflected stem, and the neighbor plumes exhibiting the largest deflections. This suggests the existence of  a horizontal mode with a finite wavenumber  during  the coarsening of the plumes. Indeed, as observed  for $L=0.04\,m$ the beginning of the secondary 
instability 
 shows the stems deflecting in a periodic form along the horizontal direction. 
 
 During the coarsening, the solid/liquid front loses the periodicity characteristic of the linear regime becoming irregular in some regions. This loss of symmetry happens with
 a smooth increase in the amplitude of the modulations of the interface.  Once  the coarsening stage is completed, a new periodic solid/liquid interface is reached with lower wavenumber. 

\begin{figure}[!h]
\centering
\includegraphics[scale=0.5]{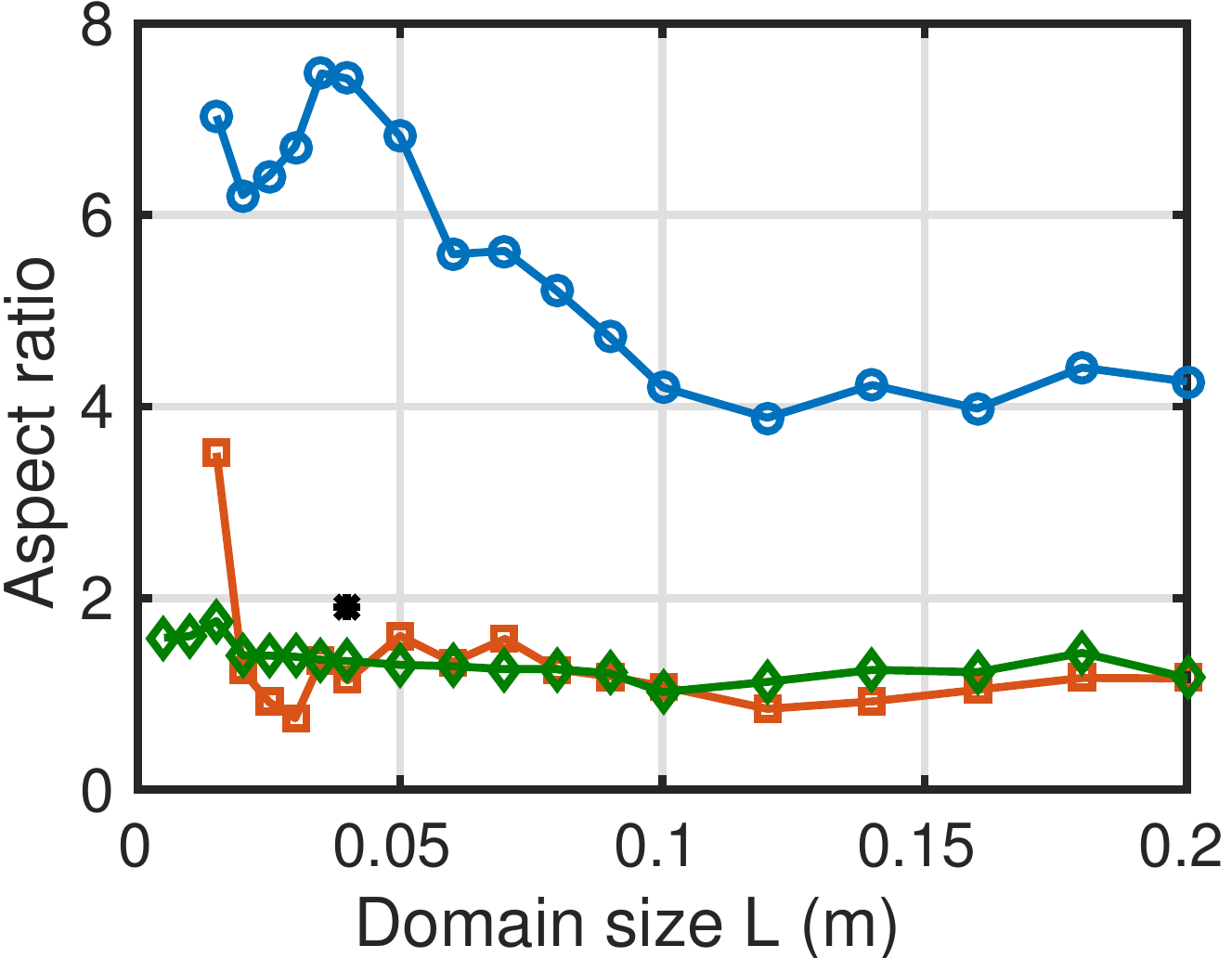}
\caption{ Aspect ratio (defined as the ratio of the average depth of the convective cells to the domain size $L$) of the convective cells before  the first coarsening  (blue circles) and once it is completed (red squares)  versus the domain
 side $L$.  The aspect ratio of the convective cells appearing after  the first (Rayleigh-B\'enard) instability is shown with green diamonds. The lonely 
black point corresponds to the aspect ratio after the second coarsening (third  instability) of  $L=0.04\,m$. 
\label{fig:aspect-ratio_L}}
%\JC{Aspect ratio of the convective cells before (blue circles, mean= 5.9852) and once the first coarsening  is completed (red squares, mean= 1.3674, 
%mean without first point =  1.2020)  versus the grid size $L$.
% The aspect ratio before the first instability is shown as green diamonds (mean=1.3658). The lonely black star corresponds to the aspect ratio after the second 
%coarsening (third  instability) of $L=4\,cm$}} \label{fig:aspect-ratio_L}
\end{figure}

  Remarkably,  the completion of the coarsening phase occurs with the cells  recovering  an aspect ratio similar to the convective cells at the onset of the Rayleigh-B\'enard instability. This is illustrated in Fig. \ref{fig:aspect-ratio_L}, where 
 the aspect ratio of the convective cells  is plotted  at the threshold of the coarsening regime (circles) and after the coarsening regime is completed (squares).  The maximum aspect ratio reached
by the convective cells  is bounded between $4.7$ and $7.5$ for all domain sizes $L$. Although this maximum ratio does not decrease monotonously with $L$, there is an overall decreasing trend with higher $L$.
After merging is completed we obtain an averaged value of the aspect ratio of $1.4$. The same average  value of the aspect ratio of the convective cells at the origin of the first instability,  shown  as well in Fig. \ref{fig:aspect-ratio_L}  with green diamonds. However  the aspect ratio with respect to $L$ after the coarsening exhibits much larger fluctuations about the mean than after the first instability. 
 
The evolution after the coarsening regime depends on the domain size. For large $L$, there is room for additional interplay between phase separation and convective motion before complete melting, and we 
generally find  a transition towards a turbulent state. However, in some case such as $L=0.04\,m$ a third instability occurs developing a second coarsening 
(c.f. Fig. \ref{fig:coarsening_regime_004}(right)). This creates  a behavior similar to a coarsening
cascade. After this second coarsening the number of cells is reduced to one to recover again an aspect ratio close to the cells of the first instability. This aspect ratio is shown in Fig. \ref{fig:aspect-ratio_L}  as a black star.

While no systematic studies have been carried out on this state in the context  of melting of PCM heated from below  
on the scaling of $Nu$, boundary layers or statistics of plume, we notice that  using  related models  Prudhome  \cite{Prudhome1991} showed  snapshots of merged convective cells in a cylinder
 heated from below, and  Gong \etal \cite{Gong1998} showed the merging of convective cells and plumes at different
times  within square geometries heated from below taking the domain size as the characteristic length for the  Rayleigh number.\\

%From these observations, we suggest that coarsening from the secondary instability is induced by a finite wavenumber horizontal mode that pushes the convective cells to recover an aspect ratio close to that of the first convective cells that appear in the system. 

\subsubsection{Turbulent regime}
\begin{figure}[!h]
\begin{tabular}{ccccc}
 &  a) $t=320s$  &  b) $t=400s$ &  c) $t=450s$ & d) $t=500s$\\
  \rotatebox{90}{ Stream Function}&
  \includegraphics[scale=0.25]{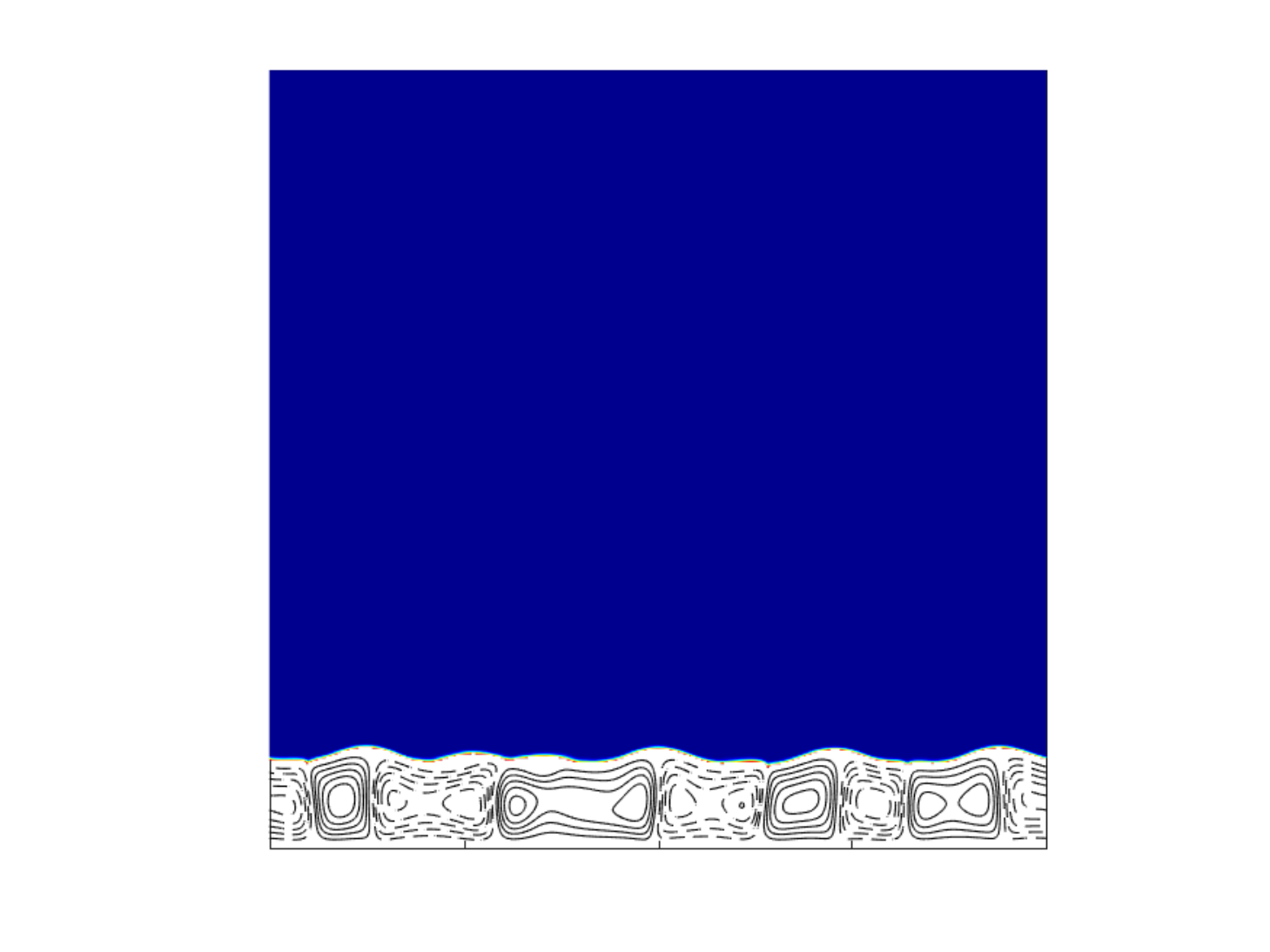}&  
  \includegraphics[scale=0.25]{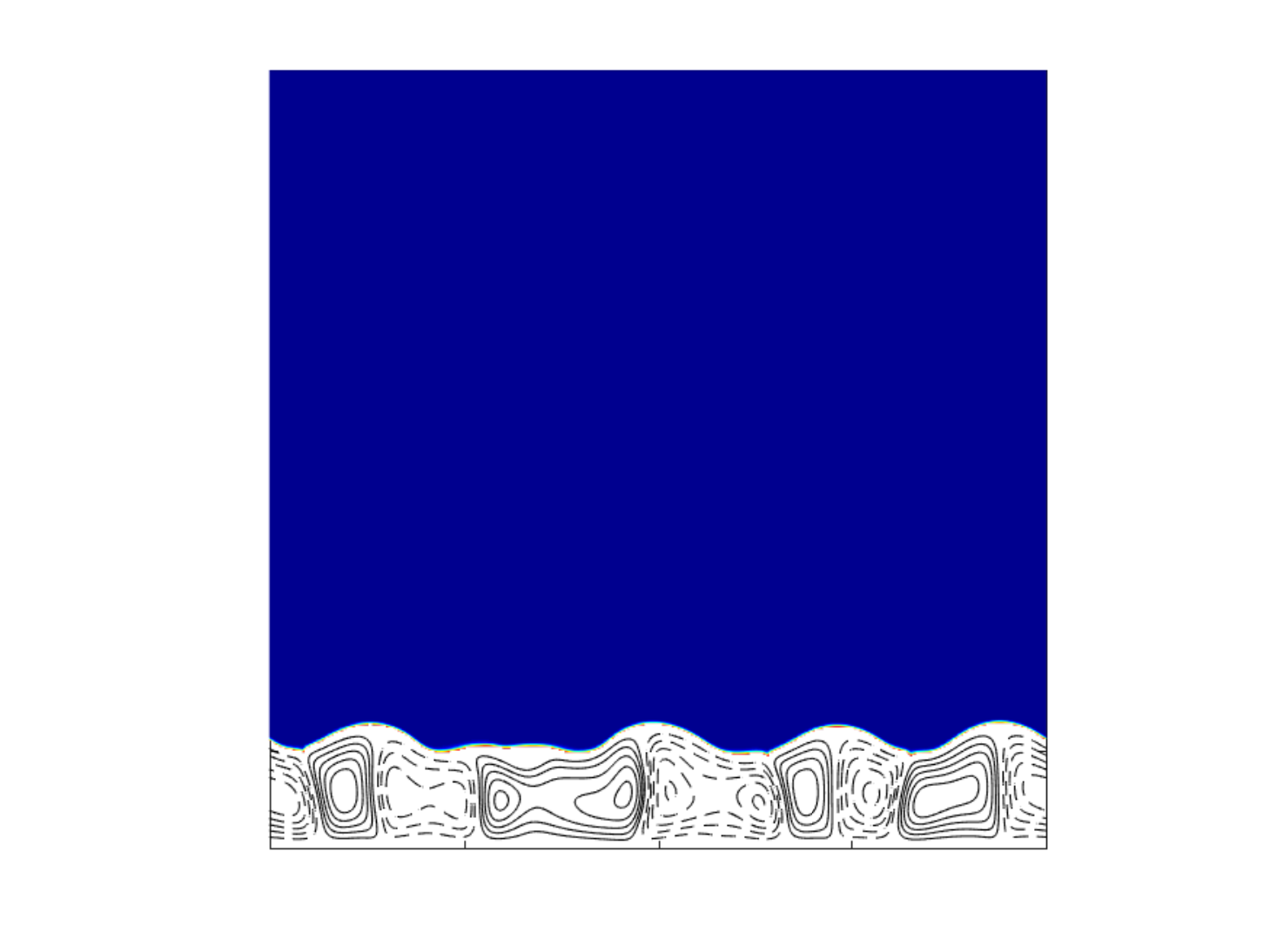}&
  \includegraphics[scale=0.25]{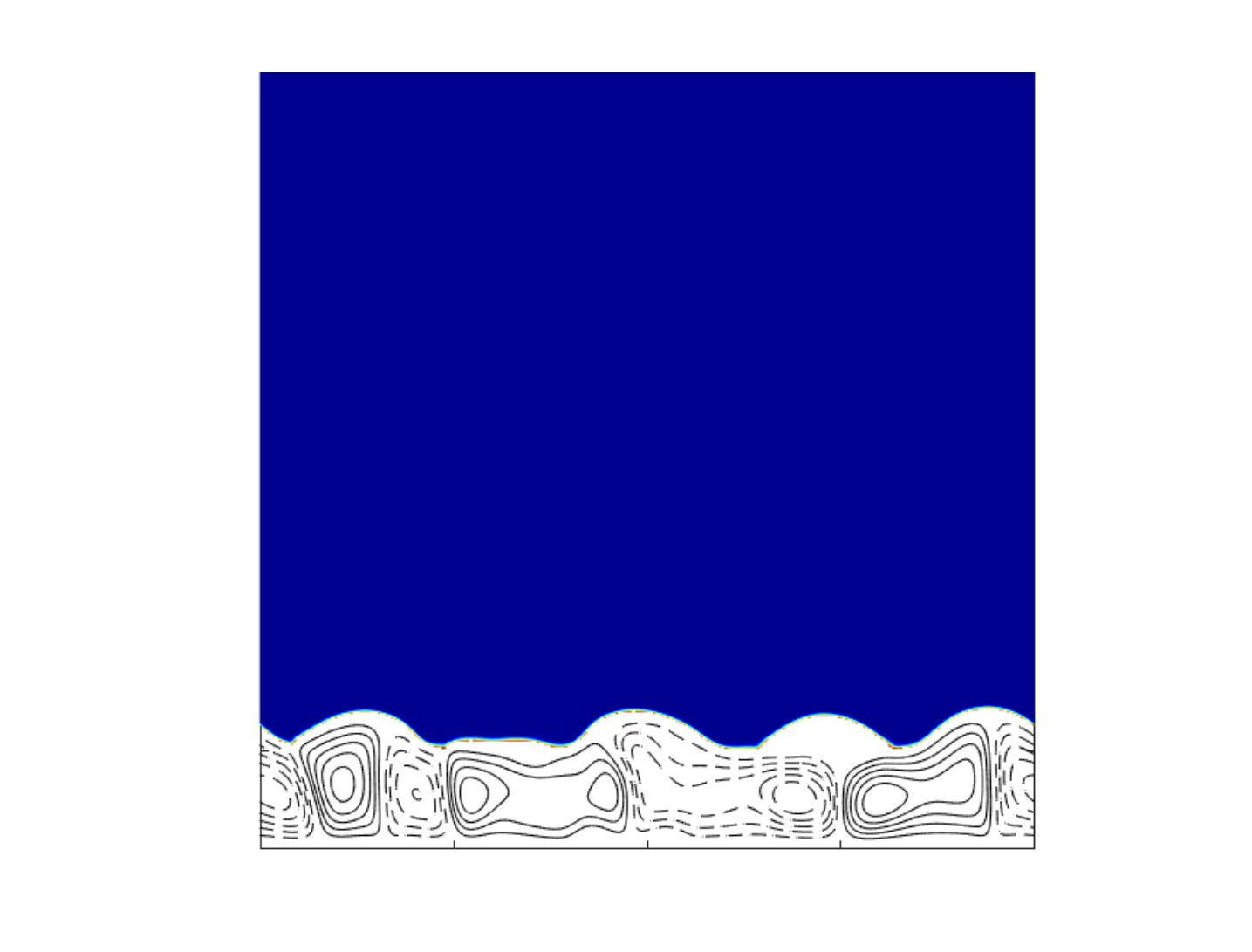}&
  \includegraphics[scale=0.25]{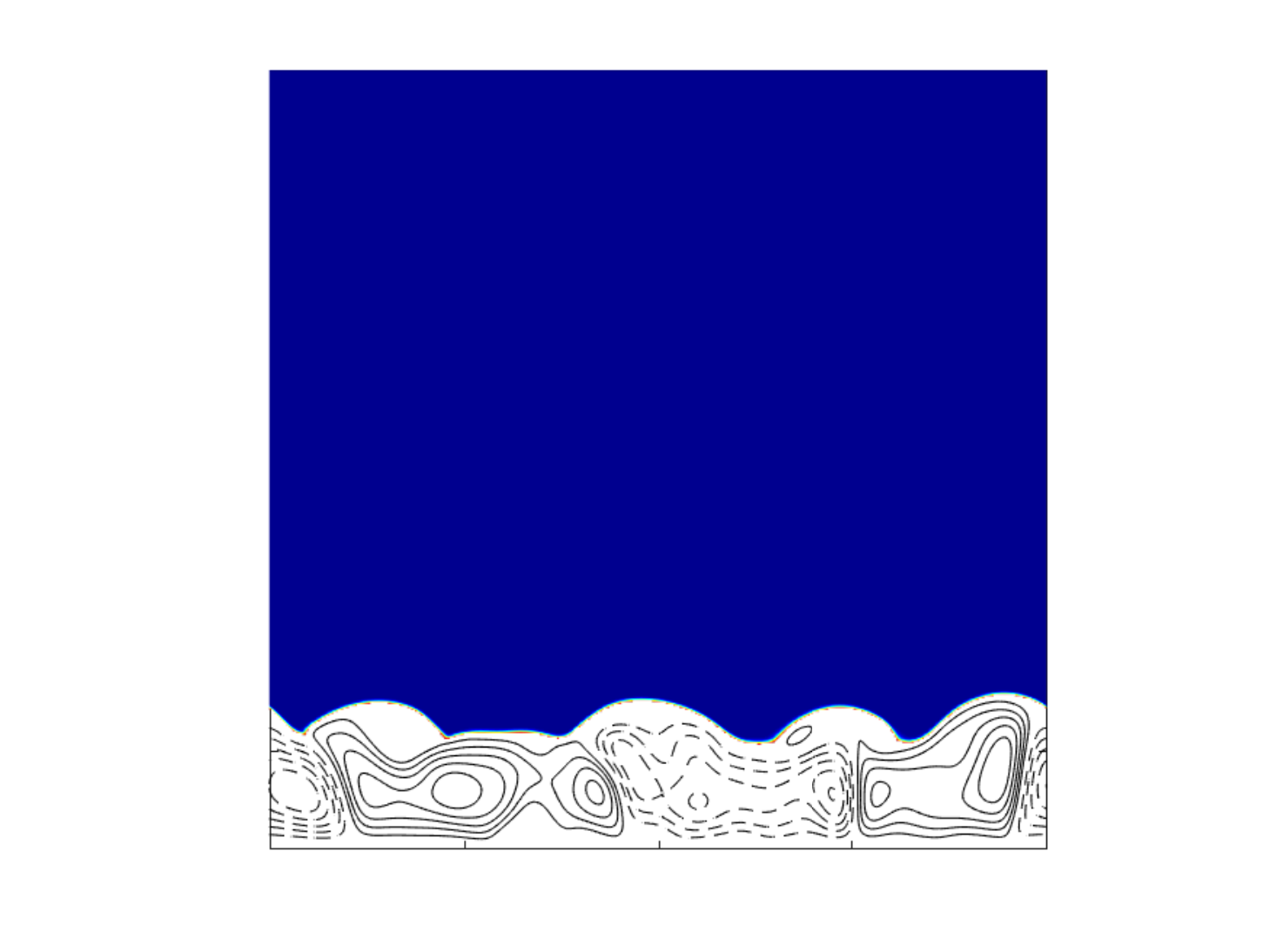}\\
   \rotatebox{90}{\large Temperature field}&
   \includegraphics[scale=0.25]{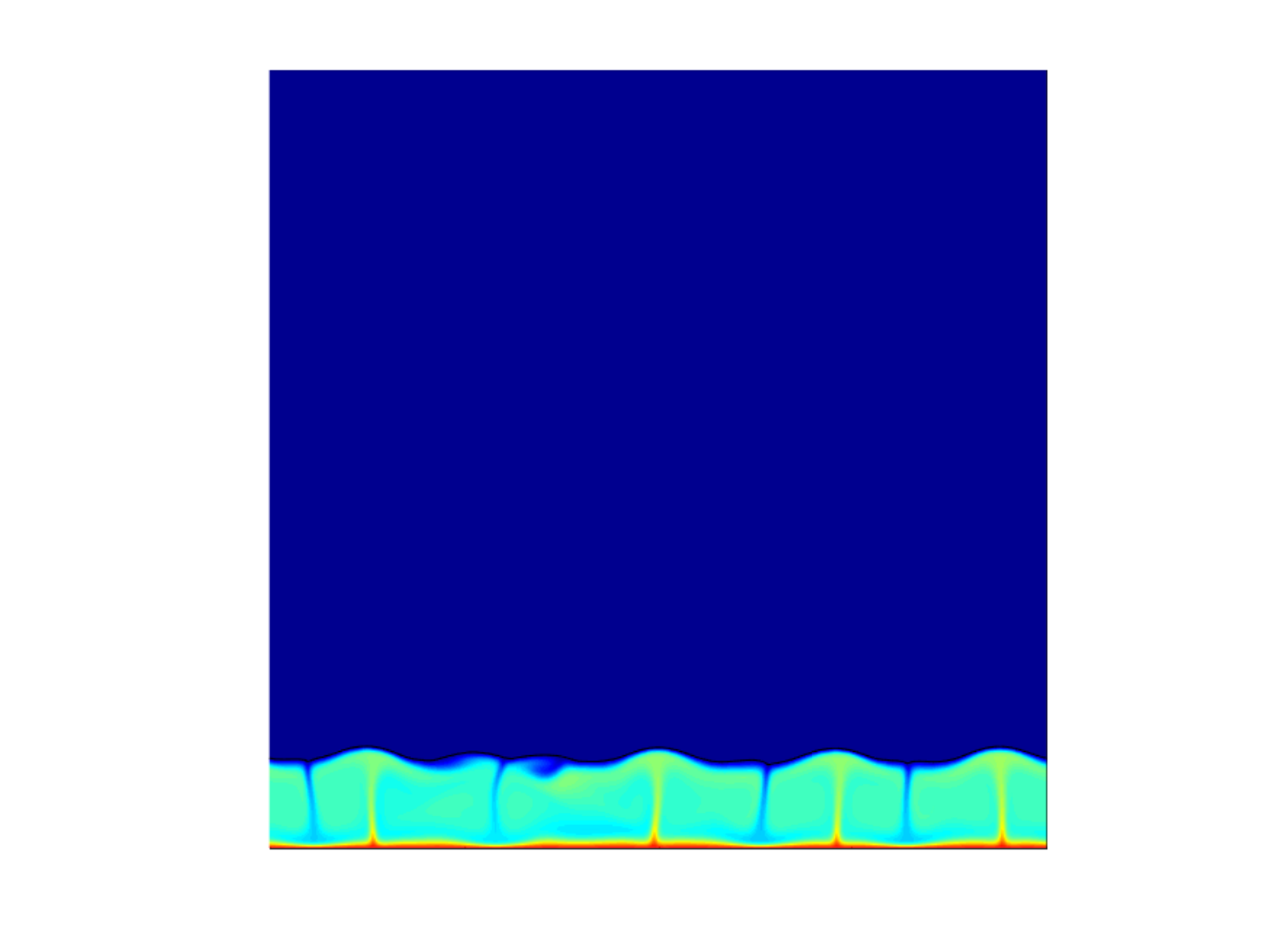}&
  \includegraphics[scale=0.25]{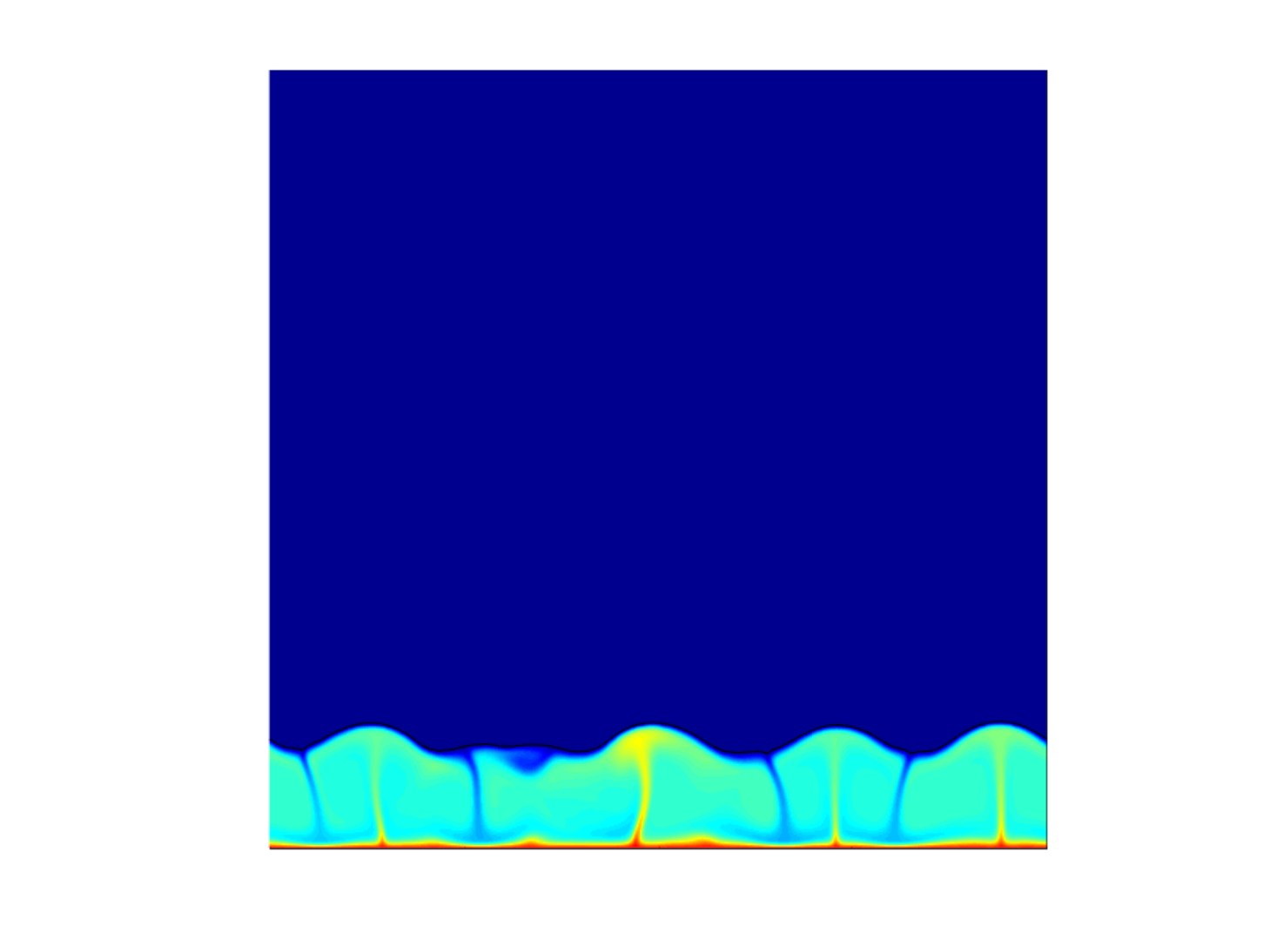}&
  \includegraphics[scale=0.25]{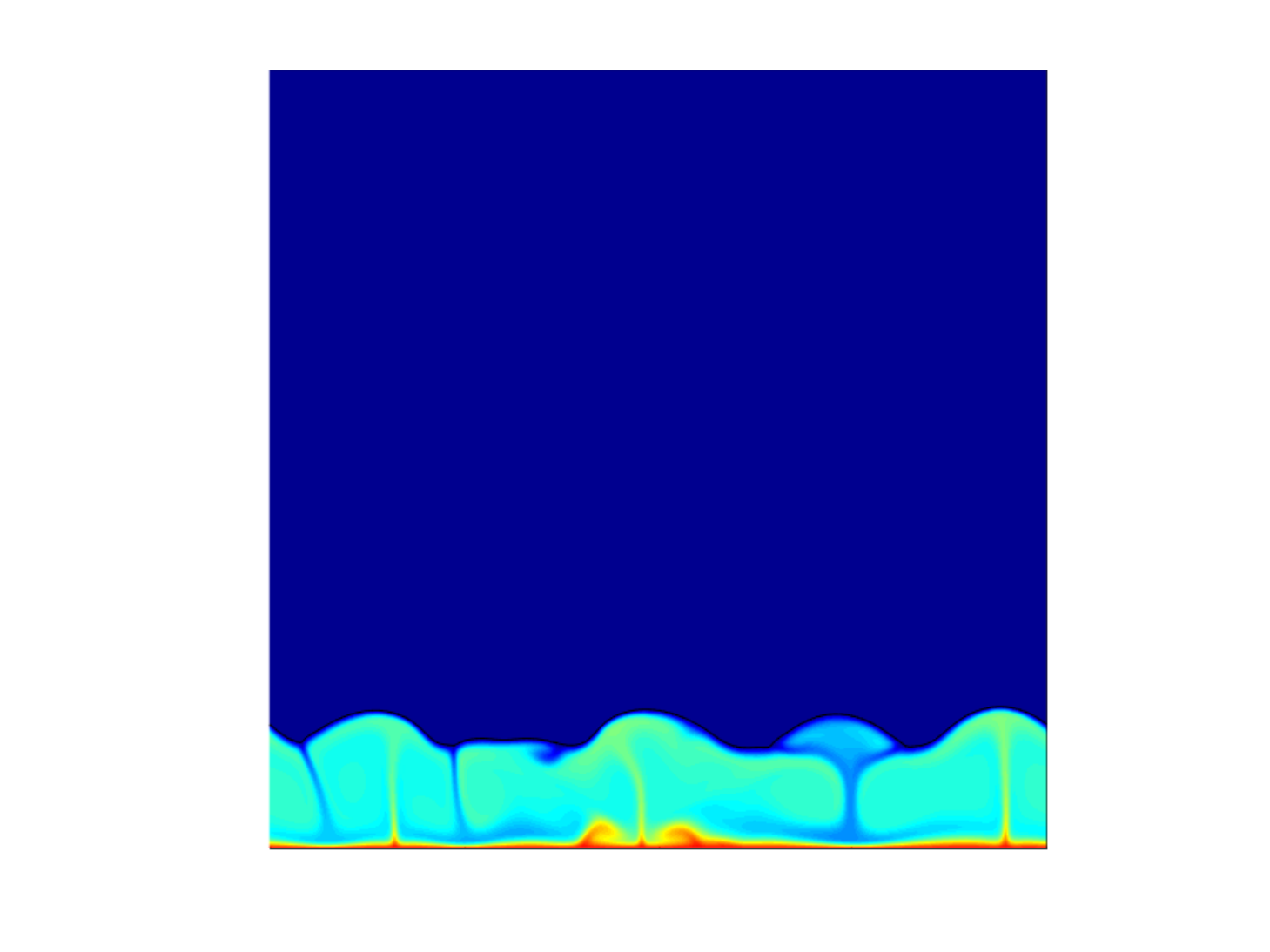}&
  \includegraphics[scale=0.25]{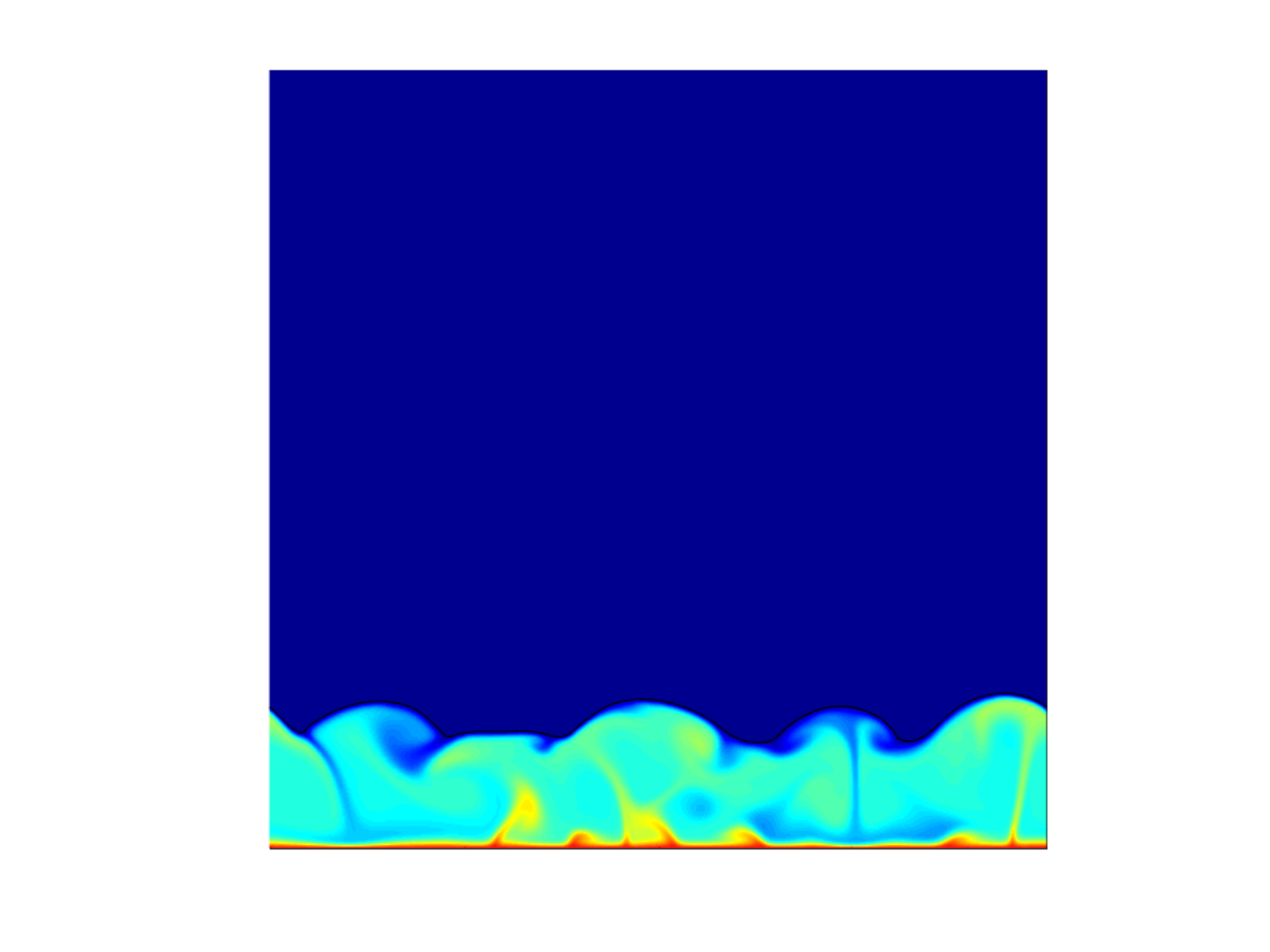}\\
  \hline \\
  & e) $t=550s$ &  f) $t=600s$ &\Large  c) $t=750s$ &\Large  d) $t=1000s$ \\
   \rotatebox{90}{ Stream Function}&
   \includegraphics[scale=0.25]{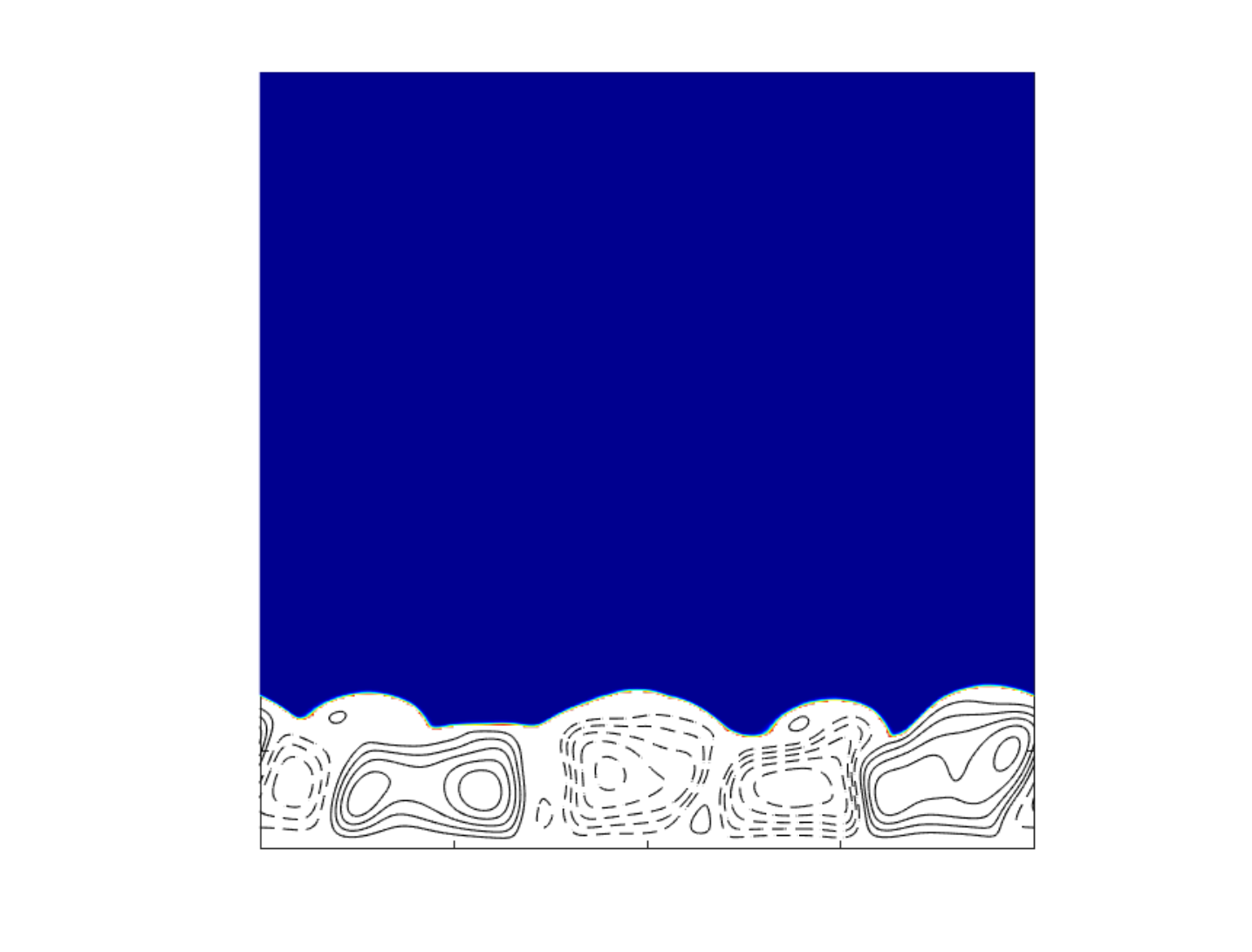}&
   \includegraphics[scale=0.25]{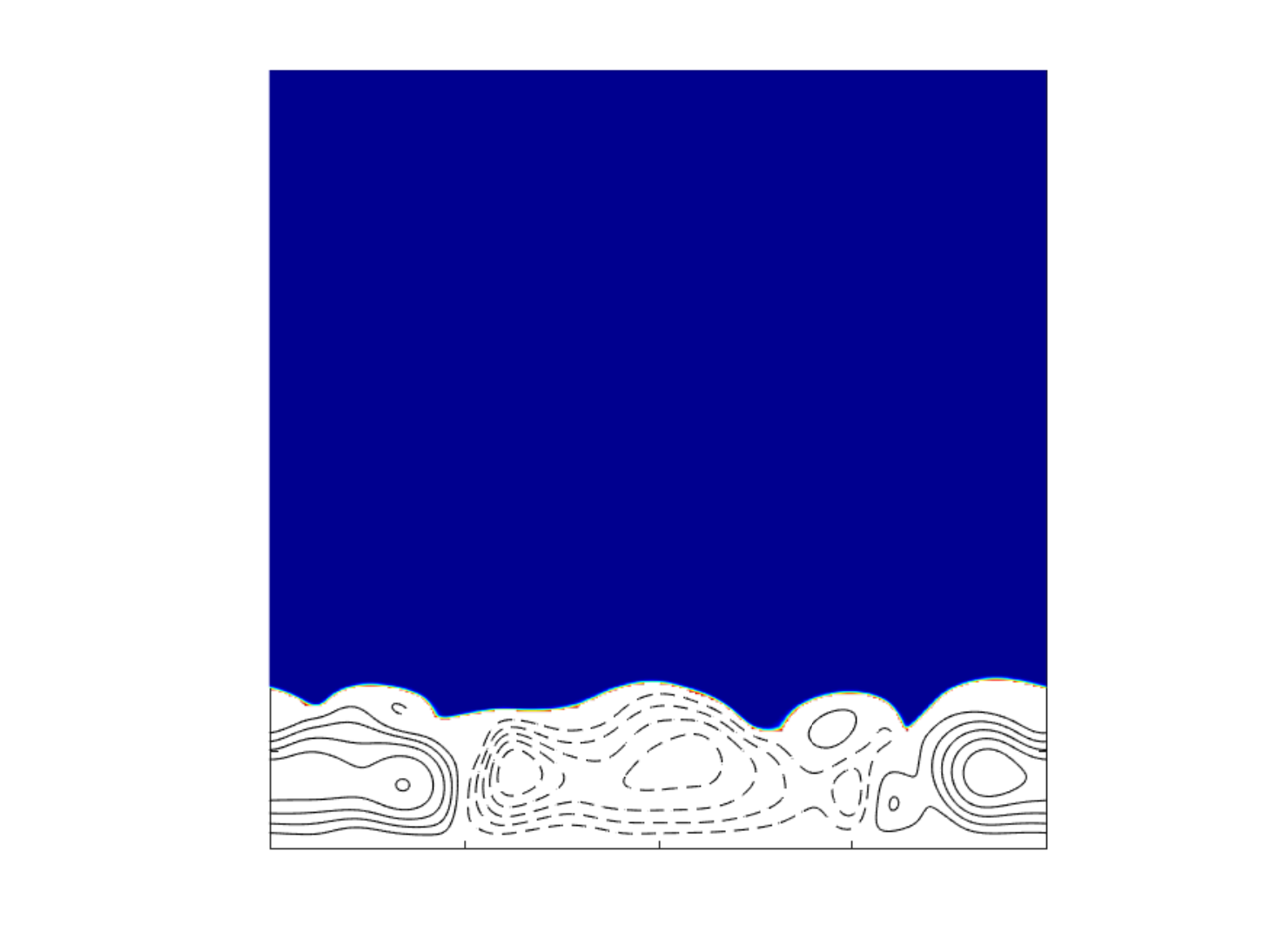}&
   \includegraphics[scale=0.25]{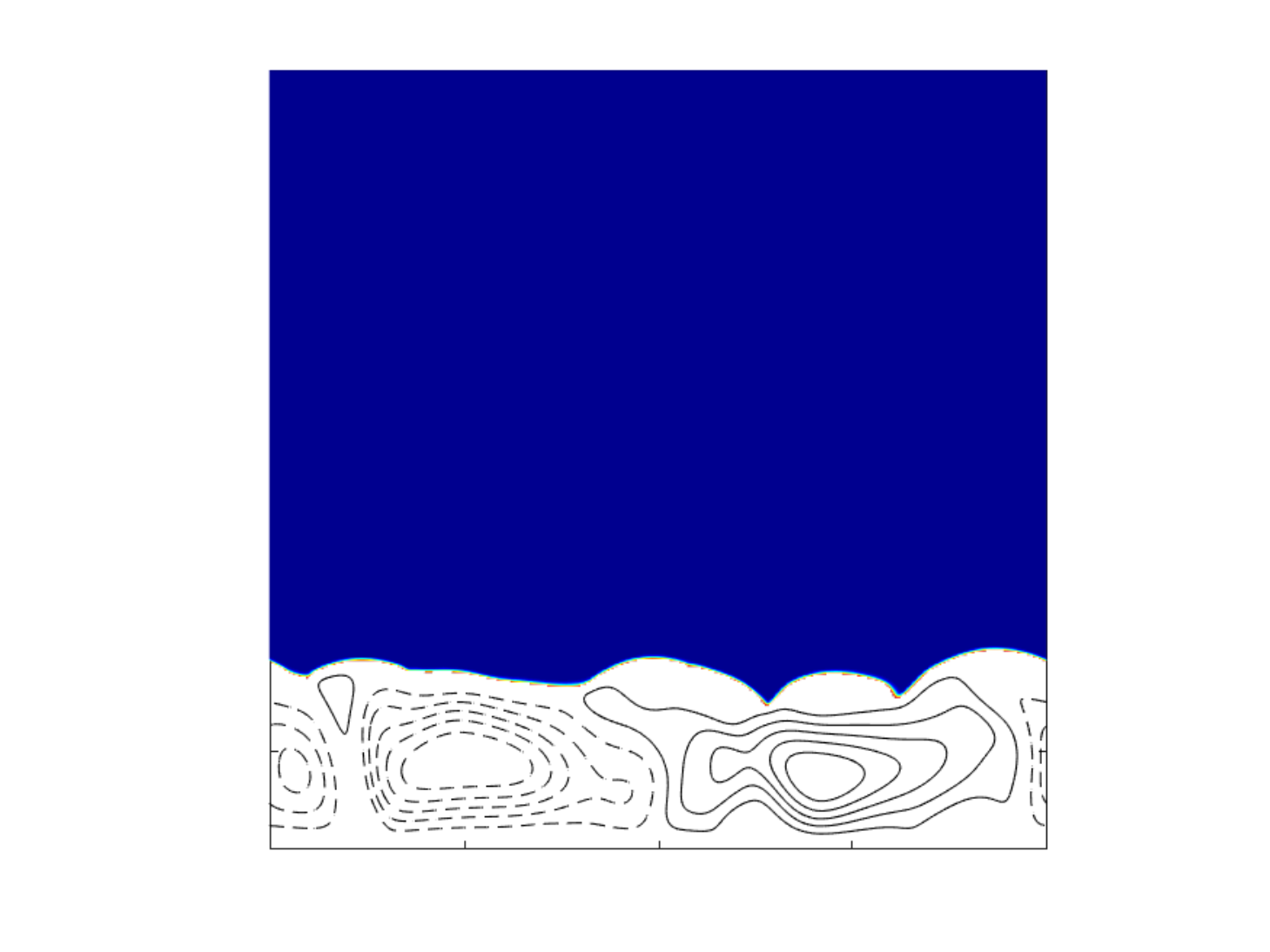}&
   \includegraphics[scale=0.25]{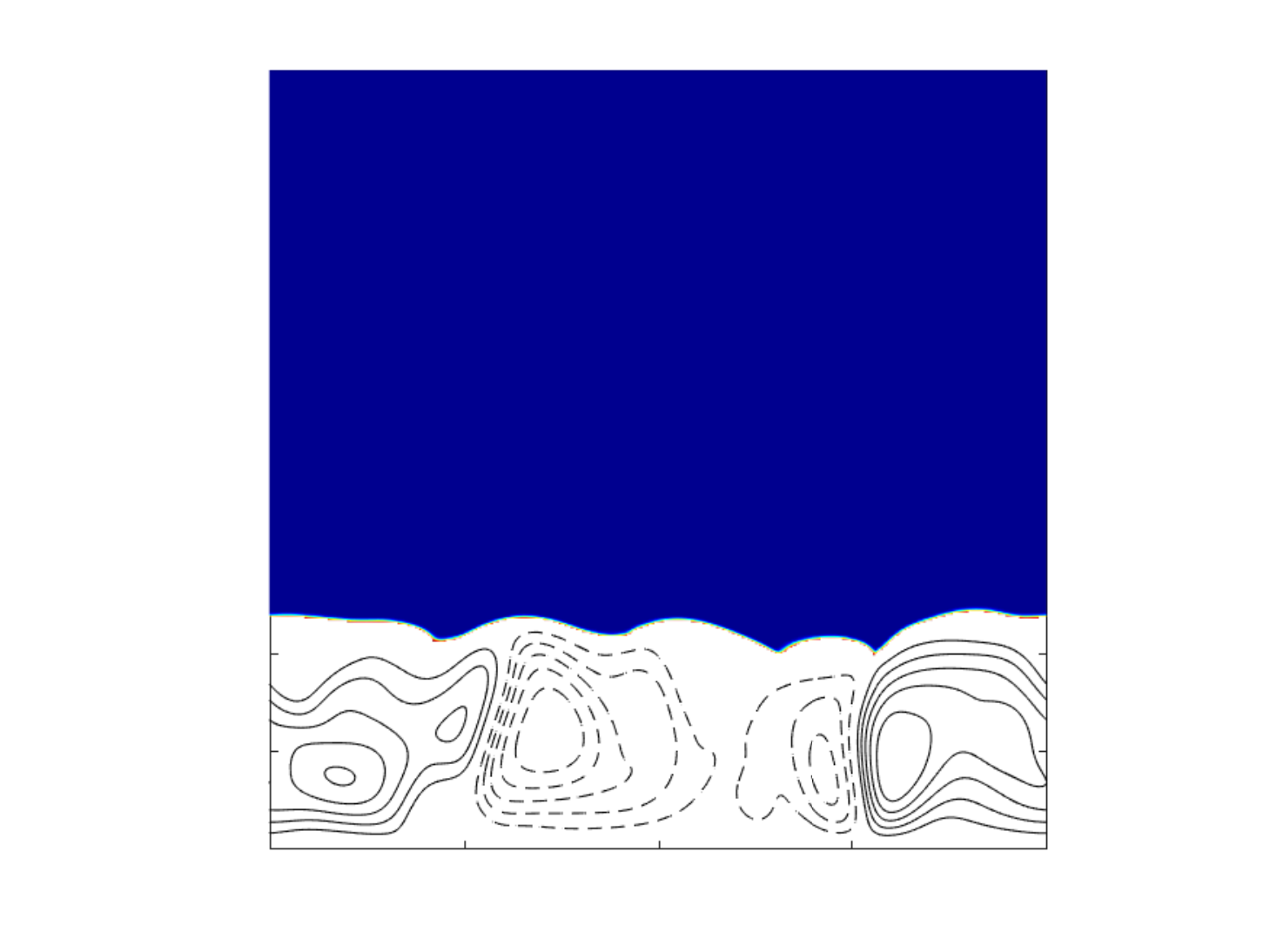}\\
  \rotatebox{90}{ Temperature field}&  
   \includegraphics[scale=0.25]{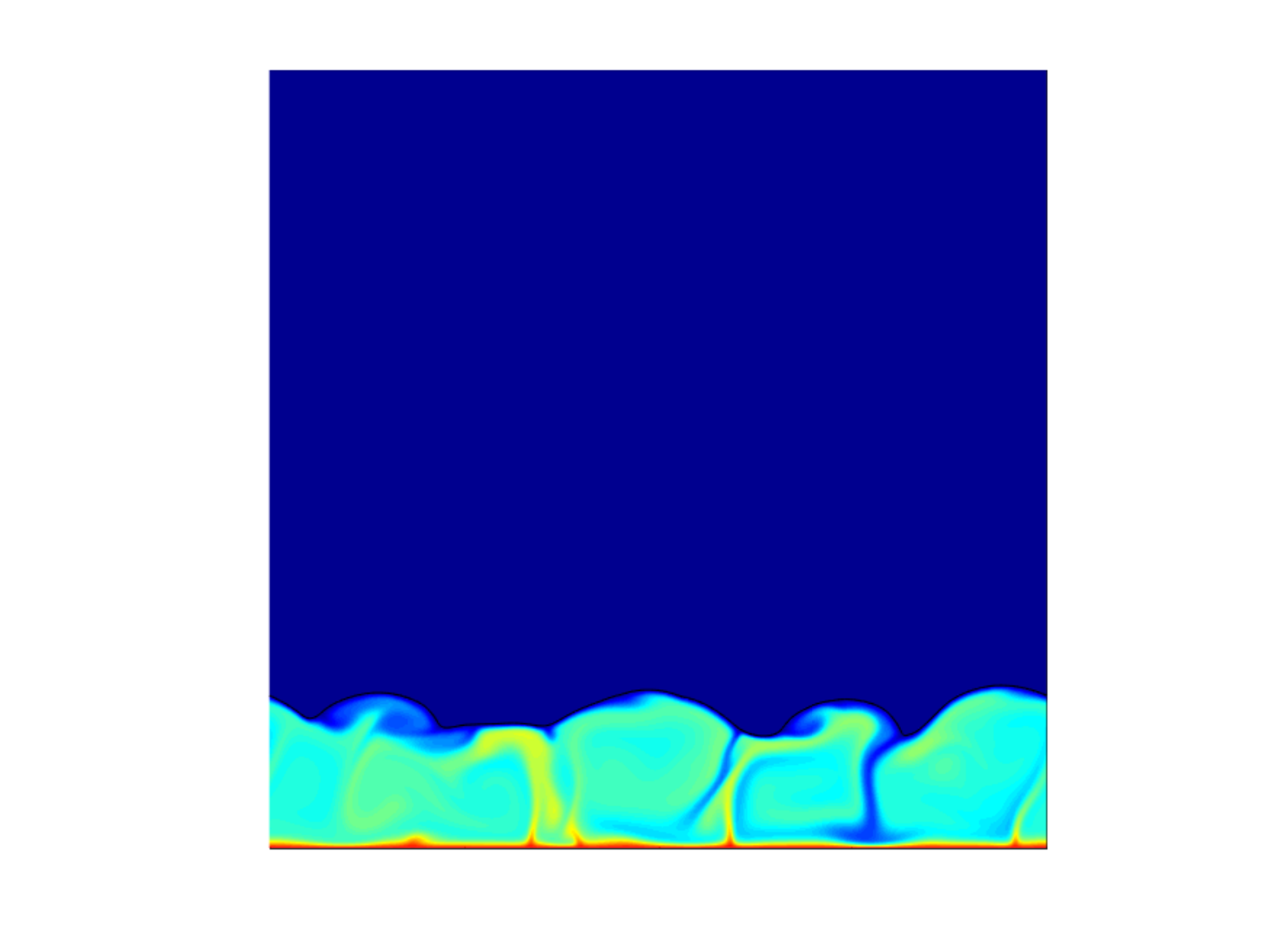}&
  \includegraphics[scale=0.25]{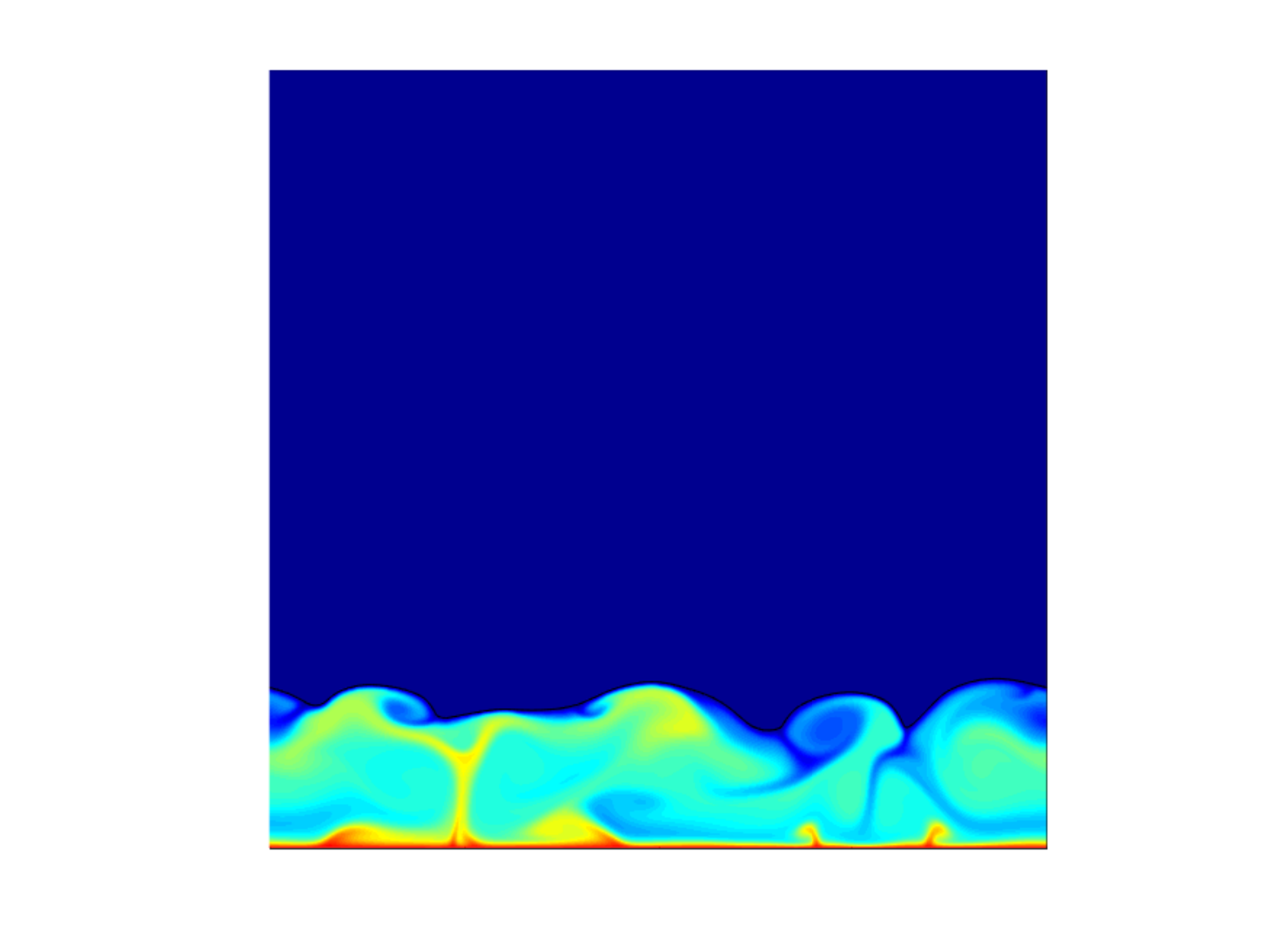}&
   \includegraphics[scale=0.25]{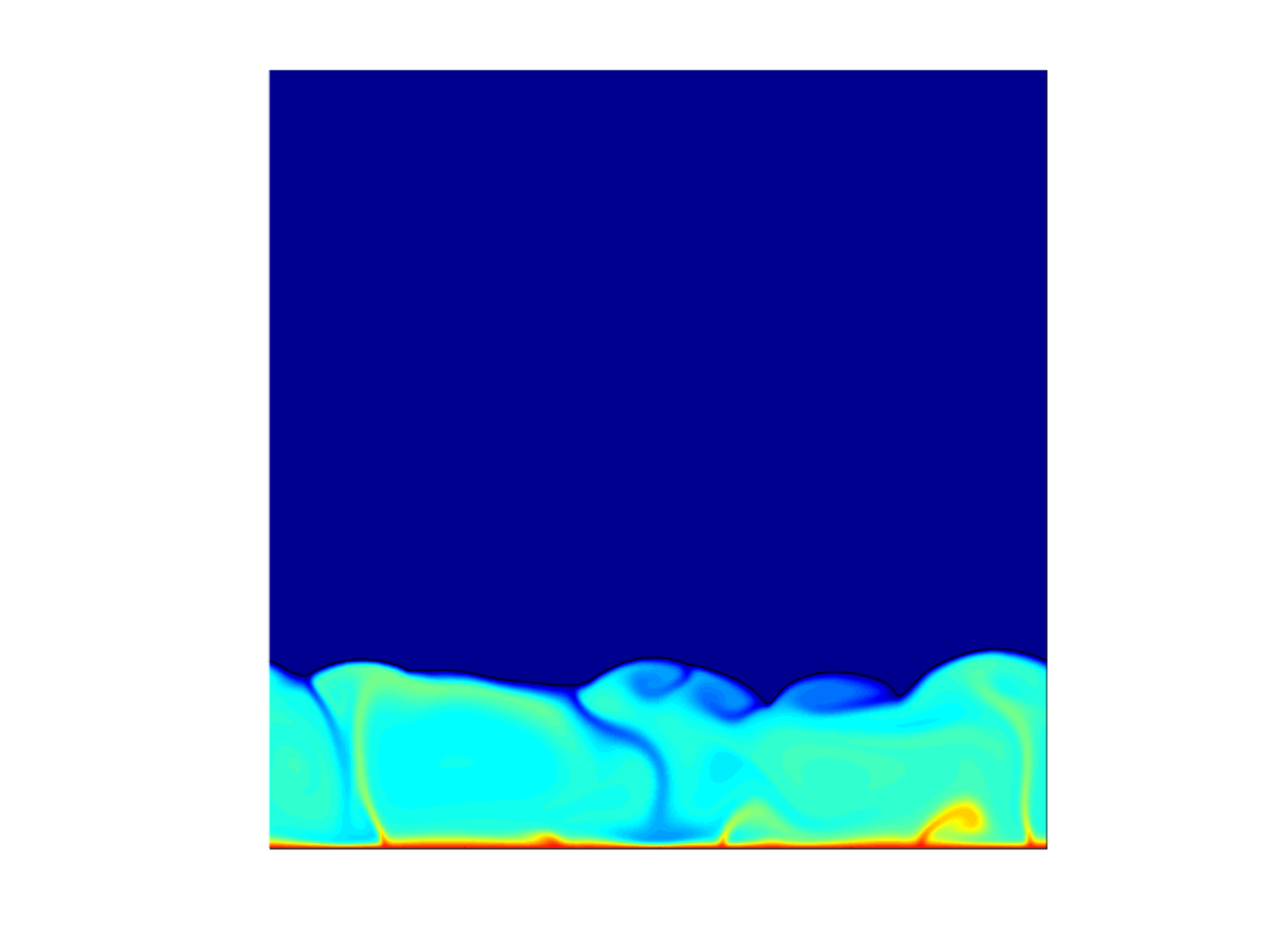}&
    \includegraphics[scale=0.25]{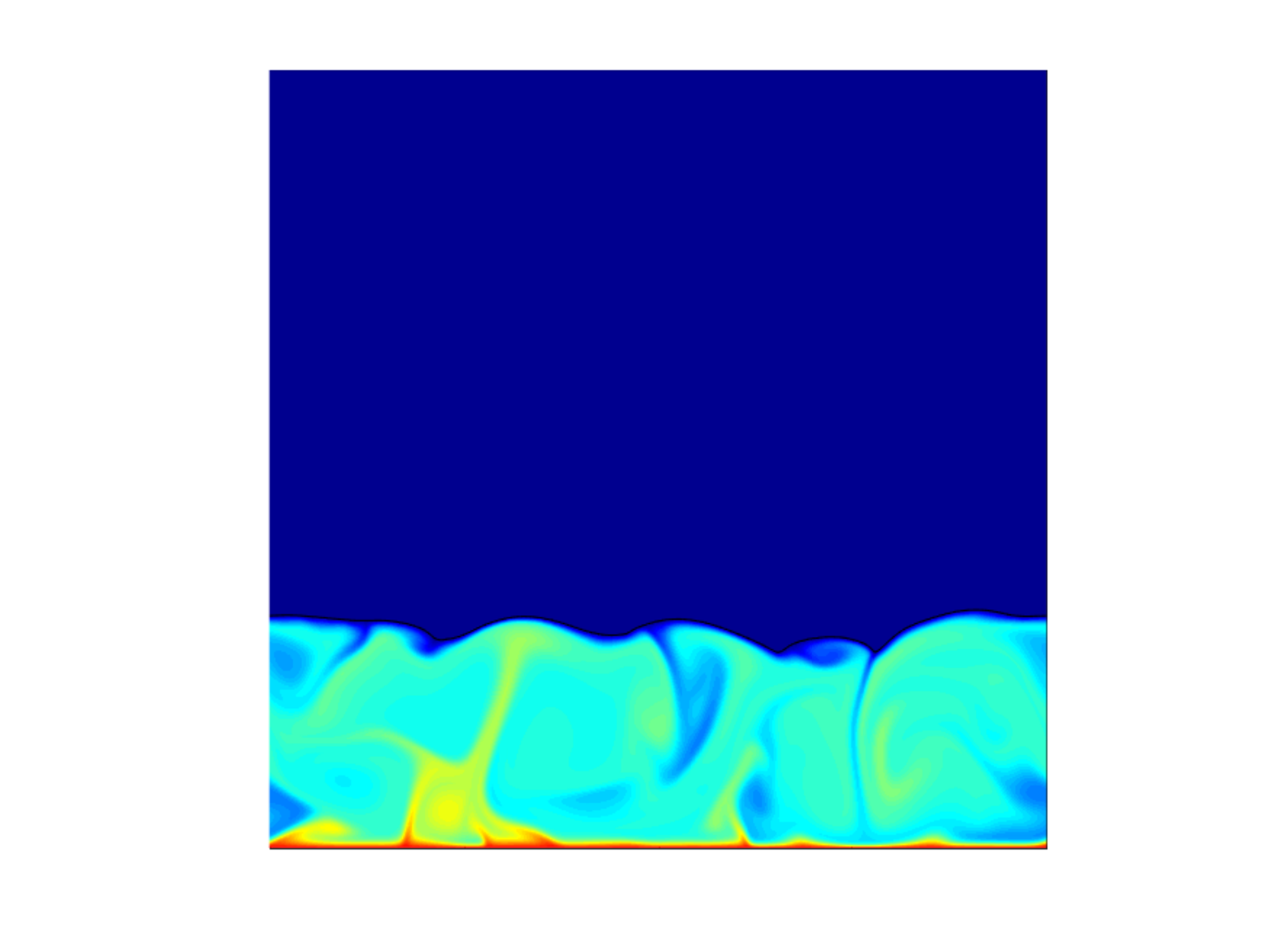}
\end{tabular}
\caption{Snapshots of the turbulent regime at  times $t=320,400,450,500,550,600,750,1000\,s$ for the stream function (top row) and temperature field (bottom 
row) for a domain of side  $L=0.08\,m$. Strong variations in the thermal boundary layer, erratic motion and small plumes arising from the hot wall are 
observed. \label{fig:turbulent_regime_008}}
\end{figure}

 Finally, for large enough domains ($L>0.035\,m$ in the geometries tested) a fourth regime appears as the latest stage of melting after the coarsening 
regime. It is characterized by a high deformation of the thermal boundary  layer on the hot wall and the emergence of new plumes no related to coarsening 
phenomena. Flow fields and plume dynamic are erratic in this regime.  This is the regime where most of the melting occurs for  
the larger domains. Fig. \ref{fig:turbulent_regime_008} at $L=0.08\,m$
shows a sequence of snapshots of the temperature and velocity fields  for this regime from $320\,s$.  It can be observed how small hot plumes emerge  from the hot wall and strong variations of the thickness  occur on the thermal 
boundary layer. As we will see based on scaling arguments, this regime  is similar to the classic turbulent regime observed in high Rayleigh number convection.

\subsection{Evolution of the Rayleigh number}

\begin{figure}[!h]
\centering
 \includegraphics[scale=0.4]{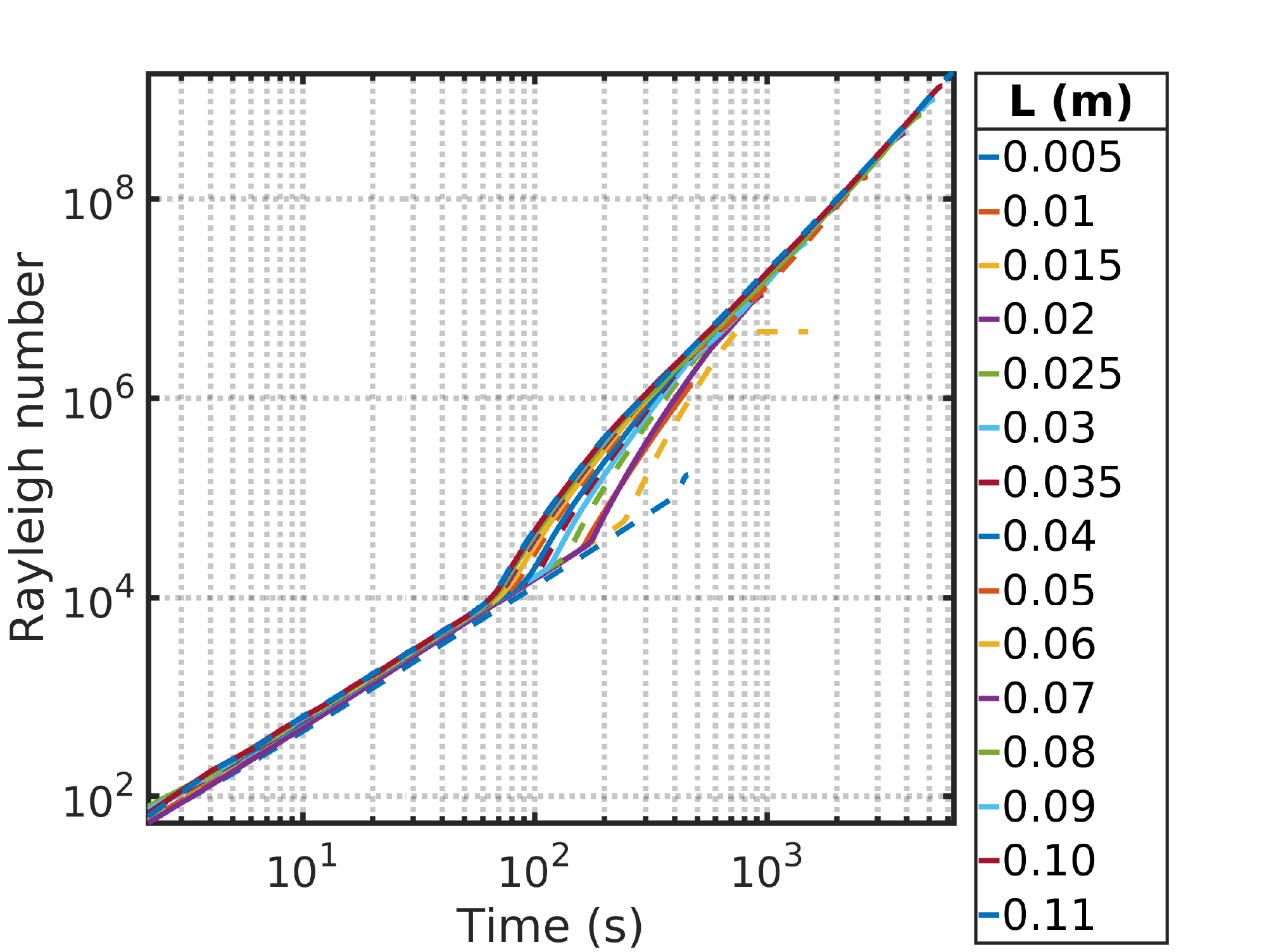}
\includegraphics[scale=0.4]{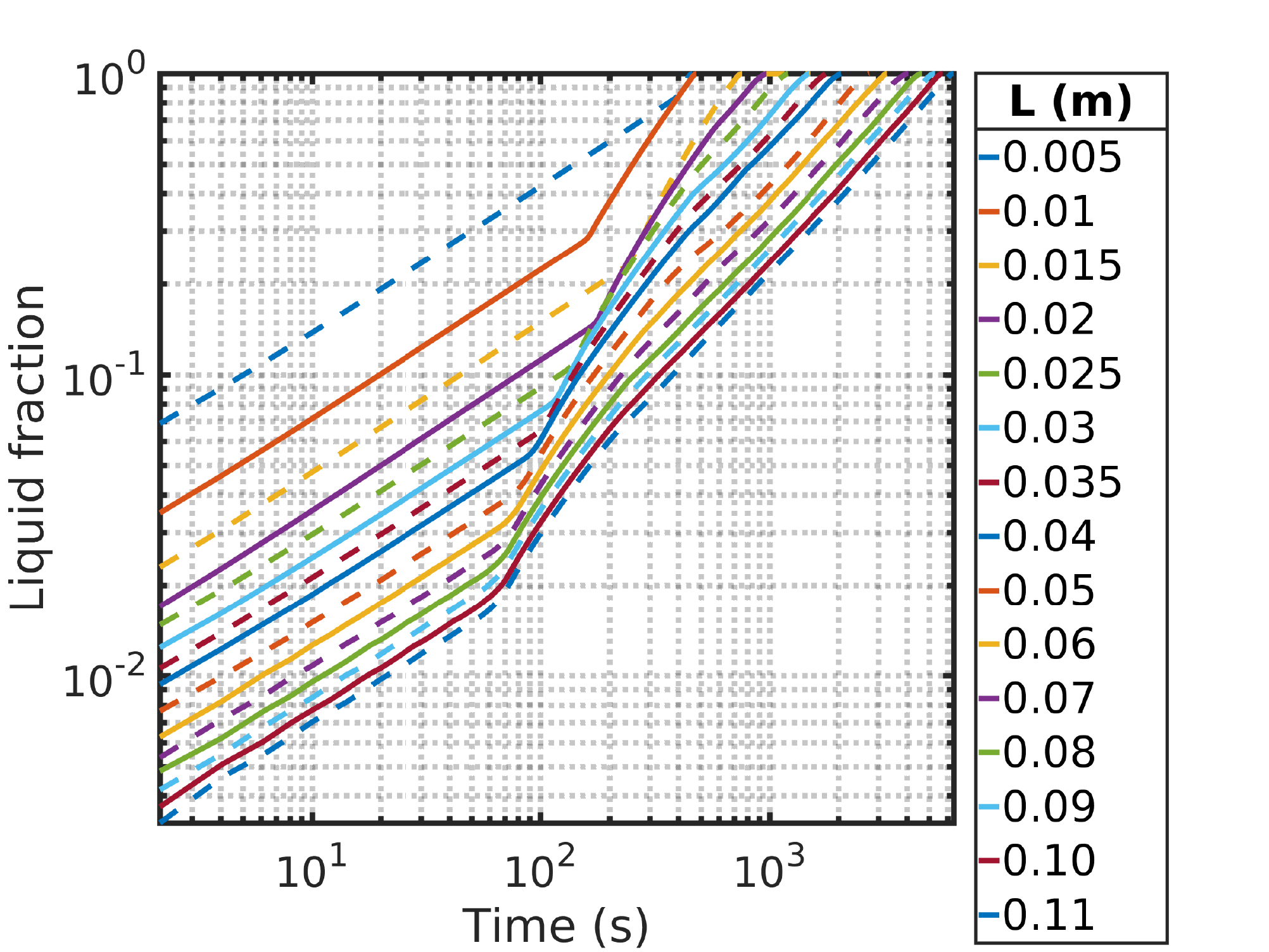}
  \caption{Rayleigh number as a function of time for all domain sizes (left). Liquid fraction as a function of time for all domain sizes considered (right)} 
\label{fig:Ra_vs_time}
  \end{figure}

We now characterize the regimes described previously with measures related to the Rayleigh number, the Nusselt number, the number of plumes and the thickness of the thermal and kinetic boundary layers.\\

Fig. \ref{fig:Ra_vs_time}(a) shows a log--log plot of the Rayleigh number as a function of the time for all domain sizes.  Every curve stops when melting of the 
PCM is  complete   for the corresponding domain size. The curves start as straight lines  in the log-log plot following a power law averaged over all 
the domain sizes  $Ra\sim t^{1.47}$. % exponente ok
The exponent provides  an averaged front height dependence on time as $h \sim t^{0.49}$.  % exponente ok
This region is 
within the conductive  melting regime, and this time
dependence of the Rayleigh number  extends  up to the beginning of the linear regime. After entering into the linear regime,  the simple power law of the conductive regime is lost and the melting rate is accelerated with a stronger increase of $Ra$ with time. Indeed, the linear regime with lengthening plumes barely interacting between them, exhibits the fastest advance of the solid/liquid interface during the melting of the PCM.

For domain sizes $L>0.035\,m$ 
the system is large enough to access the turbulent regime. Interestingly, at this regime a simpler power law behavior is recovered at the late stages of 
melting with averaged $Ra\sim t^{2.46}$. 
Fig. \ref{fig:Ra_vs_time}(b) shows a log-log representation of $Ra$ with respect to the liquid fraction  that 
allows to distinguish better the different 
melting regimes.  The growth of $Ra$ with time during the turbulent regime is lower than during the linear regime, but higher than in the  conductive regime. The smallest 
growth rate during the conductive regime indicates how the convective motion accelerate the melting rate. 
 The beginning of the coarsening and turbulent regime 
is not clearly identified from the $Ra$ curves, however we will show how measures based on the Nusselt number are more suitable to indentify these.

\begin{figure}[!h]

 a)\includegraphics[scale=0.45]{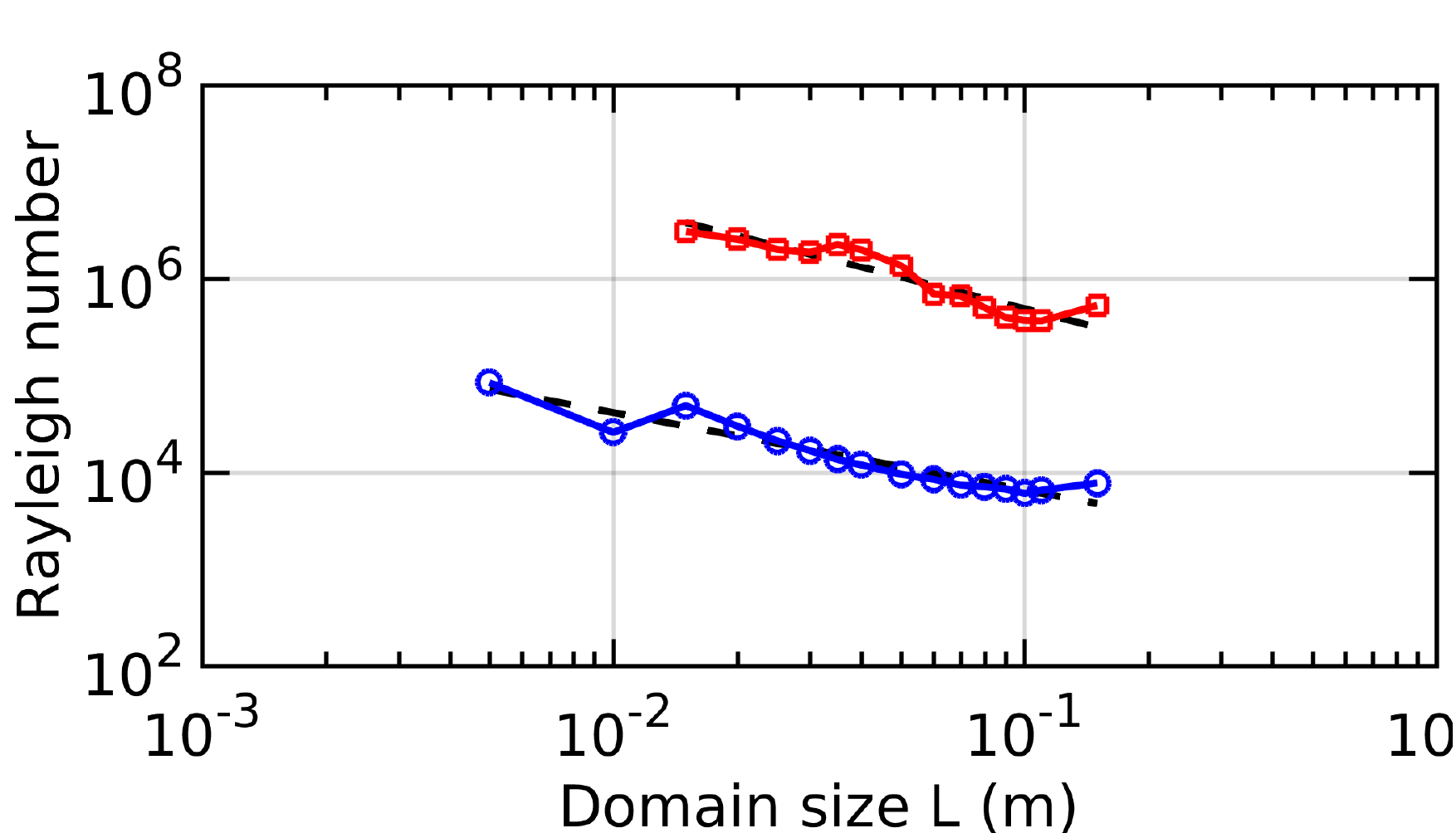}\quad \quad
 b)\includegraphics[scale=0.45]{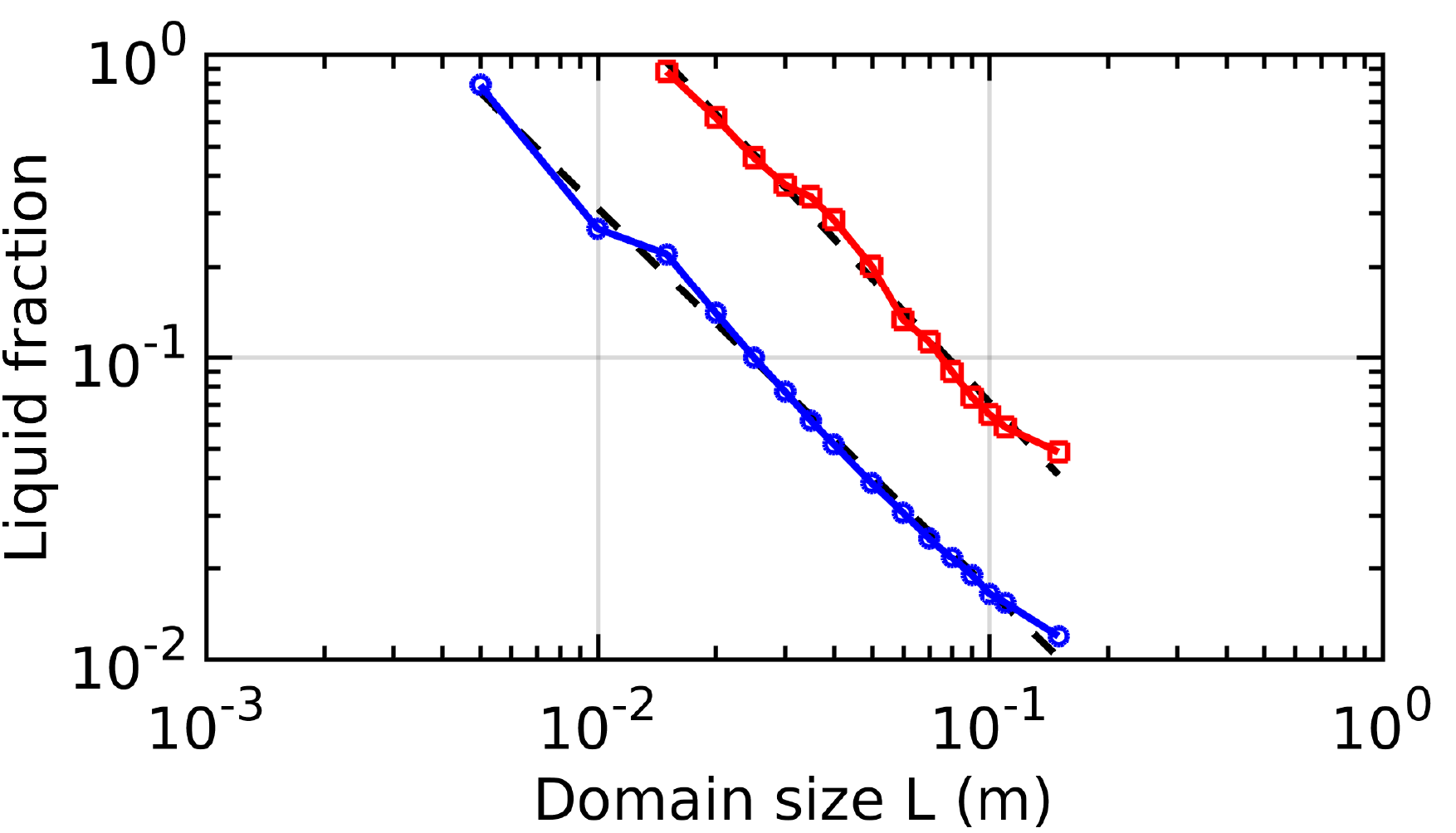}\\ \quad \\
 \centering
c)\includegraphics[scale=0.45]{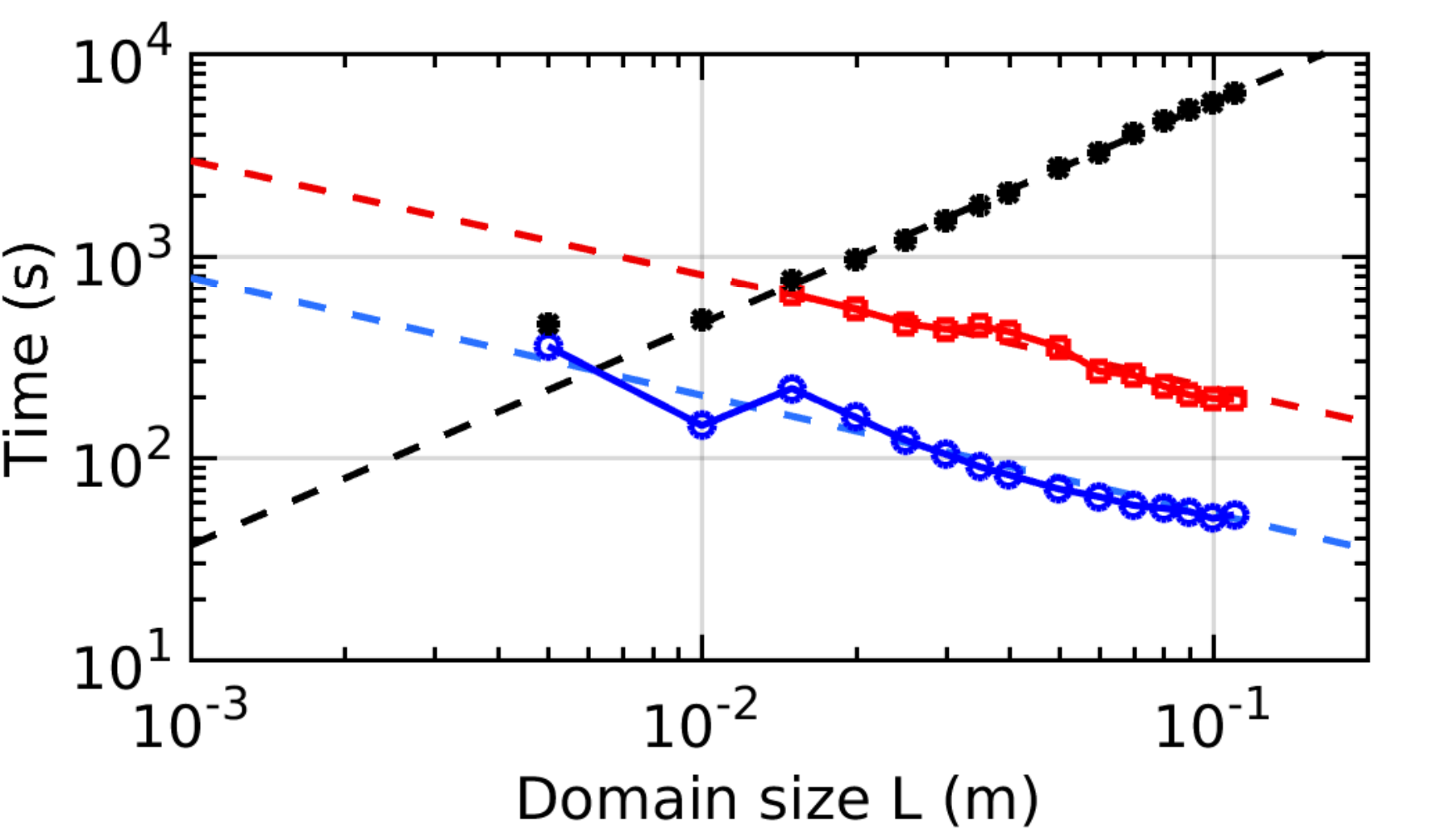}
 \caption{Left panel (a): Rayleigh number with respect to the domain size $L$ at the onset  the first (blue circles) and second (red squares) instabilities. 
The 
dashed black lines correspond to  power laws $Ra \sim L^{-0.85}$ and $Ra \sim L^{-0.87}$ for the first and second instability, respectively. 
\newline
         Right panel (b): Same thresholds represented with the overall liquid fraction, fitted with $f_{l}\sim L^{-1.31}$ and 
$f_{l}\sim \,L^{ -1.27}$
 for the first and second instability, respectively.  
Bottom panel (c): time of complete PCM melting as a function of the domain size (black dots). The black dashed line corresponds to a fit
$ t \sim L^{1.1}\,s$. The times of first and second instability are included as well as blue circles and red squares, respectively and
their corresponding fits are  $t \sim \,L^{-0.59 }$ and $t \sim  L^{-0.57}$.}
\label{F:Ra_time_instability-1and2}
 \end{figure}

The onset of the linear and coarsenining regimes is anticipated with  the domain size. Each point of Fig. \ref{F:Ra_time_instability-1and2}(a) corresponds to the Rayleigh number at the onset 
of the first (blue circles) and second (red squares) instabilities for each domain size $L$. The Rayleigh number at the onset of the the first instability 
follows a power law as $Ra \sim L^{-0.85}$%
while the onset of the coarsening regime can be fitted with $Ra \sim L^{-0.87}$,  indicating a similar dependence on the domain size of both regimes.
Likewise, the overall liquid fraction for the first and second instabilities follows as well a power law (c.f. 
Fig. \ref{F:Ra_time_instability-1and2}(b)), with $f_{l}\sim L^{-1.31 }$ for the first instability and 
$f_{l}\sim L^{-1.27}$ for the second.
A representation of the time of these instabilities is additionally illustrated in Fig. 
\ref{F:Ra_time_instability-1and2}(c), where is
added the time when complete PCM melting occurs. The time for complete melting follows $t\sim L^{1.1}$, indicating a progressively longer time for $L$ 
respect to a linear relationship. 
Thus a configuration of smaller domains containing the same total volume and subjected to the same boundary conditions would melt faster than a larger
domain with that  PCM volume.

In short, the Rayleigh number at the conductive and tuburlent regime exhibits power laws.  However, the linear regime, with a simple plume lengthening dynamics, exhibits the fastest melting rate
with a more involved  dependence of $Ra$ with time.  Besides, the onset of the first and second instabilities, and the total melting time exhibit  power laws with the domain side. 
The coarsening and onset of the turbulent regime are not clearly distinguished on the evolutive Rayleigh curves.

\subsection{Number of plumes}

\begin{figure}[!h]
\centering
 \includegraphics[scale=0.55]{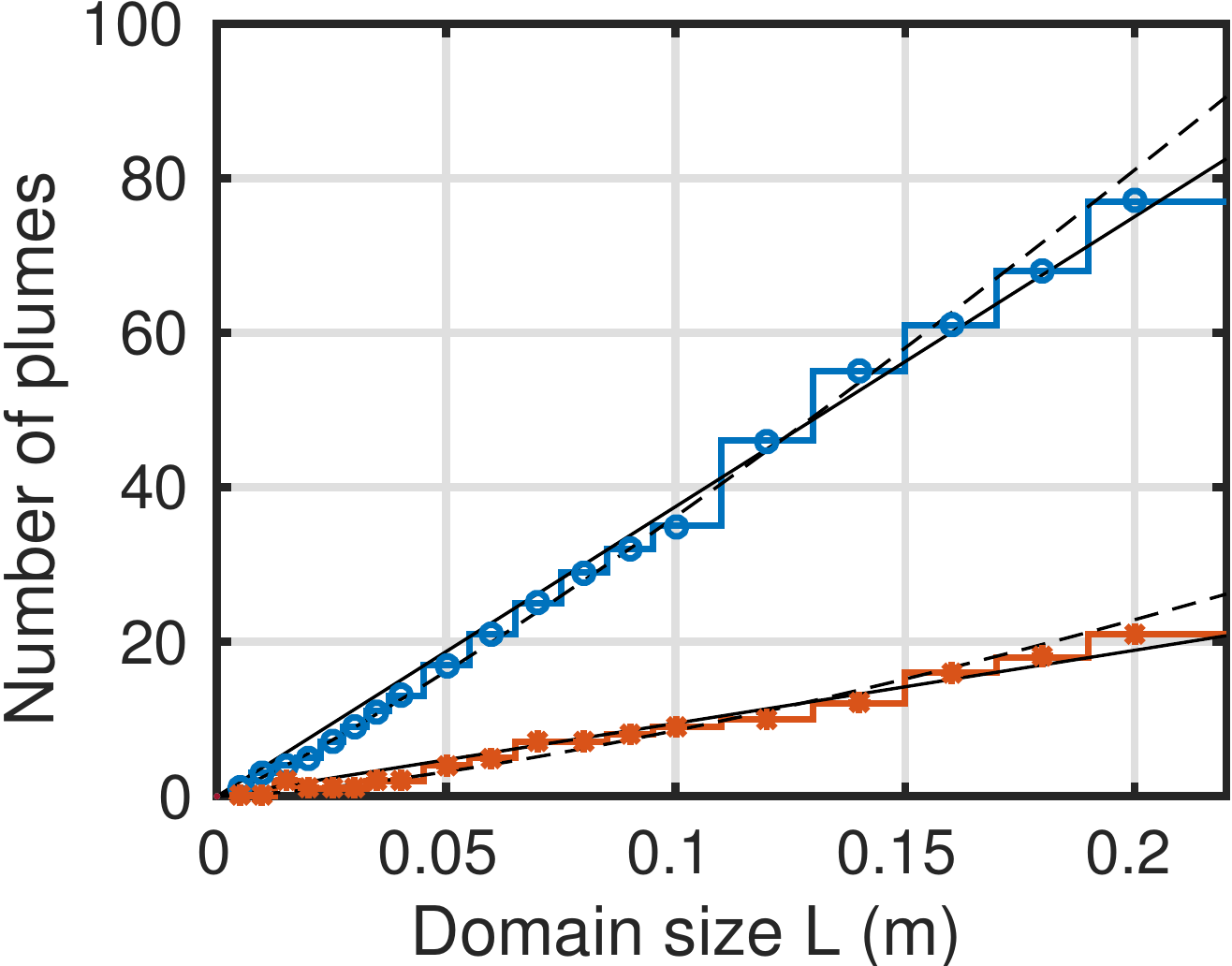}
\caption{Number of thermal plumes  after the primary (blue symbols) and secondary (red symbols) instability as a function of  the size of the domain in meters 
$L$. The solid lines correspond to the best linear fits $N_{primary}=375\, L$ and $N_{secondary}=94.54\, L$, and the dashed lines  correspond to the best power 
law fits  $N_{primary}=455.7\, L^{1.1}$ and $N_{secondary}=151.8\,L^{1.24}$. }
\label{F:size_Nplumes}
\end{figure}

The number of plumes emerging from the hot wall after the linear Rayleigh-B\'enard instability depends on the domain size. This number fits  a power law  
$N_{primary}=455.7\, L^{1.1}$   as seen  in Fig. \ref{F:size_Nplumes}.  Since this figure is only provided for a set of sizes, a 
step function (solid blue line) has been superposed to delimitate the  boundaries between  consecutive sizes.
 We have compared  the quality of fits between  straight lines and power laws. While for small $L$  the difference between  both fits is small, it becomes meaningful for large $L$ where the power law matches 
better the number of plumes.  
%Besides, panel \ref{F:size_Nplumes}(b) shows how the residuals of the power law are less biased along all the sizes. This 
%indicates a super-linear relationship between the number of plumes at the first instability. 
Furthermore, Fig. \ref{F:size_Nplumes} shows with squares the number of plumes after suffering the secondary instability and completed the coarsening  as a function of $L$. In this case, the difference between the linear and power law fits is smaller but the residuals for the  power law fit are less skewed. Thus the  power law 
%however a look at  Fig. \ref{F:size_Nplumes}(b)
matches better the numeric data  with  $N_{secondary}=151.8\,L^{1.24}$.   Comparing the exponents of these power  laws,   we 
obtain a weakening trend towards
the formation of plumes after the secondary instability with $L$. Unfortunately we have not identified an adequate number of cases for the secondary coarsening
to come to a conclusion whether this trend is obeyed for a tertiary instability.

%\begin{figure}[!h]
% \includegraphics[scale=0.55]{Figuras/size_Number_of_plumas_new.pdf}\quad \quad 
%\includegraphics[scale=0.55]{Figuras/residual_L_Number_of_plumas_new.pdf}
%\caption{Number of thermal plumes  after the primary (blue symbols) and secondary (red symbols) instability as a function of  the size of the domain in meters 
%$L$. The solid lines correspond to the best linear fits $N_{primary}=375\, L$ and $N_{secondary}=94.54\, L$, and the dashed lines  correspond to the best power 
%law fits  $N_{primary}=455.7\, L^{1.1}$ and $N_{secondary}=151.8\,L^{1.24}$. }
%\label{F:size_Nplumes}
%\end{figure}

\subsection{Evolution of the Nusselt number}

\begin{figure}[t]
 \includegraphics[scale=0.4]{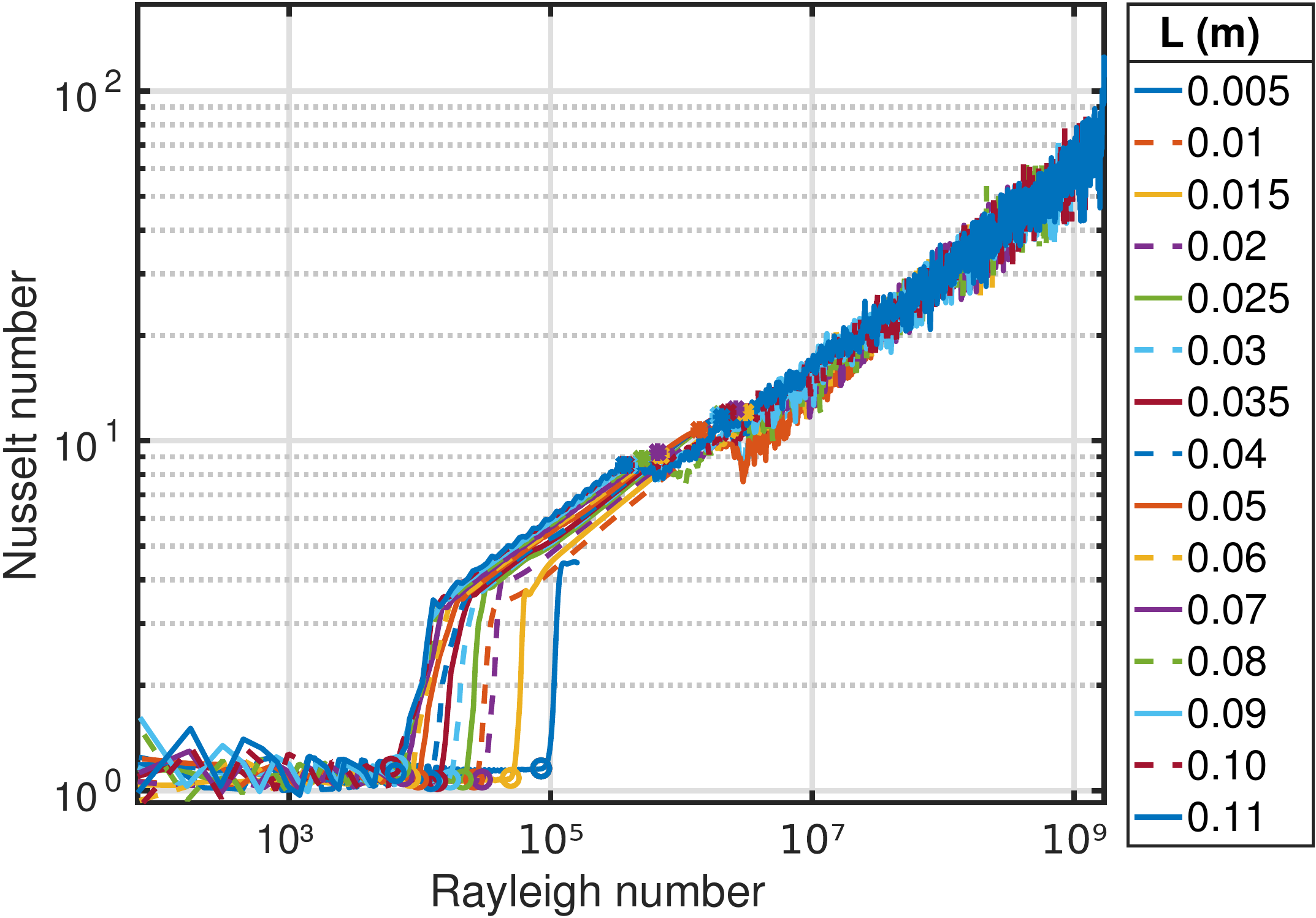}
 \includegraphics[scale=0.4]{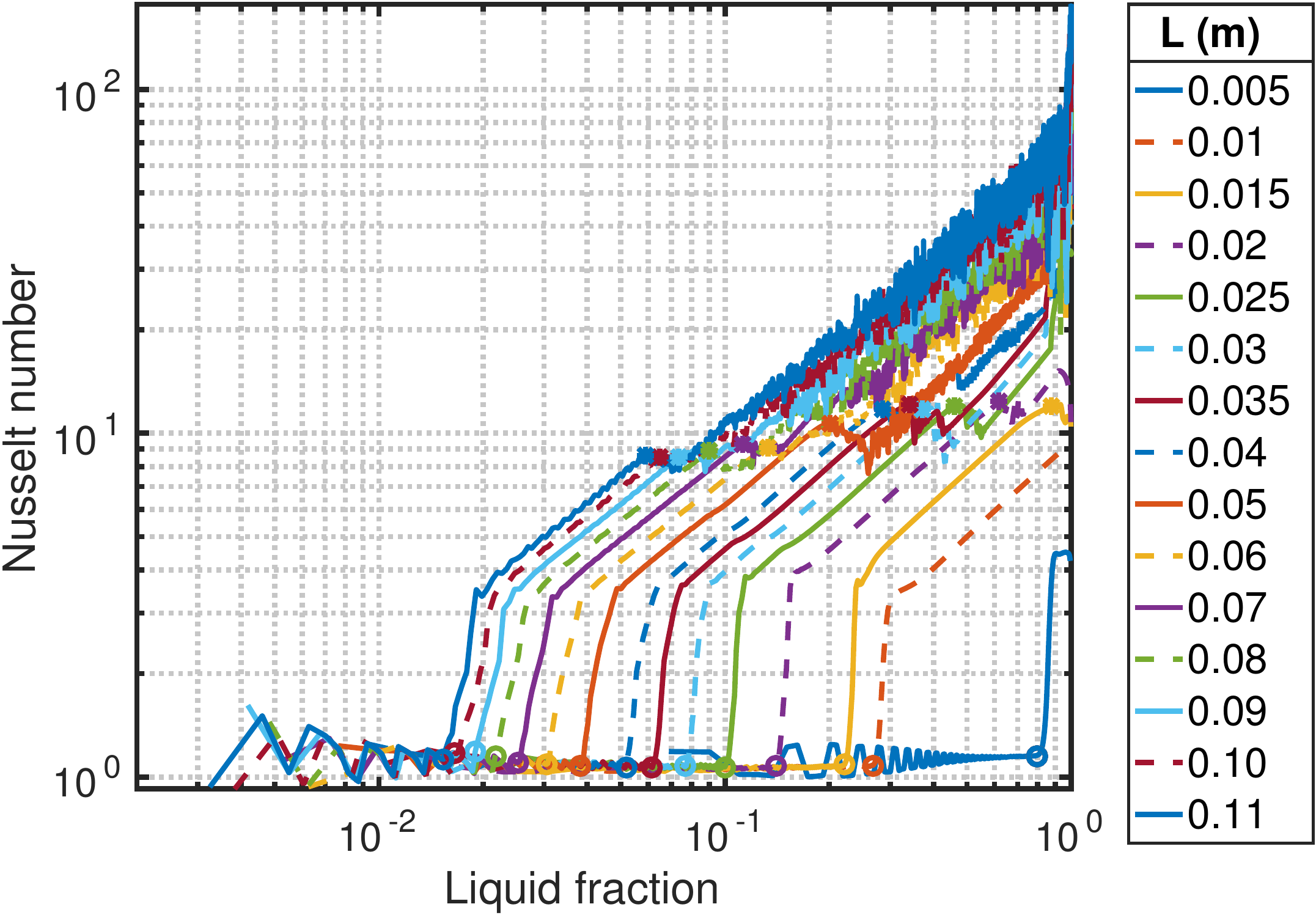}\\
\caption{ Nusselt number as a function of Rayleigh number for all the domain sizes considered in this work (left).
 Nusselt number as a function of the overall liquid fraction (right).
 Circles mark the threshold of the first instability and stars the second instability.
} \label{fig:Nu_vs_Ra}
\end{figure}

We investigate the relation between the heat flux  represented by the Nusselt number, which is calculated as described in Sec. \ref{sec:dimensional_numbers}, and the 
strength of buoyancy represented by the Rayleigh number in Fig. \ref{fig:Nu_vs_Ra}(a). 
Since $T_h$ is fixed and we have a moving solid/liquid interface, the curves $Nu$ v.s. $Ra$ are effectively $Nu$ v.s. $h^3$.
The conductive, linear, coarsening and turbulent regimes are remarkably  well distinguished in this representation of the melting dynamic. 
The quick destabilization of the conductive regime after the Rayleigh-B\'enard instability leading to the formation of the convective cells 
 exhibits the fastest rate of heat flux with a very strong dependency of the  Nusselt  number with the Rayleigh number. 
This destabilization was not captured clearly in the previous charts and appears in Fig. \ref{fig:Nu_vs_Ra} as almost vertical lines with  an averaged scaling of
 $Nu\sim Ra^{2.81}$.
 After the creation of the convective cells appears the linear regime following a power law with averaged smaller exponent $Nu\sim 
Ra^{0.29 }$. 

As time progresses,  the plumes begin to merge entering into the coarsening regime. This regime is short lived and surfaces in Fig. \ref{fig:Nu_vs_Ra} as a break of the previous
power law  with $Nu$ fluctuating for all $L$, without a clear trend.  However, $Nu$ v.s. $Ra$ representation, even better $Nu$ v.s. liquid fraction of Fig. \ref{fig:Nu_vs_Ra}(b), allows 
for a clear separation between the linear and coarsening regime. Table \ref{tab:exponents} lists the individual values of the exponents of the linear regime as a function of size.
For very small domain sizes $L<0.05\,m$ the effect of the confinement is strong and exponents are higher that once they settle to an average $Nu\sim Ra^{0.27}$ 
for $L>0.05\,m$, where the exponent is not strongly affected by the system size.   This exponent is slighlty lower than $2/7$ reported in a broad set of 
literature in high Rayleigh number convection \cite{Castaing1989,Shraiman1990,Cioni1997}. This is probably due to the fact that this regime
is characterized by the vertical development of plumes without meaningful interaction between them or chaotic behavior.

For large domains $L>0.035\,m$ the system supports new instabilities and the development of turbulent states. 
Nusselt curves at the turbulent states are characterized by high fluctuations, whose amplitude tends to escalate with $Ra$. In spite of these fluctuations, these curves 
follow a  power law  $Nu\sim Ra^{0.29}$, after averaging with all the domain sizes (c.f. Table \ref{tab:exponents}).  
This averaged exponent is slightly above than $2/7$, and get closer to this number for the largest domain sizes. 
Thus the turbulent state of the melted PCM exhibits a behavior
for the heat flux very close to body of literature on high Rayleigh convection that reports and exponent $2/7$ in theoretical and numerical models.
This is a remarkable result since the effect of heat release during melting and corrugated and moving  solid/liquid interface does not affect strongly to the scalinng between bouyoancy and heat flux. 

The exponent  $0.9$ found in this work agrees as well with a group of experimental results on convection at high $Ra$ numbers  \cite{Castaing1989,Chilla1993,Cioni1997,Glazier1999}.
Interestingly, a recent experimental article by Ditze and Scharf \cite{Ditze2017} on melting of ingots of pure metals, metal alloys and ice on their own melts  reports a scaling law as $Nu \sim Ra^{0.291}$. The exponent  meaningfully deviates from the predictions of literature based on lateral heating as emphasized by those authors. In this work, we find a numerical value of $Nu \sim Ra^{0.90}$ in strong agreement with this experimental results, indicating the relevance of vertical gradients of temperature in this poblem.

A representation of the Nusselt number with respect to the overall liquid fraction  provides a  clearer
separation between the different regimes (c.f. Fig. \ref{fig:Nu_vs_Ra}(b)).  We identify the threshold of  the first instability  with circles and the 
threshold of the second  instability  with stars 
%\SM{Put squares at the threshold of the turbulent states, not at the threshold of the coarsening regime} 
in Fig. \ref{fig:Nu_vs_Ra}(b).  As the overall liquid fraction for the geometry of this work corresponds to a bounded dimensionless form of the average 
thickness $h$, Fig. \ref{fig:Nu_vs_Ra}(b) provides the dependence of the heat flux with the position of the solid/liquid interface. As a consequence $Nu$ v.s. 
$f_l$ has the same exponents as discussed above for $Nu$ v.s. $Ra$  divided by three. \\

\begin{table}
\begin{center}
\begin{tabular}{c|cc}
 $Nu$ - $Ra$& \multicolumn{2}{c}{Exponents power laws}\\
\hline
     L (m)  &  Linear regime & Turbulent regime\\
\hline
     0.01   &   0.30  &       -\\
    0.015   &  0.30  &        -\\
     0.02   &  0.30   &       -\\
    0.025   &  0.30  &        -\\
     0.03   &  0.29  &        -\\
    0.035   &  0.28  &        -\\
     0.04   &  0.28  & 0.31\\% 0.43032\\
     0.05   &  0.27  & 0.31 \\%0.40888\\
     0.06   &  0.27  &  0.27\\%0.38613\\
     0.07   &  0.27  & 0.30 \\%0.36753\\
     0.08   &  0.27  & 0.29 \\%0.40635\\
     0.09   &  0.27  &  0.28 \\%0.35685\\
      0.1   &  0.27  & 0.29 \\%0.32094\\
     0.11   &  0.27  &  0.30 \\%0.3615\\
     0.15   &  0.26  &   0.28\\%0.26154\\
     \hline
     mean:  &    0.28 & 0.29
     
\end{tabular}

\end{center}
\caption{Exponent of the power laws of the linear regime (first column) and turbulent regime (second column).}\label{tab:exponents}
\end{table}

\subsection{Thermal and kinetic boundary layers}

The thickness of the thermal boundary layer  $\delta_T$ can be defined in various ways. 
We define the thickness $\delta_T$ as the distance between the intersection of the tangent of the horizontally averaged temperature profile at the hot wall $\bar{T}=T_h+\left.d\,T/d\,y\right|_{y=0}\,y$ with the mean temperature  between the hot wall and the interface   $T_m=\frac{T_h+(T_s+T_l)/2}{2}$. Notice how the temperature of the interface is fixed at $(T_s+T_l)/2$, but its position is moving upwards, 
providing a $T_m$ that depends on increasingly distant points.   As pointed out in Ref. \cite{Ng2015} these
definitions roughly correspond to the crossover point between mean dissipation and turbulent dissipation.  From the definition of the averaged Nusselt number in Sec. \ref{sec:dimensional_numbers} the thickness of the thermal boundary layer can be expressed as well as 
\be
\delta_T=\frac{h}{2\,Nu}
\ee
which is a more practical expression if the Nusselt curves are known.

\begin{figure}
\includegraphics[scale=0.4]{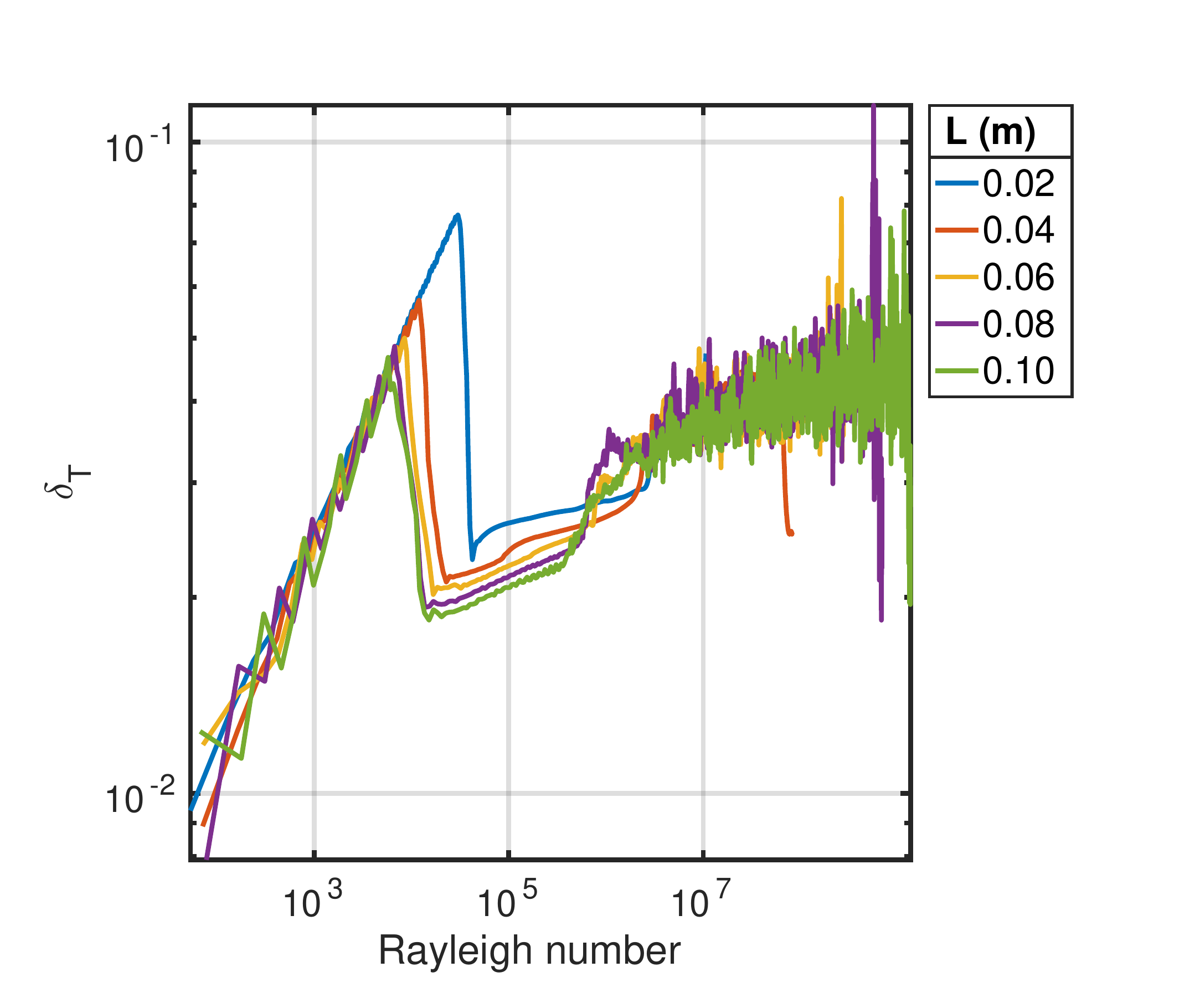}
\includegraphics[scale=0.4]{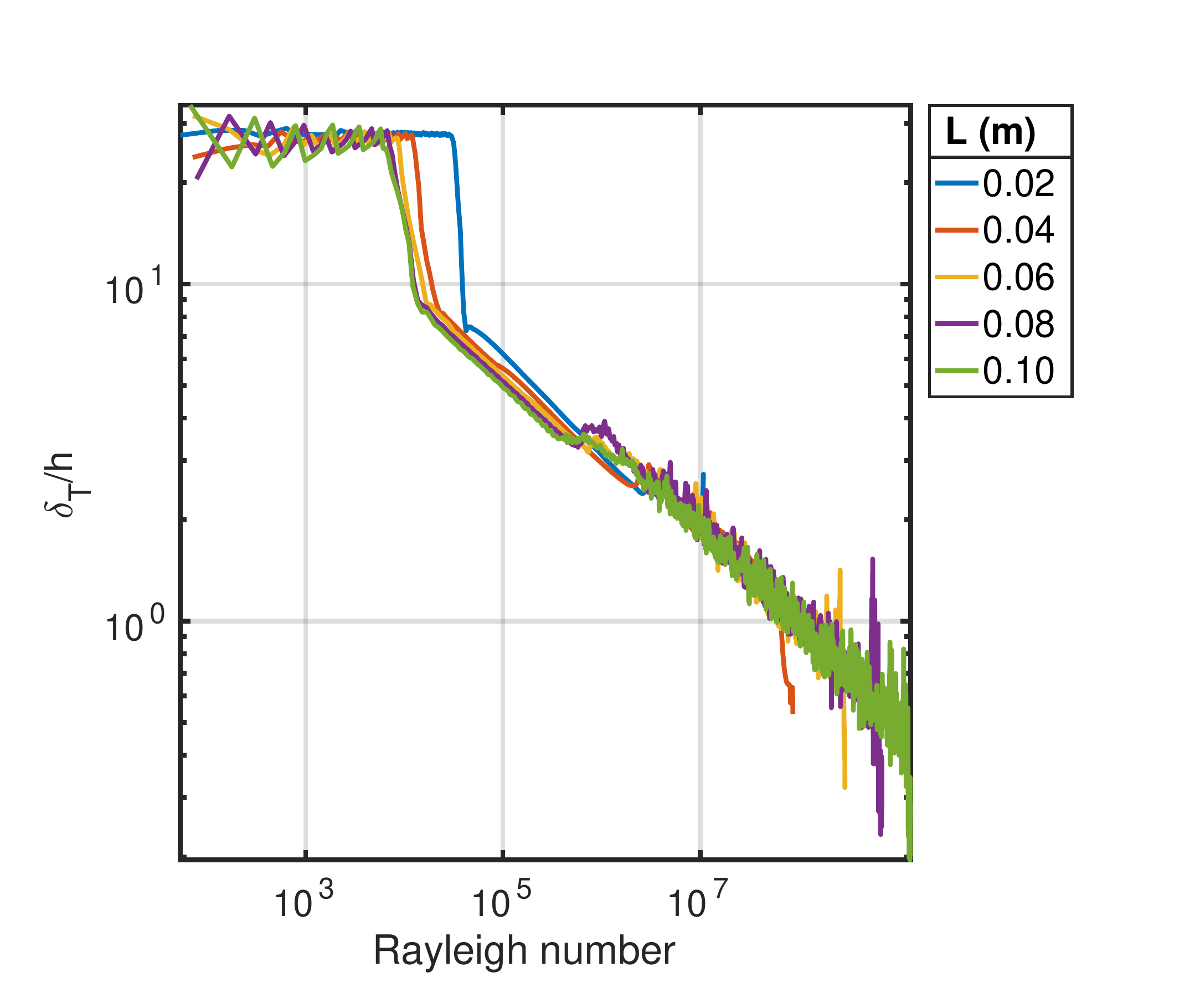}
\caption{ Left: Thickness of the thermal boundary layer $\delta_T$ as a function of the Rayleigh number in log-log scale.
          Right:   Thickness of the thermal boundary layer $\delta_T$ scaled by the average depth of the melted PCM  $h$ as a function of the Rayleigh number in log-log scale.
} \label{fig:thermal_bl_thickness}
\end{figure}

The thickness of the thermal boundary layer $\delta_T$ on the hot wall as a function of $Ra$  for different $L$ is plotted in  Fig. \ref{fig:thermal_bl_thickness}(a).  The curves display clearly the four melting regimes:
(i) the conductive regime appears as almost  straight lines with positive slope in the log-log plot, (ii) the destabilization of the conductive regime produces a strong decrease of $\delta_T$ due to the quick enhancement of the heat flux at the bottom, (iii) the linear regime with its characteristic high heat flux exhibits the lowest values for $\delta_T$ once convection sets in, (iv) the turbulence shows very large fluctuations of $\delta_T$ with increasing amplitude when $Ra$ growths. 

Interestingly, on the contrary to the  problem turbulent Rayleigh-B\'enard convection, where the liquid domain is fixed,  the value of $\delta_T$ for the melted 
PCM increases with $Ra$ instead of decreasing  \cite{Ahlers2009}. Thus, for instance, we obtain $\delta_T \sim Ra^{0.05}$ for $L=0.1\,m$. 
However, we can compare with the classic Rayleigh-B\'enard convection representing $\delta_T$  normalized with the average  depth of the melted PCM, i.e.  $\delta_T/h$ 
v.s.  $Ra$, as shown in Fig. \ref{fig:thermal_bl_thickness}(b). Now  we obtain decreasing curves with $Ra$, which follow power laws at the turbulent regime. For 
instance,  $\delta_T/h \sim Ra^{-0.29}$  for $L=0.10\,m$. This value is above the Prandtl-Blassius theoretical prediction of $-0.25$. However, it agrees with values of numerical simulations of $-0.29$ for $Pr=0.7$ in a cylindrical cell with aspect ratios $1$  within a  lower range de $Ra$ used in this work \cite{Verzicco1999}, and for with $-0.31$ obtained for an aspect ratio $1/2$ with a wider range of $Ra$ \cite{Verzicco2003}. There are as well experimental
works in water that report an exponent $-0.33$ \cite{Sun2008}, above as well of  Prandtl-Blassius  theory. 

\begin{figure}[!h]
\includegraphics[scale=0.4]{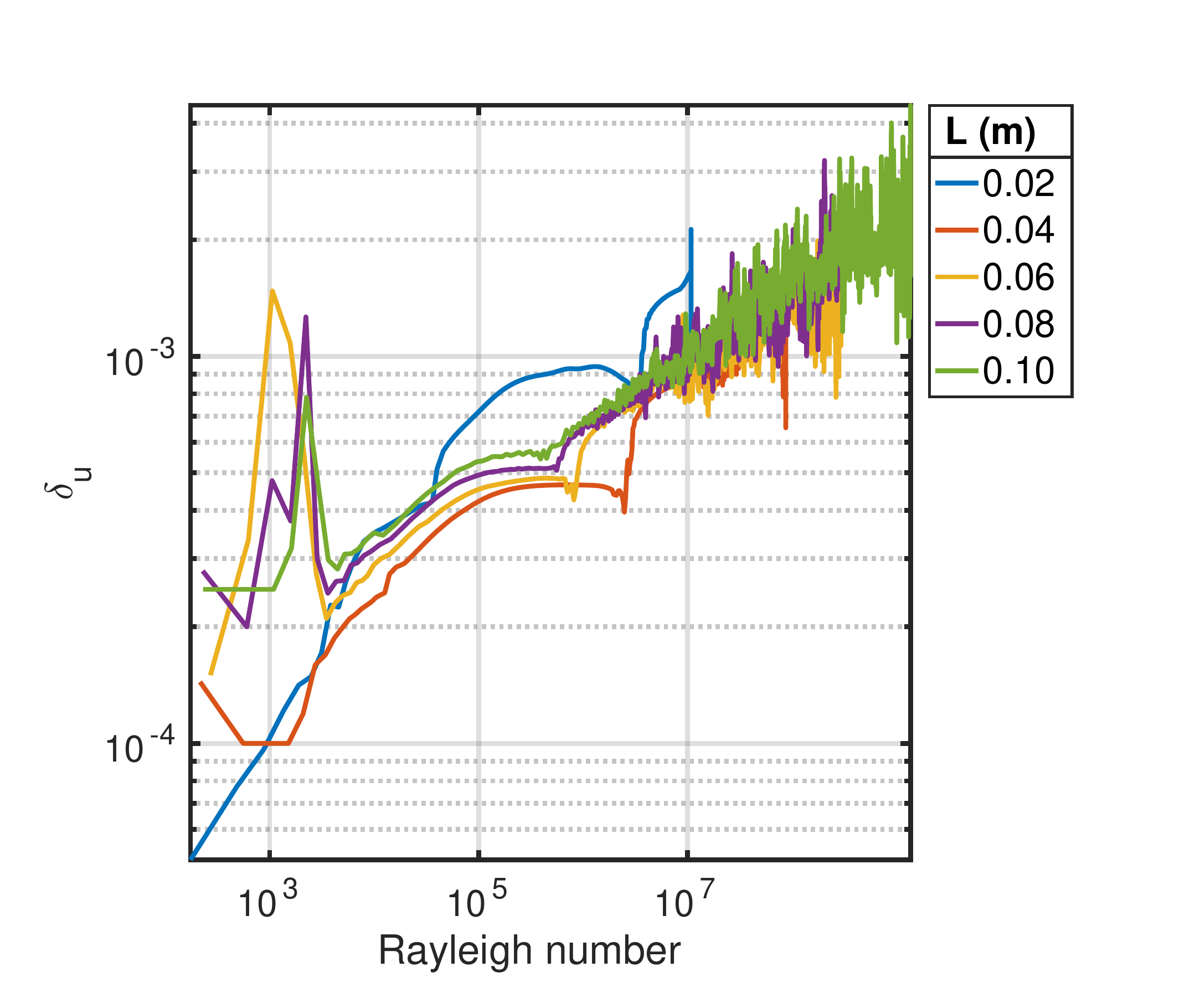}
\includegraphics[scale=0.4]{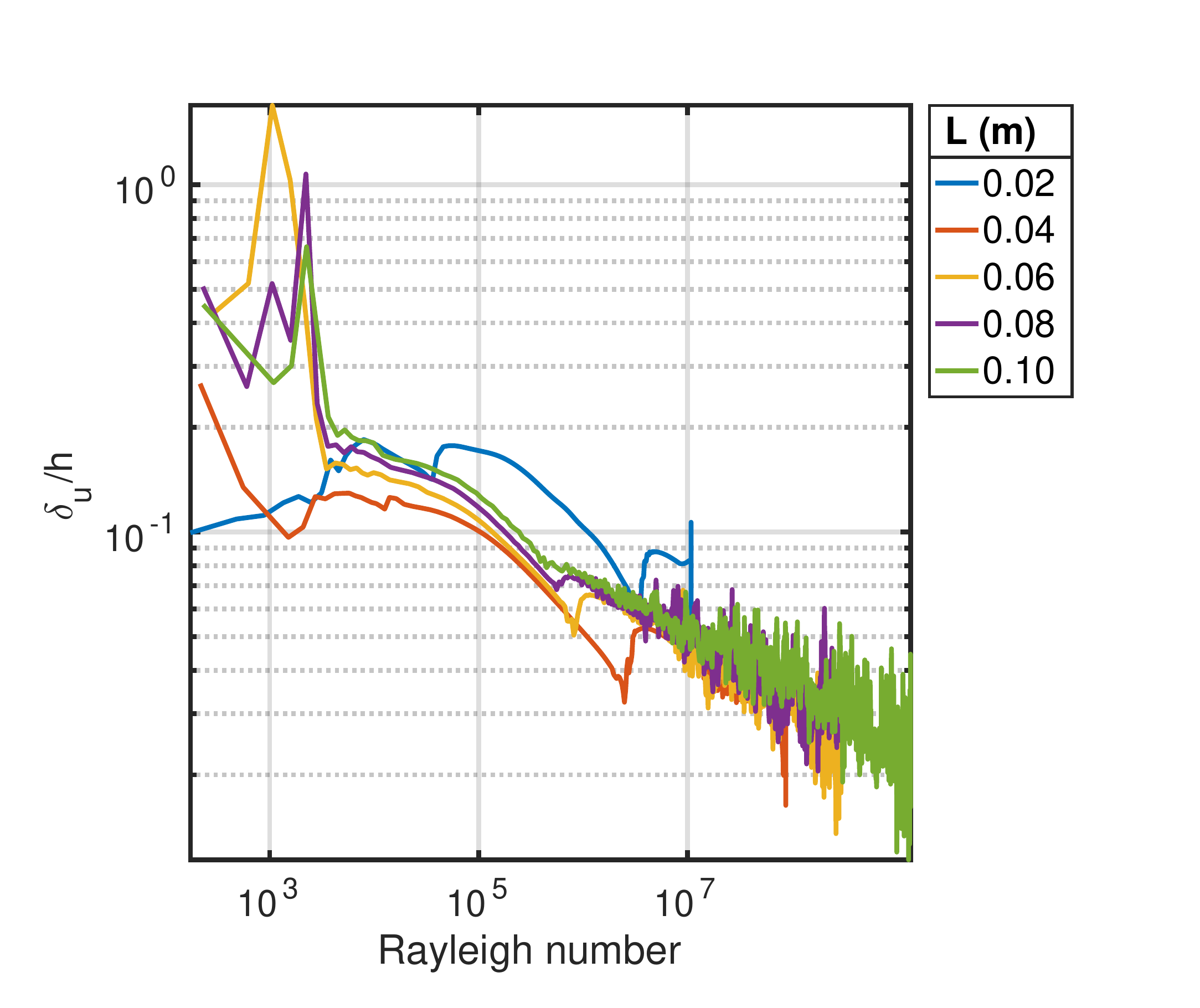}\\
\caption{Left: Thickness of the kinetic boundary layer $\delta_u$ from $L=0.02\, m$ to $L=0.1\, m$ as a function of the Rayleigh number in log-log scale. The kinetic boundary 
layer is defined as the wall distance to the intercept of $max(\overline{u})$ and $\left.\frac{d\,\overline{u}}{dy}̣\right|_{y=0}y.$
        Left: Thickness of the kinetic boundary layer $\delta_u$  scaled by the average depth of the melted PCM  $h$ as a function of the Rayleigh number in log-log scale. 
} \label{fig:kinetic_bl_thickness}
\end{figure}

We follow  \cite{Ng2015} to define the thickness of the kinetic boundary layer $\delta_u$ as the  distance from the hot wall to the intercept of 
$\overline{u}=\left. \frac{d\,\overline{u}}{d\,y}\right|_{y=0}\,y$ and  $\overline{u}=max(\overline{u})$, where $\overline{u}$ is the spatial average along the 
horizontal axis of the module of the velocity. 
We observe in Fig. \ref{fig:kinetic_bl_thickness}(a)  how $\delta_u$  and its fluctuations increase with $Ra$; similarly to $\delta_T$ in the turbulent regime. 
However, $\delta_u/h$ v.s $Ra$ decreases, following a power law in the turbulent regime, for example  $\delta_u/h\sim Ra^{-0.16}$ for $L=0.1\,m$. This 
scaling does not agree with the prediction of an exponent $-0.25$ from the Prandtl-Blasius theory,  or with numerical and experimental results in very low 
Prandtl fluids that exhibit exponents close to Prandtl-Blasius \cite{LiShi2012,Scheel2012}. However, we find good agreement between our results in some 
experiments for liquids in cylindrical and cubic geometries covering a wide range of Prandtl numbers, reporting an exponent about $-0.16$ for the  kinematic boundary layer at 
the bottom plate \cite{XinXia1996,XinXia1997,Lam2002}.  Similarly to the excellent agreement for $\delta_T/h$,   Verzicco {\em et al.}  \cite{Verzicco1999} 
obtain an exponent $-0.18 $ for the kinematic boundary layer in simulations with $Pr=0.7$.

\section{Summary}\label{S:4}

We have presented numerical results for the melting dynamics of n-octadecane within square geometries with periodic boundary conditions along the horizontal direction heated from below with sizes ranging from $0.005\,m$ to 
$0.2\,m$. 
The n-octadecane is a viscous paraffin with Prandtl $Pr=60.3$, and we have chosen a Stefan number for the melting process of $0.49$, and reached Rayleigh numbers up to $\sim 10^9$. 
We have carried out simulations using a finite volume code with a grid able to resolve the thermal and kinetic boundary layers above the hot wall. 
The most appealing feature of melting in this geometry is that we scan a very wide range of Rayleigh numbers for the liquid phase of the PCM, from characteristic values  of 
conductive transport to values associated with turbulence, without changing any external parameter of the problem.

We identify four different regimes as time evolves: (i) conductive regime, (ii) linear regime, (iii) coarsening regime and (iv) turbulent regime. 
Notice that Jany and Bejam \cite{Jany1988a}  described four regimes in melting induced by lateral heating,
and we have reported in this work as well four different regimes but with the coarsening regime not appearing in  lateral melting.
The first two regimes appear at all domain sizes. However the third and fourth regimes require a long advance of the solid/liquid interface to develop, and we observe them only for
$L>0.035\,m$. Interestingly, the transition to turbulence takes places with a deformation and coarsening of the plumes, up  to recover an aspect 
ratio close to that of the cells appearing after the Rayleigh-B\'enard instability.
  For domain size $L=0.04\,m$ we observe a second coarsening of plumes before reaching the turbulent regime. This suggests that we can 
expect, depending on factors such as size, initial conditions, boundary conditions, etc. situations in which the transition to turbulence  take place through a 
coarsening cascade.

Each one of the regimes creates a distinctive front. Thus the conductive regime generates a front whose shape is dictated by the geometry. The second regime leads to the formation of a periodic front whose peaks correspond to the position of plumes. The coarsening regime induces longer periodic variations with higher amplitude. Finally,  the turbulent regime leads to a very irregular
solid/liquid interface with  strong fluctuations of the amplitude.

We observe that most of the representations of the melting process are ruled by power laws, although not all of them. Thus the number of plumes, some regimes of the Rayleigh number as a function of time, the number of plumes after the primary and secondary instability, the onset of the first and second instabilities, the thermal and kinetic boundary layers at the turbulent regime follow simple power laws. 

As turbulence is well established we find an averaged relation between the Nusselt and Rayleigh number as $Nu \sim Ra^{0.29}$.  This exponent
fits well with a broad body of numerical and experimental works of the classic turbulent Rayleigh-B\'enard literature. Remarkably, we find this agreement in spite
of the moving solid/liquid interface that becomes highly corrugated in the turbulent regime, and the varying aspect ratio of the depth of melted PCM with respect to the horizontal size
when melting progress. Other turbulent exponents, like the scaling of the thermal and kinematic boundary layers, $\delta_T/h \sim Ra^{-0.29}$ and  $\delta_u/h\sim Ra^{-0.16}$ respectively,  do not fit well with the predictions of the 
Blasius-Prandtl scaling. However, we notice that this disagreement appears as well  in other numerical and experimental works that report similar values to this work. 

Finally, while a comprehensive study on the onset of the coarsening, dependence on system size, and plumes statistics has been carried out,
   the relationship between aspect ratio and trigger of this secondary instability suggests a destabilization of the liquid column within the plumes
   due to gravity forces as the responsible of this instability. However a precise mechanism of this instability is lacking and is  a direction 
   of future work. Also, this work has been carried out in 2D geometries, and whether the scaling exponents  are held or not in  3D geometries is still an open question that deserves further study.

\section*{Acknowledgement}

Santiago Madruga acknowledges support by Erasmus Mundus EASED programme (Grant 2012-5538/004-001) coordinated by Centrale Sup\'elec, the Spanish {\em Ministerio
  de Econom\'{\i}a  y Competitividad} under Projects No.  TRA2016-75075-R, No. ESP2013-45432-P and No. ESP2015-70458-P, and  the computer resources and technical assistance provided by the Centro de Supercomputaci\'on y Visualizaci\'on de Madrid (CeSViMa).
 
Jezabel Curbelo  acknowledges support by the MINECO under grant No. MTM2014-56392-R and ICMAT under  Severo Ochoa project No. SEV-2011-0087, and ICMAT for computing facilities.

%\bibliography{local}
\bibliographystyle{unsrt}
\bibliography{MadrugaCurbelo_resubmitted}
%\bibliography{/home/santiago/.bibfiles/library_MadrugaJezabel,/home/santiago/.bibfiles/pcm,/home/santiago/.bibfiles/madruga}
%\bibliography{bib_MaCu17}

\end{document}